\documentclass[varenna]{cimento2} 
\usepackage{epsfig}
\def\lambdabar{\protect\@lambdabar}
\def\@lambdabar{%
\relax
\bgroup
\def\@tempa{\hbox{\raise.73\ht0
\hbox to0pt{\kern.25\wd0\vrule width.5\wd0
height.1pt depth.1pt\hss}\box0}}%
\mathchoice{\setbox0\hbox{$\displaystyle\lambda$}\@tempa}%
{\setbox0\hbox{$\textstyle\lambda$}\@tempa}%
{\setbox0\hbox{$\scriptstyle\lambda$}\@tempa}%
{\setbox0\hbox{$\scriptscriptstyle\lambda$}\@tempa}%
\egroup
}

\title{Making, probing and understanding Bose-Einstein condensates}
\author{W. Ketterle, D.S. Durfee, 
     \atque D.M. Stamper-Kurn}
\institute{Department of Physics and Research Laboratory of
Electronics, \\
Massachusetts Institute of Technology, Cambridge, MA 02139}
\PACSes{}

\begin{document}

\maketitle

\section{Introduction}

The realization of Bose-Einstein condensation (BEC) in dilute atomic gases 
\cite{ande95,davi95bec,brad97bec,frie98,gsuhomepage2} achieved several
long-standing goals.
First, neutral atoms were cooled into the 
lowest energy 
state, thus exerting ultimate control over the motion and position of atoms, 
limited only by 
Heisenberg's uncertainty relation. 
Second, a coherent macroscopic 
sample of atoms all 
occupying the same quantum state was generated,
leading to the realization of atom lasers, devices 
which generate 
coherent matter waves.  
Third, degenerate quantum gases were produced 
with properties 
quite different 
from the quantum liquids $^{3}$He and $^{4}$He.  This provides a testing ground 
for many-body 
theories of the dilute Bose gas which were developed many decades ago but never 
tested 
experimentally \cite{huan64}.  BEC of dilute atomic gases is a macroscopic 
quantum phenomenon 
with similarities to superfluidity, superconductivity and the laser 
\cite{grif95}.

The rapid development over the last few years in the BEC field has been 
breathtaking.  It took 
only one year to advance from the first observations of evaporative cooling in 
alkalis to the 
achievement of BEC. Since then, the developments have exceeded even our most 
optimistic 
expectations.  Almost every month, new topics related to BEC emerge making BEC 
more than just a 
phenomenon of statistical physics: it is a new window into the quantum world.  
This excitement 
was felt throughout the ``Enrico Fermi'' summer school.

At Varenna, one of the authors gave four lectures
describing the experimental techniques used to study BEC, and the results 
obtained so far.  
These notes cover the same topics.  However, rather than giving a full account 
of this 
burgeoning field, we have selected a few aspects which we treat in depth.  This 
includes an 
extensive treatment of magnetic trapping (sect.~\ref{sec:magtrap}), 
various imaging techniques 
(sect.~\ref{probesection}),
image analysis (sect.~\ref{dataanalysis}), and a comprehensive discussion 
of sound 
(sect.~\ref{sound})  and coherence (sect.~\ref{coherencesection}).
Many aspects and details of these 
sections cannot be 
found elsewhere.  Other topics are covered in less detail, and we will refer the 
reader to the 
other contributions in this volume as well as to other relevant literature.

A major part of this review paper is based on our previous research publications 
and several 
previous review papers: ref.~\cite{kett96phsc} 
summarizes the first 
experiments with Bose condensates, refs.~\cite{vand96,town96tops,town97icap} 
contain a more 
complete account on the cooling and trapping techniques and describe the 
progress during the 
summer of 1996, ref.~\cite{andr98} includes the 
experiments on sound 
propagation and the atom laser, ref.~\cite{sten98odt} describes experiments 
done with 
optically trapped Bose condensates, and ref.~\cite{mies98rev} gives an 
overview of 
techniques and experiments performed through the end of 1997.  The 
technique of evaporative 
cooling is reviewed in \cite{kett96evap}.  
Refs.~\cite{kett96leos,town97phwo,durf98,kett99atla} are more popular papers, with 
ref.~\cite{durf98} containing many animated movies of experimental data and 
ref.~\cite{kett99atla} 
discussing the concept of an atom laser.

\subsection{Basic features of Bose-Einstein condensation}

BEC in an ideal gas, described in various textbooks (e.g. \cite{huan87}), is a 
paradigm of 
quantum statistical mechanics which offers profound insight into macroscopic 
quantum 
phenomena.  We want to focus here on selected aspects of BEC 
pertaining to 
current experiments 
in trapped Bose gases.

\subsubsection{Length and energy scales}

\begin{figure}
\epsfxsize = 7cm
\centerline{\epsfbox[52 101 471 590]{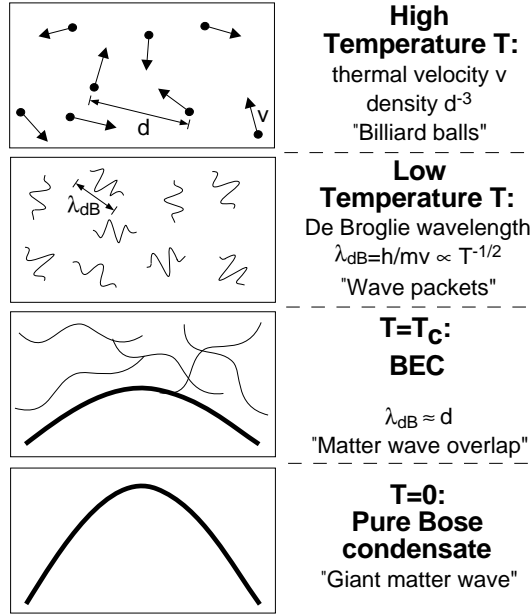}}
\caption[Criterion for Bose-Einstein condensation]
	{Criterion for Bose-Einstein condensation.  At high 
temperatures, a weakly interacting gas can be treated as a system of 
``billiard balls.''  In a simplified quantum description, the atoms can 
be regarded as wavepackets with an extension $ \lambda_{dB}$.  
At the BEC transition temperature, 
$\lambda_{dB}$ becomes comparable to the distance between atoms, and a 
Bose condensate forms.  As the temperature 
approaches zero, the thermal cloud disappears leaving a pure 
Bose condensate.}
\label{whatisbec}
\end{figure}

Bose-Einstein condensation is based on the indistinguishability and wave nature 
of particles, 
both of which are at the heart of quantum mechanics.  In a simplified picture, 
atoms in a 
gas may be regarded as quantum-mechanical wavepackets which have an extent on 
the order of a 
thermal de Broglie wavelength $\lambda_{dB} = (2 \pi \hbar^{2} / m k_B 
T)^{1/2}$ 
where $T$ is the 
temperature and $m$ the mass of the atom.  $\lambda_{dB}$ can be regarded as the 
position 
uncertainty associated with the thermal momentum distribution.  The lower the 
temperature, the 
longer $\lambda_{dB}$.  When atoms are cooled to the point where 
$\lambda_{dB}$ is 
comparable to the interatomic separation, the atomic wavepackets 
``overlap'' and the 
indistinguishability of particles becomes important (fig.~\ref{whatisbec}).  
At this temperature, bosons undergo a 
phase transition and form a Bose-Einstein condensate, a coherent cloud of atoms 
all occupying 
the same quantum mechanical state.  The transition temperature and 
the peak 
atomic density $ n $ are related as $n \lambda_{dB}^{3} \simeq 2.612$.

Bose-Einstein condensation in gases allows for a ``first-principles'' 
theoretical description 
because there is a clear hierarchy of length and energy scales (table 
\ref{tab:lengthscales}).  In a gas, the 
separation 
between atoms $n^{-1/3}$ is much larger than the size of the atoms 
(characterized by the s-wave 
scattering length $a$), i.e.\ the quantity $na^{3} \ll 1$.  In a Bose condensed gas, 
the separation 
between atoms is equal to or smaller than the thermal de Broglie wavelength.  
The largest length 
scale is the confinement, either characterized by the size of the box potential 
or by the 
oscillator length $a_{HO} = \sqrt{\hbar / m \omega}$ which is the size of the 
ground 
state wavefunction in a harmonic oscillator potential with 
frequency $ \omega $.

With each length scale $l$ there is an associated energy scale which is the 
kinetic energy of a 
particle with a de Broglie wavelength $l$.  The energy scale associated with the 
scattering 
length is the temperature below which s-wave scattering predominates.  The 
energy scale 
associated with the separation between atoms, $n^{-1/3}$, is the BEC transition 
temperature.  
The energy associated with the size of the confining potential is the energy 
spacing between 
the lowest levels.

Atom-atom interactions are described by a mean field energy $U_{int} = 4 \pi 
\hbar^2 n a /m$.  
The length scale associated with this energy is  
the healing length
$\xi= (8 \pi n a)^{-1/2}$.  
In most experiments, $k_{B}T>U_{int}$, but the 
opposite case has 
also been realized~\cite{stam98rev,sten98odt}.  In comparison, superfluid helium is a 
strongly 
interacting quantum liquid --- the size of the atom, the healing length, the 
thermal de Broglie 
wavelength and the separation between the atoms are all comparable, creating a 
complex and rich 
situation.

\begin{table}
\caption[Energy and length scales for trapped gaseous 
Bose-Einstein condensates]{Energy and length scales for trapped gaseous 
Bose-Einstein condensates.  The hierarchy of energy and length scales 
simplifies the description of these quantum fluids.
For each energy $E$, we define a length $l$ by the relation
$E = \hbar^{2} / 2 m l^{2}$ and indicate the relation between $l$ and 
a common length scale.
Numbers are typical for sodium BEC experiments.
The values for the mean-field energy assume a density of $\sim 
10^{14} \un{cm}^{-3}$.}
\begin{tabular}{|c@{}c@{}|@{}c@{}c@{}l|}
\hline
\bf{Energy Scale} $\mathbf{E}$ & $ {= {h^2} / {2 m l^2 
}}$  \mbox{ \ } & 
	\multicolumn{3}{c}{\bf{Length Scale} }  \\
\hline 
\parbox{1.3in}{ \begin{center} limiting temperature for s-wave scattering 
\end{center} }
& 1 mK &
	\parbox{1.3in}{ \begin{center} 
	scattering length 
	\end{center} }
	& $ a = l / 2 \pi $ & = 3 nm \\
\parbox{1.3in}{ \begin{center}BEC transition temperature $T_c$ 
\end{center} }
& $ 2 \, \mu\mbox{K} $ & 
	\parbox{1.3in}{ \begin{center}
	separation between atoms 
	\end{center} }
	& \parbox{1.0in}{ \mbox{\( n^{-1/3}=\)} \\
	\mbox{ \  } \(  {l} / {\sqrt{\pi} (2.612)^{1/3}} \) } 
	& = 200 nm \\
\parbox{1.3in}{ \begin{center}single-photon recoil energy 
\end{center} }
& $ 1.2 \, \mu\mbox{K} $ &
	\parbox{1.3in}{ \begin{center}
	optical wavelength 
	\end{center} }
	& $ \lambda = l $ & = 600 nm \\
\parbox{1.3in}{ \begin{center}temperature $T$ 
\end{center} }
& $ 1 \, \mu\mbox{K} $ &
	\parbox{1.3in}{ \begin{center}
	thermal de Broglie wavelength 
	\end{center} }
	& $ \lambda_{dB} = {l} / {\sqrt{\pi}} $ &	= 300 nm \\
\parbox{1.3in}{ \begin{center}mean field energy $ \mu $ 
\end{center} }
& 300 nK &
	\parbox{1.3in}{ \begin{center}
	healing length 
	\end{center} }
	& $ \xi = {l} / {2 \pi} $ & = 200 nm \\
\parbox{1.3in}{\begin{center}harmonic oscillator level spacing $ \hbar \omega $ 
\end{center} } 
& 0.5 nK &
	\parbox{1.3in}{ \begin{center}
	oscillator length ($ \omega \simeq 2 \pi \cdot 10 \mbox{Hz} $) 
	\end{center} }
	& $ a_{HO} = {l} / {\sqrt{2} \pi} $ & 
	$ = 6.5 \, \mu\mbox{m} $ \\
\hline
	\label{tab:lengthscales}
\end{tabular}
\end{table}

\subsubsection{BEC of composite bosons} 
Atoms are bosonic if they have integer spin, or 
equivalently, if the total number of electrons, protons, and neutrons they 
contain is even~\cite{ehre31,free80}.  However, in the context of BEC, we regard these composite particles as 
pointlike particles obeying Bose-Einstein statistics.  
Under what 
conditions will this assumption break down and the composite nature of the particles affect the 
properties of the system?

The composite nature manifests itself in internal excitations.  If the energy necessary for an 
internal excitation is much larger than $k_{B}T$, then the internal degree of freedom is frozen 
out and inconsequential for describing thermodynamics at that 
temperature.
For molecules of mass $m$ and size $a$, the lowest rotational levels are 
spaced by 
$\hbar^{2}/m a^{2}$.
The first electronically excited state is at 
$\hbar^{2}/m_{e} a^{2}$ where $m_{e}$ 
is the electron mass.  In any case, the condition of diluteness $na^{3} 
\ll 1$ guarantees that 
$k_{B}T_{c}$ is much smaller than the internal excitation energy.  Therefore, the composite 
nature of particles cannot affect the properties of a dilute Bose condensate.  
Ref.~\cite{nozi95}  briefly discusses what happens at higher densities when the Fermi energy becomes 
comparable to the binding energy of the composite boson.

Since the basic fermions have spin, the composite boson can have spin structure, which 
can result in several hyperfine ground states.  
In magnetic traps, atoms are usually trapped 
in only one hyperfine state (or a selected few~\cite{myat97}),
whereas in optical traps, new phenomena can be 
explored with condensates populating several hyperfine states (sect.~\ref{opticalsection}).  Since the 
number of hyperfine states is finite, hyperfine structure can lead to multi-component 
condensates, but will not prevent BEC.

\subsubsection{Bose-Einstein condensation as thermal equilibrium state}

Bose-Einstein condensation occurs in thermal equilibrium when 
entropy is maximized
by putting a macroscopic population of atoms into the ground state of the 
system.  It might 
appear counter-intuitive that an apparently highly ordered state as the Bose 
condensate 
maximizes entropy.  However, only the particles in excited states contribute to 
the entropy.
Their contribution is maximized at a given total energy by forming a Bose 
condensate in the 
ground state and distributing the remaining atoms among higher energy states.

A macroscopic population of atoms in the ground state of the system is achieved 
simply by 
lowering the temperature of the sample.  This is in contrast to the optical 
laser where a 
non-equilibrium process is necessary to place a macroscopic population of 
photons in a single 
mode of the electromagnetic field.
This is due to the fact that, unlike photons, the number of atoms is 
conserved.
For bosonic atoms, the 
lowest entropy 
state below a certain temperature includes a macroscopic population of the 
ground state.  In 
contrast, when one cools down a blackbody cavity, the cavity empties.  Photons 
do not Bose 
condense into the ground state of the cavity, but are absorbed by the 
walls
The absorbed energy leads to a
larger entropy than forming a Bose condensate.  The laser phenomenon 
requires 
inversion of the active medium characterized by a ``negative'' temperature.  In 
that sense, 
``lasing'' of atoms is a simpler phenomenon than lasing of light --- all you need 
to do is cool a 
gas!

However, if a photon gas would thermalize while the number of photons 
is conserved, it would be described by a Bose-Einstein distribution 
with non-zero chemical potential and could form a Bose condensate.
Thermalization with number conservation is possible, for example, by 
Compton scattering with a thermal electron gas \cite{zeld72}.

\subsubsection{Macroscopic wavefunction}

In an ideal gas, Bose condensed atoms all occupy the same 
single-particle ground-state wavefunction. 
The many-body ground-state wavefunction is then 
the product of 
$N$ identical single-particle ground-state wavefunctions.
This single-particle wavefunction is therefore called
the 
condensate 
wavefunction or macroscopic wavefunction.  This picture 
retains 
validity
even when we include weak interactions.  The ground-state many-body 
wavefunction is still, 
to a very good approximation, a product of $N$ single-particle wavefunctions 
which are now 
obtained from the solution of a non-linear Schr{\"o}dinger equation.  The admixture 
of other 
configurations into the ground state is called quantum depletion.  
In Bogoliubov 
theory, the quantum depletion is $(8/3 \pi^{1/2}) \sqrt{n a^3}$, typically 1\% 
or less for the 
alkali condensates.  This means that even for the interacting gases, we can, 
with 99\% 
accuracy, regard all the atoms to have the same single-particle wavefunction.  
This is in 
contrast to liquid helium in which the quantum depletion is about 90\% 
\cite{penr56}.

The density distribution of a condensate can be directly observed in a 
non-destructive 
way (sect.~\ref{sec:nondest}).  Such observations can be regarded as a direct 
visualization of the 
magnitude of the macroscopic wavefunction.  The time evolution of the 
wavefunction of a 
single system has even been recorded in real time~\cite{andr97prop,mies98form,stam98coll}.  A 
wavefunction is a probabilistic description of a system in the sense that it 
determines the 
distribution of measurements if many identical wavefunctions are probed.  In 
BEC, one 
simultaneously realizes millions of identical copies of the same wavefunction, 
and thus the 
entire wavefunction can be measured while affecting only a small fraction of the 
condensed 
atoms.  The resulting dramatic visualizations of wavefunctions are an appealing aspect 
of experimental 
studies of BEC.

	
\subsection{BEC 1925-1995}

\subsubsection{BEC and condensed-matter physics}

Bose-Einstein condensation is one of the most intriguing phenomena 
predicted by 
quantum statistical mechanics.  The history of the theory of BEC is very 
interesting, and is 
nicely described in the biographies of Einstein \cite{pais82} and London 
\cite{gavr95} 
and reviewed by A. Griffin in these proceedings.  For instance, Einstein made 
his predictions 
before quantum theory had been fully developed, and before the differences 
between bosons and 
fermions had been revealed \cite{eins25qua2}.  After Einstein, important 
contributions were 
made by, most notably, London, Landau, Tisza, Bogoliubov, Penrose, Onsager, 
Feynman, Lee, Yang, 
Huang, Beliaev and Pitaevskii.  An important issue has always been the 
relationship between 
BEC and superfluidity in liquid helium, an issue which was highly 
controversial between 
London and Landau (see ref.~\cite{gavr95}).  Works by Bogoliubov, Beliaev, Griffin and others showed that 
Bose-Einstein 
condensation gives the microscopic picture behind Landau's ``quantum 
hydrodynamics.''  BEC is 
closely related to superconductivity, which can be described as being due to 
Bose-Einstein 
condensation of Cooper pairs.  Thus 
Bose-Einstein 
condensation is at the heart of several macroscopic quantum phenomena.

BEC is unique in that it is a purely quantum-statistical phase transition, 
i.e.\
it occurs even 
in the absence of interactions (Einstein described the transition as 
condensation ``without 
attractive forces''~\cite{eins25qua2}).  This makes BEC an important paradigm of 
statistical 
mechanics, which has been discussed in a variety of contexts in 
condensed-matter, nuclear, 
particle and astrophysics \cite{grif95}.  On the other hand, real-life particles 
will 
always interact, and even the weakly-interacting Bose gas behaves qualitatively 
differently 
from the ideal Bose gas \cite{huan87}. It was believed for quite some time that 
interactions 
would always lead to ``ordinary'' condensation (into a solid) before Bose-Einstein
condensation 
would happen.  Liquid helium was the only counter-example, where the light mass 
and concomitant 
large zero-point kinetic energy prevents solidification down to zero kelvin.

The quest to realize BEC in a dilute weakly interacting gas was pursued in at 
least three 
different directions: liquid helium, excitons and atomic gases.  
Experimental~\cite{croo83,repp84} and theoretical work~\cite{raso84}
showed that the onset of 
superfluidity for liquid helium in Vycor shows features of dilute-gas 
Bose-Einstein condensation.  At sufficiently low coverage, the helium adsorbed 
on the porous 
sponge-like glass behaved like a dilute three-dimensional gas.

Excitons, which consist of weakly-bound electron-hole pairs, are composite 
bosons.  The 
physics of excitons in semiconductors is very rich and includes the formation of 
an electron-hole 
liquid and biexcitons.  As nicely discussed in 
refs.~\cite{wolf95,fort95}, there are systems, most notably Cu$_{2}$O, where 
excitons form a weakly 
interacting gas with a lifetime long enough to equilibrate to a Bose-Einstein 
distribution and to 
show evidence for Bose-Einstein condensation \cite{shen97,lin93}.

\subsubsection{Spin-polarized hydrogen}

Dilute atomic gases are distinguished from the condensed-matter systems 
discussed above by the 
absence of strong or complex interactions.  Interactions at the density of a 
liquid or a solid 
considerably modify and complicate the nature of the phase transition.  Hecht 
\cite{hech59} and 
Stwalley and Nosanow \cite{stwa76} used the quantum theory of corresponding 
states to conclude 
that spin polarized hydrogen would remain gaseous down to zero temperature and 
should be a good 
candidate to realize Bose-Einstein condensation in a dilute atomic gas.  These 
suggestions 
triggered several experimental efforts, most notably by Greytak and Kleppner at 
MIT and Silvera 
and Walraven in Amsterdam.  The stabilization of a spin-polarized hydrogen gas 
\cite{silv80,clin80} created great excitement about the prospects of 
exploring 
quantum-degenerate gases.  Experiments were first done by filling cryogenic 
cells with 
the spin-polarized gas, then by compressing the gas, and since 1985, by magnetic 
trapping and 
evaporative cooling.  
BEC was 
finally accomplished in 1998 by Kleppner, Greytak and collaborators 
\cite{frie98}.  See 
refs.~\cite{grey95bec,grey84,silv86,walr96} and the contribution of 
Kleppner and Greytak to this volume for a full 
account of the pursuit of Bose-Einstein condensation in 
atomic hydrogen.

It is interesting to look at the unique role which was given to spin-polarized 
atomic hydrogen 
in the early suggestions \cite{hech59,stwa76,walr96}.  In the quantum 
theory of 
corresponding states, one defines a dimensionless parameter $\eta$ which is 
related to the 
ratio of the zero-point energy to the molecular binding energy.
This parameter determines 
whether the 
system will be gaseous down to zero temperature.  For large $\eta$, the zero-point motion 
dominates and the system is gaseous; for small $\eta$, it condenses into a 
liquid or solid.  The critical $\eta$ value is 0.46, and only spin-
polarized hydrogen 
with $\eta=0.55$ exceeds this value \cite{walr96}; alkali vapors have $\eta$ 
values in 
the range $10^{-5}$ to $10^{-3}$.

In reality, all spin-polarized gases are only \textit{metastable} at $T=0$ due 
to 
depolarization processes.  The lifetime of the gas is limited by three-body 
recombination.  
Since the triplet potential of molecular hydrogen supports no bound 
states, spin-polarized 
hydrogen can 
only recombine into the singlet state with a spin-flip.  In contrast, alkali 
atoms have both 
bound singlet and triplet molecular states, and their three-body recombination 
coefficient is 
ten orders of magnitude larger than for spin-polarized hydrogen.  However, the 
rate of 
three-body processes depends on the square of the atomic density,  
is 
suppressed at 
sufficiently low density, and is almost negligible during the cooling 
to BEC.
For 
magnetically trapped atoms, dipolar relaxation is an additional loss process, 
and hydrogen and 
the alkalis have comparable rate coefficients.  So in hindsight, the unique 
benefits of hydrogen 
over other gases were not crucial for gaseous BEC. Although spin-polarized 
hydrogen has been 
called the only ``true quantum gas,'' the difference from alkali vapors is just the 
range of 
densities and lifetimes of the metastable gaseous phase.

The work in alkali atoms is based on the work in spin-polarized hydrogen in 
several respects:
\begin{itemize}
	\item Studies of spin-polarized hydrogen showed that systems can remain 
in a metastable gaseous 
state close to BEC conditions.  The challenge was then to find the window in 
density and 
temperature where this metastability is sufficient to realize BEC.

\item Many aspects of BEC in an inhomogeneous 
potential~\cite{gold81,huse82,oliv89}, and 
the theory of cold collision processes (see e.g.~\cite{stoo88})
developed in the 
`80s for hydrogen could be applied directly to the alkali systems.

\item The technique of evaporative cooling was developed first for hydrogen 
\cite{hess86,masu88} and then used for alkali atoms (sect.~\ref{sec:evapcool}).
\end{itemize}

Major efforts have also been underway to reach quantum degeneracy in a 
two-dimensional gas of 
spin-polarized hydrogen.  In two dimensions, a ``true'' Bose-Einstein condensate 
with long 
range order is not stable against phase fluctuations.  Still, below a certain 
temperature, at 
the Kosterlitz-Thouless transition, the gas becomes superfluid and shows local 
Bose-Einstein 
condensation, i.e.\
a macroscopic population of the lowest state.  Work in 
two dimensions has 
been pursued at Harvard \cite{silv95bec}, in Amsterdam \cite{mosk98}, Kyoto 
\cite{mats95}, 
and at the University of Turku, where evidence for the 2D phase transition was 
reported in 1998 
\cite{safo98prl}.

\subsubsection{Alkali atoms}

Laser cooling opened a new route to ultralow temperature physics.  Laser cooling 
experiments, 
with room temperature vacuum chambers and easy optical access, look very 
different from 
cryogenic cells with multi-layer shielding around them.  Also, the number of atomic 
species which 
can be studied at ultralow temperatures was greatly extended from helium and 
hydrogen to all of 
the alkali atoms, metastable rare gases, several 
earth-alkali atoms, and others (the 
list of laser 
cooled atomic species is still growing).  We will summarize the relevant laser 
cooling 
techniques in sect.~\ref{lasercooling}.  A full account of their development 
is given in refs.~\cite{arim92,metc94,adam97} and in the Nobel lectures of Chu,
Cohen-Tannoudji and Phillips 
\cite{chu98nob,cohe98nob,phil98nob}.  Here we make some comments on the 
specific developments 
which led to the successful observation of Bose-Einstein condensation in 1995.

Some papers and proposals written in the early and mid `80s, before and during 
the 
developments of the basic cooling and trapping techniques, listed 
 quantum degeneracy is a gas as a visionary goal for this new emerging field 
\cite{leto80,chu85,prit86trap}.  However, major limitations of laser 
cooling and trapping were 
soon identified.  Although there is no fundamental low temperature 
limit, the
final temperature provided by  polarization gradient cooling --- 
about ten times the recoil energy --- was regarded as a practical
limit.
Sub-recoil cooling techniques, especially in three dimensions, are 
harder to implement, and require long cooling times.  The 
number and density of atoms was limited by inelastic, light-induced collisions 
(leading to 
trap loss \cite{walk94,wein95}) and by absorption of scattered laser 
light \cite{walk90} which results in an outward radiation pressure (weakening the 
trapping potential and 
limiting the density).  Indeed, even the most advanced optical techniques 
\cite{adam95,boir98,depu98} achieved only a factor of twenty improvement 
over the density of 
the first optical trapping experiment ($5 \times 10^{11} \un{cm}^{-3}$ \cite{chu86}).  Further, 
since the 
lowest temperatures could not be achieved at the highest densities 
\cite{drew94,town95,town96spot}, most trapping and cooling techniques reached 
a 
maximum phase-space density around $10^{-5}$.

In the end, the successful approach was to use laser cooling only as pre-cooling 
for magnetic 
trapping and evaporative cooling.  Evaporative cooling proved to work much 
better for alkali 
atoms than for hydrogen, for which the technique was originally developed.  This 
didn't come as 
a big surprise.  Already in 1986, Pritchard correctly estimated the rate 
constants of elastic 
and inelastic collisions for alkali atoms \cite{prit86trap}.  From these 
estimates one could 
easily predict that for alkali atoms, in contrast to hydrogen, the so-called good 
collisions (elastic 
collisions necessary for the evaporation process) would clearly dominate over 
the so-called bad 
collisions (inelastic two- and three-body collisions); therefore, evaporative 
cooling in 
alkalis would probably not be limited by intrinsic loss and heating processes.  
However, 
pessimism \cite{vigu86} and skepticism remained, and researchers at both 
Boulder and 
 MIT 
explored  experimentally \cite{corn91} and theoretically 
\cite{kett92stat} the 
possibility of 
confining atoms in the lowest, strong-field seeking hyperfine state.  Trapping 
atoms in that state would 
eliminate inelastic two-body collisions which had limited progress toward BEC in 
atomic 
hydrogen.  As estimated~\cite{prit86trap}, however, these collisions turned out to be negligible 
for sodium and 
rubidium.

When one of the authors (W.K.) teamed up with Dave Pritchard at MIT in 1990, 
evaporative cooling was discussed as a way to break through the limits in 
temperature and 
density.  Following the example of the spin-polarized hydrogen experiment at 
MIT, evaporation 
should be done in a magnetic trap using rf induced spin-flips, as suggested by 
Pritchard and 
collaborators in 1989 \cite{prit89} (sect.~\ref{sec:evapcool}).  Magnetic traps and laser 
cooling had 
 already been used simultaneously in the first experiments on magnetic trapping 
at NIST~\cite{migd85} and 
MIT~\cite{bagn87stop}, and on Doppler cooling of magnetically trapped 
atoms at MIT~\cite{prit89,helm92cool}.  In 1990, a magnetic trap was loaded 
from a magneto-optical trap and optical molasses in Boulder \cite{monr90}.  So 
most of the 
pieces were known in 1990, but there was doubt about whether they 
would fit 
together.  The laser cooling route to BEC was summarized by Monroe, 
Cornell and Wieman in ref.~\cite{monr92}.

The efforts at Boulder and at MIT both faced the challenge of simultaneously 
achieving effective 
laser cooling and trapping, which work best at low atomic densities,
and efficient evaporative cooling, which requires high densities.  
The problem was to avoid excessive scattering of photons, with a 
resonant cross section of $\sim 10^{-9} \un{cm}^{2}$, while achieving 
sufficient scattering between atoms at a thousand-times smaller cross 
section.
This shifted 
the emphasis of the optical techniques from attaining low temperatures and high 
phase-space 
density towards achieving high elastic collision rates.  
The first major improvement towards this goal was the invention of the Dark SPOT 
trap in 1992 
\cite{kett93spot}, which turned out to be crucial to the BEC work both at 
Boulder \cite{ande95} 
and at MIT \cite{davi95bec}.  Beginning in 1991, Dave Pritchard and W.K.\
investigated several schemes  to 
improve the limits 
of optical traps based on coherent dark state cooling and trapping.
The solution --- the Dark SPOT --- turned out to be an 
incoherent optical 
pumping scheme.

The only requirement for evaporative 
cooling to commence is a 
collisional re-thermalization time much shorter than the lifetime of an atom in 
the trap.
In the summer of 1992, the elastic collision rate in the Dark SPOT 
trap was 100 Hz \cite{kett93spot}.
The remaining obstacle to evaporative cooling was achieving a vacuum 
in the range of $10^{-11}$ mbar to reduce the background gas 
collisions. 
Evaporative cooling of alkali atoms was demonstrated at MIT and in Boulder in 1994 
\cite{davi94icap,petr94icap}. 
Both groups employed rf-induced evaporation 
(sect.~\ref{sec:evapcool}), a highly efficient evaporation technique.
In 1995, after improvements in the magnetic traps (sect.~\ref{sec:magtrap}), the 
breakthrough came with the observations of BEC in Boulder in early June 
\cite{ande95} and at MIT in late September \cite{davi95bec}.  The Rice group 
obtained evidence for 
reaching the quantum-degenerate regime in July \cite{brad95bec}.  

\begin{figure}
\epsfxsize = 7cm
\centerline{\epsfbox[30 47 681 540]{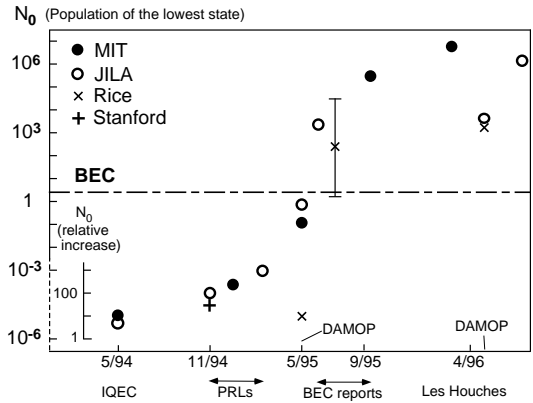}}
\caption[Progress in evaporative cooling of alkali atoms up to 1996]
{Progress in evaporative cooling of alkali atoms up to 1996.
The number of atoms in the lowest quantum
 state is proportional to the phase-space density, and has to exceed
 a critical number of 2.612 to achieve Bose-Einstein condensation.  For
 $N_0<10^{-3}$, the increase in phase-space density due to evaporation is plotted. 
 For the Rice
 result of July 1995 see ref.~\cite{brad95bec} and 
 the erratum~\cite{brad97erra}.}
 \label{racetobec}
 \end{figure}

Figure  \ref{racetobec}  
shows how 
dramatic the progress was after laser and evaporative cooling were combined.  
Within less than 
two years, the number of alkali atoms in a quantum state was increased by about 
twelve orders 
of magnitude --- a true singularity demonstrating that a phase transition was 
achieved!

Techniques such as the Dark SPOT, compressed MOT \cite{petr94cmot}, 
the TOP trap and the optically plugged trap were critical for first 
demonstrating BEC, 
but by no means 
indispensable.  This is best illustrated by the experiment at Rice which used 
only Doppler 
cooling to load the magnetic trap ---  a technique which had been 
developed in 
the `80s.  The collision rate was slow, but an excellent vacuum made a very 
slow 
evaporation process possible \cite{brad95bec}.  So in hindsight, BEC 
in alkali gases did not 
require major 
innovations in cooling and trapping.  It merely required enough 
optimism to risk a few years in the attempt to combine laser and evaporative cooling.  
The major 
advances in cooling and trapping that were developed along the way were not 
indispensable for 
BEC.
However, they have permitted the rapid developments in 
the field over 
the last years.

%


\section{Cooling, trapping, and manipulation techniques}

In order to create a Bose condensate in a dilute gas, atoms must be cooled and 
compressed in a trap until the thermal de Broglie wavelength is on the order of 
the spacing between atoms.  In addition, the atoms must be thermally isolated 
from all material walls.
This is done by trapping atoms with magnetic fields or with laser light 
inside ultrahigh vacuum chambers.  Such traps can store atoms for seconds 
or even minutes, 
which is enough time to cool them.  Pre-cooling is a prerequisite for 
trapping
because conservative atom traps can only confine neutral atoms with a 
maximum energy of one kelvin at best 
(and in many cases the trap depth is just a few millikelvin).  The pre-cooling is done by 
laser cooling, and the final cooling by evaporation.  
Table \ref{tab:coolingstages}  shows how these 
cooling techniques together reduce the temperature of the atoms by a factor of a 
billion.  The density at the transition temperature is close to the density in the atomic 
beam oven.  The phase space density enhancement is almost equally distributed 
between laser 
cooling and evaporative cooling, providing about six orders of magnitude each.  
Bose-Einstein condensation can be regarded as ``free cooling,'' as it increases the quantum 
occupancy by another factor of about a million without any extra effort. 
This reflects one 
important aspect of BEC: the fractional population of the ground state is 
no longer inversely proportional to the number of states with energies smaller than 
$k_{B}T$, but quickly approaches unity when the sample is cooled below the transition 
temperature.

\begin{table}[htbf]
\begin{center}
\caption[Multi-stage cooling to BEC]{Multi-stage cooling to BEC in 
the MIT experiment.
Through a combination of optical and evaporative cooling, the 
temperature of a gas is reduced by a factor of $10^{9}$, while the 
density at the BEC transition is similar to the initial density in the 
atomic oven (all numbers are approximate).  In each step shown, the ground 
state population increases by about $10^{6}$.}
\begin{tabular}{|l|c|c|c|} \hline
 & Temperature & Density (cm$^{-3}$) & Phase-space density \\ \hline
 Oven & 500 K & $10^{14}$ & $10^{-13}$ \\
 Laser cooling & 50 $\mu$K & $10^{11}$ & $10^{-6}$ \\
 Evaporative cooling & 500 nK & $10^{14}$ & 1 \\
 BEC & & & $10^{7}$ \\ \hline
\label{tab:coolingstages}
\end{tabular}
\end{center}\vskip-\lastskip
\end{table}

\subsection{Pre-cooling}

\subsubsection{Standard laser cooling techniques}
\label{lasercooling}

Several steps of laser cooling are applied before the sample is transferred into 
a magnetic 
trap.  We only briefly summarize these steps and refer to the many books and 
review 
articles for more details 
\cite{arim92,metc94,adam97}.

\begin{itemize}
	
	\item {\it Zeeman slowing:}  Atomic beams can be decelerated by resonant 
radiation pressure. Among the many methods of slowing atomic beams 
\cite[and references therein]{phil85,phil92}, Zeeman slowing (first demonstrated 
by Phillips {\it et al.\ }  
\cite{prod82}) gives the highest flux of slow atoms.  
In this technique, an inhomogeneous magnetic field 
compensates for the changing Doppler shift as atoms slow down and ensures 
that the atoms stay in resonance during the deceleration.  Typically 
a Zeeman-slowed sodium beam has a velocity of 
30 m/s, corresponding to a kinetic energy of 1 K. This is sufficiently 
cold to capture atoms in a magneto-optical trap (MOT).  The distinguishing feature of our 
Zeeman	slower is the high flux of up to $10^{12} $ slow atoms per second 
\cite{joff93}, which enables more than $10^{10} $ atoms to be loaded into 
the MOT in one or two seconds.

	\item {\it Doppler molasses:}  The slow atomic beam is stopped by optical 
molasses, first reported in \cite{chu85} and reviewed in \cite{chu92}.  
The usual arrangement for 
molasses consists of six laser beams intersecting in the center of 
the vacuum chamber.  Doppler molasses reduces the temperature of the atoms 
to one millikelvin or below.
	
	\item {\it Magneto-optical trap:} Cooling in optical molasses is one feature of 
the magneto-optical trap.  In addition to cooling the atoms, it also confines 
the atoms and compresses them to higher densities (typically between $10^{10}$ cm$^{-
3}$ and $10^{12}$ cm$^{-3}$).  
The magneto-optical trap was first realized in 1987 \cite{raab87} (for a 
review, see \cite{prit92,chu92}).

	\item {\it Dark SPOT trap:}  High densities for a large number of atoms are 
reached in a variant of the magneto-optical trap, in which the atoms are kept in a dark 
hyperfine state which does not absorb the laser light necessary for cooling and 
trapping \cite{kett93spot}.  This avoids the limitations of the ordinary 
magneto-optical trap due to light absorption and trap loss caused by 
laser-induced collisions~\cite{ande94}.  
The  necessary confinement and cooling are provided by occasionally cycling the 
atoms back	into the bright hyperfine state for a short time.  Optimization of the 
population in each of these two states gave rise to a density increase of almost two 
orders of magnitude.  The dark SPOT trap was a key 
technique in the BEC experiments 
at JILA \cite{ande95}, MIT \cite{davi95bec} and 
elsewhere~\cite[etc.\ ]{han98,hau98,ande98pra}.

	\item {\it Polarization gradient cooling:}  Even colder temperatures (between 1 
and 50 	$\mu$K) are reached with polarization gradient cooling (first demonstrated 
in 	\cite{lett88} and reviewed in \cite{lett89}).  This cooling mechanism is 
already present in the center of the magneto-optical trap, but colder temperatures 
are usually reached by switching off the MOT's magnetic coils and adding a short cycle 
(a few ms) 	of optimized polarization gradient cooling.  We usually reach 
temperatures for sodium
between 50 and 100 $\mu$K, somewhat higher than the lowest temperatures reported 
\cite{lett89,weis89}, probably because of the large number of atoms and the high 
atomic density of our samples \cite{coop94}.

	\item {\it Vapor cell trap:}  In many experiments, the first step (atomic beam 
slowing) is replaced by directly loading atoms from the low energy 
tail of a thermal vapor into the 
magneto-optical	trap \cite{cabl90,monr90}.  However, the stringent UHV requirement of 
magnetic trapping and evaporative cooling limits the vapor pressure and therefore 
the loading	rate.  Many experiments use the Òdouble MOT techniqueÓ 
\cite{stea95,myat96,gibb95}, where one MOT at high vapor pressure
collects atoms which are periodically
sent to another MOT operated at UHV conditions.  Another option is 
to use a slow atom beam produced in a vapor cell 
\cite{lu96lvis,swan96,diec98}.

\end{itemize}
	
In almost all BEC experiments, pre-cooling is done in a 
magneto-optical trap loaded from a slowed beam which is either 
extracted from a vapor cell MOT or generated by a Zeeman slower.

With these conventional optical cooling and trapping methods, temperatures are limited 
by heating due to spontaneous emission while densities are limited by radiation trapping effects and trap 
loss due to excited-state collisions.  More sophisticated cooling methods (``sub-recoil'' 
techniques) have been developed such as Raman cooling \cite{kase92,reic95} 
and velocity-selective coherent population trapping \cite{aspe88,lawa95}, but they 
have not been used so far for realizing BEC. The 
highest phase-space density achieved with standard optical techniques
is five orders of magnitude short of BEC,
while sub-recoil cooling was used to come 
within a factor of 300 \cite{lee96,lee98}.  Given the ease and high efficiency of evaporative 
cooling (sect.~\ref{sec:evapcool}), the objective for the laser pre-cooling is no longer to achieve 
very low temperatures or high phase-space densities, but rather to ensure high elastic 
collision rates to obtain efficient evaporation.  Another objective 
for the pre-cooling is the collection of a large number 
of atoms.
This is 
crucial for several reasons.
(1) It is much easier to transfer  a larger cloud into a magnetic 
trap without losing density.  
 (2) Evaporative cooling can be performed faster 
at the expense of expelling more atoms.
(3) Many studies of BEC benefit from large condensates and their 
improved signal-to-noise ratio.


The standard techniques of laser 
cooling are very forgiving, and are fairly insensitive to laser
polarization, frequency, and  power. Most BEC 
experiments have about one order of magnitude reserve, so once you have 
optimized the production of condensates, you can get away with 
sub-optimal optical alignment, trap loading, transfer to the magnetic trap, 
evaporation etc., but the best starting point is to peak everything!  
For our apparatus, our experience has been that consistent loading of a large 
number of atoms into the MOT guarantees successful evaporative cooling to BEC.

\subsubsection{Cryogenic pre-cooling}

The bottleneck in the number of atoms in a Bose condensate is the number of 
atoms which can be laser cooled.  The Dark SPOT trap
mitigates trap loss and absorption of trapping light by ``hiding'' 
most of the atoms in a dark hyperfine state.  But 
ultimately 
the usual limitations of the magneto-optical trap apply to 
the Dark SPOT, albeit at two orders of magnitude higher densities.
A hard limit to laser cooling 
might be set by the number of available laser photons.  It takes about $10^4$ photons 
to slow an atom in a Zeeman slower.  A slowing beam with 10 mW of laser power is completely 
used up by slowing $3 \times 10^{12}$ atoms per second!

Cryogenic pre-cooling does not suffer from these limitations, and can lead to much 
larger sample sizes.  This technique was applied to hydrogen at MIT 
\cite{hess87} and in 
Amsterdam \cite{vanr88} to load magnetic traps.  
Bose-Einstein condensates of $10^9$ hydrogen 
atoms (forty times larger than the largest alkali condensates)
were produced with this 
technique \cite{frie98}.

The idea of pre-cooling by thermalization with a cryogenic environment was 
recently extended to a large class of paramagnetic atoms and molecules by Doyle and 
collaborators \cite{wein98, kim97}.  In these experiments molecules are pre-cooled 
by a buffer gas of $^{3}$He and settle into a magnetic trap.  After the loading process, the $^{3}$He is 
pumped out by lowering the cell temperature, thereby reducing the vapor pressure 
of $^{3}$He to negligible levels.  Subsequent evaporative cooling could then be used to 
reduce the temperature of the trapped gas.

\subsection{Conservative atom traps}

Conservative atom traps fulfill two essential roles in BEC: they keep the atoms 
tightly compressed during cooling, and hold the condensate for 
study.  In principle, any trap that has a 
sufficiently small heating rate could be used.   
Conservative trapping potentials have been 
realized with dc magnetic fields (sect.~\ref{sec:magtrap}), ac magnetic 
fields \cite{corn91}, microwave fields \cite{spre94} and 
far-off-resonant laser beams (sect.~\ref{opticalsection}).

The requirements for the trap during cooling are more 
stringent than they are for holding condensates.  
First, the time for cooling (typically thirty seconds for 
evaporative cooling) is usually much longer than the time for 
performing experiments on BEC, requiring low heating and trap loss 
rates.
Furthermore, for cooling, the trap 
needs a sufficiently high trap depth and trapping volume to hold 
the initial (pre-cooled) cloud, and must accommodate 
a cooling scheme able to reach BEC temperatures.  So far, only the 
combination of magnetic trapping (sect.~\ref{sec:magtrap}) and evaporative 
cooling (sect.~\ref{sec:evapcool}) has 
accomplished this.  Evaporative cooling has also been observed in an 
optical dipole trap \cite{adam95}, but with only a factor of 30 gain in
phase-space density.
Rf-induced evaporation is particularly effective and simple to
implement in a magnetic trap, and perhaps this combination will 
become the 
workhorse of the nanokelvin temperature range, just as the MOT has in the 
microkelvin range. 
However, there is 
always room for improvement, in particular for atoms which don't have 
the collisional properties necessary for evaporative cooling in a 
magnetic trap (sect.~\ref{sec:atomsforbec}).  
Considerable progress is being made on 
far-off resonant trapping using blue-detuned \cite{ovch97,lee98,ovch98}, 
near-infrared \cite{boir98}, and CO$_2$ 
lasers \cite{take95cs,frie98co2}.

After cooling, trap requirements are different, and other options 
are available for holding the condensate.  
Our recent experiments on optically confined BEC (sect.~\ref{opticalsection}) 
demonstrate that optical dipole traps confine 
Bose-Einstein condensates more easily than much hotter 
atoms.  Traps for condensates can be much weaker than for laser 
cooled atoms, making them easier to implement and greatly 
reducing heating due to beam jitter, intensity fluctuations, and 
spontaneous emission \cite{sava97}.  

There are several traps 
which have so far not been pursued beyond their first demonstration: A 
microwave trap originally suggested for hydrogen atoms \cite{agos89} 
has been realized with laser cooled cesium atoms \cite{spre94}.  An 
ac magnetic trap for strong-field seeking atoms was suggested in 
1985 \cite{love85}, and realized in 1991 \cite{corn91}.  
Finally, ac electric fields offer another possibility to trap 
strong-field seekers \cite{riis93,shim92}.  All these traps are 
rather weak and don't seem to offer obvious advantages over the combination of 
optical and magnetic forces.  A recent resource letter contains many 
references on atom traps \cite{newb96}.

\subsection{Magnetic trapping\label{sec:magtrap} }
\label{magnetictraps}

The major role of the magnetic trap in 
a BEC experiment is to accommodate the pre-cooled atoms and compress 
them in order to achieve high collision rates 
and efficient evaporative cooling.  The steps that must be taken to 
obtain high collision rates, including 
``mode-matched'' transfer and anisotropic compression, are
discussed in this section.  The collision rate after compression is suggested as 
the most important figure of merit for a magnetic trap, and will be 
related to the magnetic field parameters.

Magnetic trapping of neutral atoms was first observed in 1985 
\cite{migd85}.  Shortly afterwards, orders of magnitude 
improvements in density and number of trapped atoms were achieved at 
MIT and in Amsterdam using superconducting traps and different loading 
schemes 
\cite{bagn87stop,hess87,vanr88}.  
Important aspects of magnetic trapping are discussed in 
\cite{phil85,berg87,grey95bec,walr96,kett96evap}.

Magnetic forces are strong for atoms with an unpaired electron, such as 
the alkalis, resulting in magnetic moments $ \mu_{m} $ of the order of a Bohr 
magneton.  However, it is worth pointing out that magnetic confinement 
was first observed for neutrons, despite their thousand-times smaller 
magnetic moment \cite{kugl85}.


The interaction of a magnetic dipole with an external magnetic field 
is given by  $ - \vec{\mu}_{m} \cdot 
\vec{B} = - \mu_{m} B \cos \theta $.  Classically, the angle $ \theta $ between the 
magnetic moment and the magnetic field is constant 
due to the rapid precession of $ \vec{\mu}_{m} $ around the magnetic 
field axis.  Quantum-mechanically, the energy levels in a magnetic 
field are $ E(m_F) = g \mu_{B} m_F B $,
where $ g $ is the g-factor and $ m_F $ the quantum number of the $z$-component 
of the angular momentum $ F $. The classical term $ \cos \theta $  is now 
replaced by $ m_F/F $; the classical picture of constant $ \theta $ 
is equivalent to the system remaining in one $ m_F $ quantum state.

An atom trap requires a local minimum of the magnetic potential energy 
$ E(m_F) $.  For $ g  m_F>0 $ (weak field seeking states) this 
requires a local magnetic field minimum.  Strong field seeking states 
($g m_F < 0 $) cannot be trapped by static magnetic fields, because 
Maxwell's equations don't allow a magnetic field maximum  
in free space \cite{wing84,kett92stat}. 

Because magnetic traps only confine weak field seeking states, 
atoms will be lost from the trap if they make a transition into a strong field 
seeking state.  
Such transitions can be induced by the motion in the trap because an
atom sees a field in its moving frame which 
is changing 
in magnitude and direction.  
The trap is only stable if the atom's magnetic moment adiabatically 
follows the direction of the magnetic field.  This requires that the 
rate of change of the field's direction $ \theta $ must be slower 
than the precession of the magnetic moment:

\begin{equation}
\frac{d \theta}{d t}  < \frac{\mu_{m} \left| \mathbf{B} 
\right|}{\hbar} = \omega_{Larmor} \label{eq:majorana}
\end{equation}
The upper bound for
$ d \theta / d t $ in a magnetic trap is the trapping frequency.
This adiabatic condition is violated in regions of very small magnetic 
fields, creating a region of trap loss due to  spin flips to 
untrapped states.  These spin flips are referred to as ``Majorana 
flops'' \cite{majo32}.

\subsubsection{Quadrupole-type traps}
There are two basic types of static magnetic traps: those in which the 
minimum is a zero crossing of the magnetic field, and those which have 
a minimum around a finite field \cite{berg87}.
Traps with a zero-field 
crossing usually  generate a linear potential characterized by
the gradient of the magnetic field:
$  B_x = B_x' x $, $ B_y = B_y' y $, $ B_z = B_z' z $. 
Maxwell's equations require that $ B_x' + B_y' + B_z' = 0 $.  
The case 
of axial symmetry is a spherical quadrupole field in which $ B' \equiv
B_{x}' = B_{y}' = -B_{z}' /2 $.
This configuration is 
realized with two coils in ``anti-Helmholtz'' configuration.  This was 
the configuration which was first used to trap neutral atoms 
magnetically \cite{migd85}.

A linear trap offers superior confinement compared to 
traps with a parabolic potential minimum 
(sect.~\ref{sec:iptraps}).  This follows from a simple argument. 
Coils which are a
distance \( R_{coil} \) away from the trapped cloud and generate a field 
$ B_{coil} $ at the coil typically produce a field gradient of $ B' \approx 
 B_{coil}/R_{coil} $ and a curvature of $ B'' \approx  
B_{coil}/R_{coil}^{2} $.  
A cloud of size $ r $ in a linear potential with gradient  $ B' $ would be 
confined to the same size in a parabolic potential with a 
curvature equal to $  B'/r $.  This 
allows us to define an ``effective curvature'' of linear confinement:
$ B''_{eff} = B_{coil}'/ r $.  
This exceeds the curvature of a parabolic trap by 
$ R_{coil}/r $, which is usually an order of magnitude or more.  
 \label{sec:linearvsharmonic}

When linear traps were employed for the first demonstrations of 
evaporative cooling with alkali atoms \cite{davi94icap, petr94icap}, 
trap loss due to Majorana spin 
flips~\cite{majo32,migd85, schw37, phil85, berg89} 
near the zero of the magnetic field was encountered.  
For atoms moving at a velocity $v$, the effective size
of this ``hole'' in the trap is 
$ \sqrt{ 2 \hbar v / \pi \mu_{m} B'} $ 
which is about $ 1 $ $\mu$m for $ \mu_{m} = \mu_{B}$, $ v=1 $ m/s  
and $ B'=1000 $
G/cm. As long as the hole is small 
compared to the cloud diameter, the trapping time can be long (even longer 
than a minute), and evaporative cooling in such a trap was used to 
increase phase-space density
by more than two orders of magnitude \cite{davi95evap}.  
However, as the temperature drops, the trap loss due to the hole 
becomes prohibitive for further cooling.  Although the size of the 
hole depends on the thermal velocity of the atoms, and therefore 
shrinks as the atoms are cooled, the diameter of the atom cloud shrinks even 
faster with temperature, resulting in a $ T^{-2} $ dependence of the loss rate 
\cite{davi95evap,petr95}.

Two methods have been demonstrated to plug the hole.  One solution is 
to add a rotating magnetic bias field $ B_0 $ to the spherical 
quadrupole field.  
The frequency of the rotating field is much higher than the 
orbiting frequency of the atoms, but much lower than the Larmor 
frequency.  The resulting time-averaged, orbiting potential (TOP) trap 
is harmonic, but much tighter than what could be obtained by dc 
magnets of the same size \cite{petr95}.  The time-averaged potential 
can be written as
\begin{eqnarray}
U_{TOP}  =  \frac{\mu_{m}}{2} \left( B_\rho'' \rho^{2} + B_z'' z^2 \right)\\
B_{\rho}''  =  \frac{{B'}^2}{2 B_0} \label{eq:toprad},
B_z''  =  \frac{4 {B'}^2}{B_0} \label{eq:topz}
\end{eqnarray}
where $\rho^{2} = x^{2} + y^{2}$ is the radial coordinate.

The rotating field moves the zero of the magnetic field around in a 
circle (the ``circle of death'') of radius $ r_D = B_0 / B' $.  Due to 
Majorana flops, this limits the depth of the potential to $ 
U_{TOP}(r_D) = \mu_{m} B_0 / 4 $.  A large circle of death requires either 
large $ B_0 $ or small $ B' $, both of which lead to weaker confinement.  
The TOP trap was used in the first demonstration 
of BEC \cite{ande95}. An 
interesting variant of the TOP trap has been proposed, where a 
rotating quadrupole field avoids the circle of death \cite{shap96}.

Another solution is to plug the hole using the optical dipole forces 
of a tightly focused blue-detuned laser beam  
to repel atoms from the center of the trap (see also sect.~\ref{sec:odforces})
\cite{davi95bec}.  The ``optically plugged trap'' achieves 
very tight confinement corresponding to a curvature of about $ B'/x_{0} $ 
where $ x_{0} $, the separation of the potential minimum from the zero 
of the magnetic field, is typically $ \sim 50 \mu$m.  
BEC in sodium was first achieved with this trap \cite{davi95bec}.

\subsubsection{Ioffe-Pritchard traps\label{sec:iptraps}}
The lowest order (and therefore tightest) trap which can have a bias field is 
a harmonic trap.  
A magnetic trap with finite bias field along the $z$-direction has an 
axial field of $ B_{z}=B_{0}+ B'' z^{2} /2 $.  The leading term of the 
transverse field component $ B_{x} $ is linear, $ B_{x}=B' x $.  Applying 
Maxwell's equations (and assuming axial symmetry) 
leads to the following field configuration 
\cite{berg87}:
\begin{equation}
\mathbf{B} = B_0   \left( 
\begin{array}{c} 0 \\ 0 \\ 1 \end{array} 
\right)
+ B' \left( \begin{array}{c}  x \\ -y \\ 0  \end{array} \right)
+ \frac{B''}{2}  \left( \begin{array}{c} - x z \\ - y z \\ z^2 - \frac{1}{2} 
(x^2 + y^2) 
 \end{array} \right) \label{eq:ipfields}
\end{equation}

The parabolic trap was first suggested and demonstrated
for atom trapping by Pritchard~\cite{prit83,bagn87stop}, and is similar to the Ioffe configuration discussed earlier 
for plasma confinement \cite{gott62}.  We refer to any trap 
which has this field configuration as a Ioffe-Pritchard (IP) trap.

The Ioffe-Pritchard trap 
has two different regimes.  For temperatures $ k_{B} T < \mu_{m} B_{0} $, the cloud 
experiences the potential of a 3D anisotropic harmonic oscillator.  In 
the case of $ k_{B} T > \mu_{m} B_{0} $, the potential is predominantly linear along 
the radial direction ($ U_\rho = \mu_{m} B' \rho $) and harmonic along the 
axial direction ($ U_z = \mu_{m} B'' z^{2}/2 $). 
As a consequence, the loss in confinement compared to a linear trap is not 
as severe as implied in section 
\ref{sec:linearvsharmonic}.

For small clouds (and all condensates) the trapping potential is very 
well approximated by an anisotropic harmonic oscillator potential

\begin{eqnarray}
& U  \simeq \frac{\mu_{m}}{2} \left[ B_{radial}'' \rho^{2} + B'' z^2 \right] \\
& B_{radial}''  = \frac{{B'}^2}{B_0} - \frac{B''}{2} 
\label{eq:ipeffectivecurve}
\end{eqnarray}

Beyond a certain axial displacement, the radial confinement vanishes.
This occurs because the 
radial component of the curvature term $ - B''/2 \, x z $ interferes 
destructively with the radial gradient $ B'$.  From equation 
(\ref{eq:ipfields})  we find
that this point of instability $ z_{inst} $ occurs at:

\begin{equation}
z_{inst}  = \pm \left( \frac{B'}{B''} - \frac{1}{2} \frac{B_0}{B'} 
\right) 
\label{eq:zinstab}
\end{equation}
When $ B_{radial}'' \leq 0 $, the instability is at the origin, and 
there is no radial confinement at all.  
The saddle point at $z_{inst}$ 
requires special attention when loading large clouds 
into a Ioffe-Pritchard trap, as discussed in section \ref{sec:modematching}.

\begin{figure}
\begin{center}
\epsfig{file=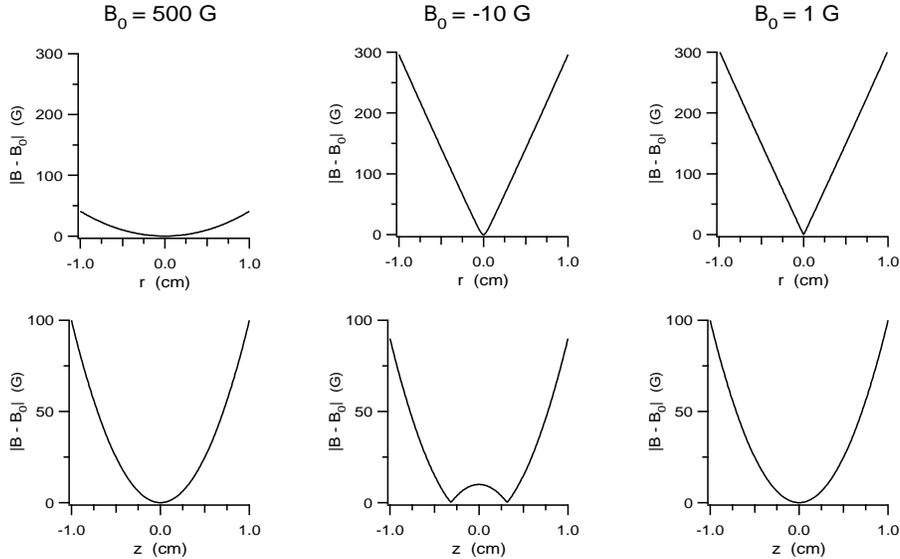,height=3in}
   \caption[Bias field compensation in an Ioffe-Pritchard trap]{Bias
   field compensation in an Ioffe-Pritchard trap is important for 
   tight radial confinement.  The magnetic field in an IP trap characterized 
   by a 
   radial gradient of 300 G/cm and an axial curvature of 200 
   G/cm$^{2}$ is shown for three bias fields 
   $B_{0}$.
   The upper row displays radial cuts, and the 
   lower row displays axial cuts of the 
   magnetic field profile.  
   In the first column, radial confinement is softened as a result of 
   the large bias 
   field. In the second column, the bias field is over-compensated, resulting in 
   a pair of zero field crossings along the axis of the trap.  
   In the third column, the bias field is tuned correctly, resulting in 
   tight radial confinement and no zero field crossings.}
   \label{fig:ipbias}
\end{center}
\end{figure}

The Ioffe-Pritchard trap has been used in many BEC experiments.
The most straightforward implementation consists of two 
pinch coils and four Ioffe bars \cite{berg87}.  The Ioffe bars 
generate the radial gradient field $ B' $, while the pinch coils 
produce a bias field and the curvature term $ B'' $.  
Most of the
bias field is usually canceled by a pair of additional 
``anti-bias'' coils.  By lowering $B_0 $, radial confinement is increased 
(eq.~\ref{eq:ipeffectivecurve}).  
The bias field should be just high enough 
to suppress Majorana flops.  A typical value is 1 G, although we 
have used values as low as 0.4 G, and smaller values are probably still stable.  
If the bias field is over-compensated, the field will cross zero and 
Majorana flops will occur (fig.~\ref{fig:ipbias}).  Since both the pinch 
coils and the anti-bias coils can produce fields on the order of 500 to 1000 
G,
the cancellation has to be carefully controlled.  We 
achieved best results when both sets of coils were powered in series 
by the same power supply, and fine bias adjustments were made by 
carefully moving a few loops of the anti-bias coils.

 Since the pinch coils and the Ioffe bars are very close to the atoms, 
 this configuration is very efficient in producing a tight 
 trapping potential.  It has been used in several BEC experiments 
 \cite{erns98,frie98}. 
 The optical access
 can be improved by elongating the pinch coils, or using a cloverleaf 
 configuration \cite{mewe96bec} (fig.~\ref{figclover})
 where the Ioffe bars are removed, 
 and the radial field gradient is created by a pair of four coils surrounding 
 each pinch coil.  This configuration is used in several experiments because 
 it has open, $360^{\circ}$ access in the $x$-$y$ plane, and the coils 
 may be 
 placed outside the vacuum chamber. 
 Some simpler winding patterns 
 (baseball \cite{myat97,berg87}, yin-yang \cite{berg87}, 
 three-coil \cite{essl98,sodi98}, four-dee \cite{hau98}) don't 
 allow for independent control of $ B' $ and $ B'' $, although the radial 
 confinement can still be varied through $ B_{0} $.   
 Other implementations of the Ioffe-Pritchard trap include permanent 
 magnet traps \cite{toll95,brad97bec}, and traps 
 with ferromagnetic pole pieces \cite{desr97}.
 
\begin{figure}
	\begin{center}
	\epsfig{file=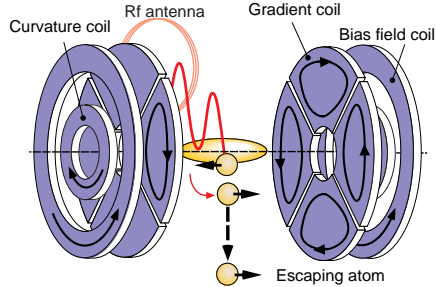, height=1.5in}
	\caption[Cloverleaf trap]{In a cloverleaf trap, Ioffe bars are 
	replaced by eight ``cloverleaf'' coils surrounding the pinch coils, 
	providing 360 degree optical access.  Evaporation is done by 
	selectively spin-flipping atoms into untrapped states with rf 
	radiation.\label{figclover}}
	\end{center}
\end{figure}

\subsubsection{Mode matching\label{sec:modematching}}
Evaporative cooling can increase phase-space density by many 
orders of magnitude as long as the atoms rethermalize quickly compared 
to the trap lifetime.  This makes the
collision rate during cooling even more important for BEC than the initial 
phase-space density of the pre-cooled atoms.
In order to maximize the collision rate, atoms are transferred into an 
optimized (``mode matched'')  magnetic trap, and then adiabatically
compressed.

``Mode matching'' between an atom cloud and a trap is achieved when
the transfer of atoms 
maximizes phase-space density.
This also optimizes the 
elastic collision rate which will be achieved after compressing the 
trap (sect.~\ref{sec:compression}).
This is because all properties of the
compressed cloud, including the collision rate, are completely 
determined 
by the number of atoms $ N $ and the phase-space 
density $ \mathcal{D} $.  Adiabatic compression conserves $ N $ and 
$ \mathcal{D} $, so a loss in phase-space
density when loading the magnetic trap corresponds to 
an equal loss in the compressed trap. 
For a power law potential $U(r) \propto r^{d/\delta}$ in $d$ dimensions,
the volume of a cloud at temperature $T$ scales as $V \propto T^{\delta}$.
The collision rate $ \Gamma_{el} $ for atoms with a collisional cross 
section $ \sigma $ and a thermal velocity $ v $ is equal to:

\begin{equation}
\Gamma_{el} = n \sigma v \propto
\mathcal{D}^{\frac{\delta - 1/2}{\delta + 3/2}} 
N^{\frac{2}{\delta + 3/2}}
\end{equation}
In particular, for a 3D harmonic trap ($ \delta = 3/2 $), the 
collision rate is proportional to $ \mathcal{D}^{1/3} $.
This implies that the loss in collision rate due to non-ideal 
transfer is much less severe than the loss in phase-space density.

Atoms are transferred into a magnetic trap by suddenly switching off 
the  MOT, applying polarization gradient cooling for a few milliseconds, 
and then suddenly turning on 
the magnetic trap and building a new potential around the 
atoms.  
When transferring a spherical, Gaussian shaped cloud of atoms with an  rms 
radius $ r_0 $ and
temperature $ T $ into a harmonic trap, 
phase-space density is conserved if the potential $ \frac{1}{2} 
\kappa r^{2} $ has a stiffness 
$ \kappa = \kappa_0=k_B T/r_0^2 $.
This will ensure that the atoms maintain their volume and 
temperature. If the 
trap is too tight, the atoms will be heated by the transfer.  If it is 
too loose, the atoms will  expand non-adiabatically.  In 
either case phase-space density will be lost.  For 
ideal loading we have used the term ``mode matching,'' in analogy 
with the coupling of a laser beam into a single mode fiber, where the 
efficiency suffers from focusing which is either too weak or too tight.

If the magnetic trap is not mode-matched, i.e. $ \kappa \neq  
\kappa_0 $, then the phase-space density $ \mathcal{D} $ is reduced from 
its initial value $ \mathcal{D}_0 $ \cite{mewe97thes}:

\begin{equation}
\frac{\mathcal{D}}{\mathcal{D}_0} = \frac{8 \left( \frac{\kappa}{\kappa_0} \right)^{3/2} 
}{ \left( 1 + \frac{\kappa}{\kappa_0} \right)^3} \label{eq:modepsdloss}
\end{equation}
This loss is only a weak function of mode mismatch.  For 
example, the 
phase-space density decreases only by half when 
$ \kappa $ is four times larger or smaller than $ 
\kappa_0 $.  More importantly, the collision rate
is still 
80\% of its maximum value.

Mode matching into a Ioffe-Pritchard trap is done with a high 
bias field ($ \mu_{m} B_0 > k_{B} T $).
This requires
(eq.~\ref{eq:ipeffectivecurve}):

\begin{eqnarray}
B'' = \frac{\kappa_0}{\mu_{m}}  \label{eq:modematchz}  \\
\frac{ {B'}^2 }{B_0}=\frac{3 \kappa_0}{2 \mu_{m}} \label{eq:modematchr}
\end{eqnarray}
It is important that the initial cloud radius $r_{0}$  be much smaller than the 
distance to the trap instabilities (eq.~\ref{eq:zinstab}), otherwise, atoms will spill out 
of the trap.  Inserting equations \ref{eq:modematchz} and \ref{eq:modematchr}
into equation \ref{eq:zinstab} gives an instability distance of
$ z_{inst} = \sqrt{B_0 (2 \mu_{m} / 3 \kappa_{0})} = B' (2 \mu_{m} / 3 \kappa_{0}) $.  
This shows that both
a large bias field and a large gradient are necessary, which is 
unfortunate because this
requires a fast rise time of the high-current 
power supplies.

For $ \mu_{m} B_{0}< k_{B} T $, the radial confinement is linear, and 
mode-matching would require $ \mu_{m} B' \approx k_{B} T/r_{0} $.  With this 
condition we find that $ z_{inst} \approx r_0  $, and therefore
mode matching cannot be done at low bias field.

\subsubsection{Adiabatic compression\label{sec:compression}}
Adiabatic compression plays a crucial role in BEC experiments because 
it increases the collision rate before evaporative cooling.  
In our 
first BEC demonstration \cite{davi95bec} adiabatic compression 
increased the collision rate by a factor of 20, resulting in BEC after
only 7 seconds of evaporative cooling! 

If, after loading, the 
potential (again characterized by a volume  $ \propto T^{\delta} $) 
is adiabatically increased 
by a factor $ \alpha $, the temperature 
rises by a factor $ \alpha^{2 \delta /(2 \delta +3)} $ and the density by $ 
\alpha^{3\delta/(2\delta+3)} $. 
Phase-space density is constant, but the elastic collision 
rate increases 
by a factor $  \alpha^{4 \delta/(2 \delta+3)} $ as long as the 
cross-section is constant.

Adiabaticity requires
\begin{equation}
\frac{d \omega_{trap}}{d t} \ll \omega_{trap}^2 \label{eq:adiabaticcond}
\end{equation}
Violation of adiabaticity, however, doesn't have severe consequences.  
If the compression is done suddenly in a harmonic trap ($\delta = 
3/2$), the collision 
rate increases by a factor  $ 2 \alpha^{3/2} / (1 + \alpha) $ rather 
than $  \alpha $.   For a sudden 
compression by a factor of five, one still reaches 75\% of the maximum 
possible collision rate.  
Note that this is the same loss we would see if we transferred atoms 
into a trap which was mode-mismatched by a factor $ \alpha $ 
(eq.~\ref{eq:modepsdloss}). 
Non-mode matched transfer is equivalent to a mode-matched transfer plus 
a sudden (de-)compression to the same potential strength.

Compression in an Ioffe-Pritchard trap involves 
one more complication. Initially, the aspect ratio is adjusted to 
$ \approx 1 $ to ensure mode-matched loading of a spherical cloud.  
This requires a high bias field $B_0$.  Subsequent compression includes 
lowering the bias field to achieve the maximum radial confinement.  
This process elongates the cloud into a cigar shape.  In addition to 
the usual adiabaticity criterion (\ref{eq:adiabaticcond}), 
the compression has to be 
slow compared to the elastic collision rate. Otherwise, the 
anisotropic compression would lead to an anisotropic temperature and 
subsequent loss of phase-space density during equilibration.

Anisotropic heating is hard to avoid when the initial collision rate is 
low. Fortunately, nature is forgiving. 
Radial compression by a factor of $ \alpha $ causes the temperature to rise 
by a factor of $ \alpha ^{1/3} $ if done adiabatically, and by a factor
of $ ( 2 \sqrt{\alpha}+1 )/3 $ (after rethermalization) if done suddenly.  
Since the elastic 
collision rate is inversely proportional to temperature in a given 3D 
harmonic oscillator potential, one still obtains 
94\% of the maximum possible collision rate when a 2D compression by
a factor of $ \alpha = 5 $
is done quickly with respect to the thermalization rate.

Furthermore, the radial compression in a Ioffe-Pritchard trap usually 
takes us to the small bias field regime  ($B_{0} < k_{B} T / \mu_{m} $),  
changing the radial confinement from harmonic 
to linear.  As pointed out by Pinkse \etal~\cite{pink97} and 
discussed in sect.~\ref{revers}, this 
increases the phase-space density by a factor of $e$.  This increase is 
due to collisions which change the population of energy levels, but 
conserve entropy.  

When the bias field $B_{0}$ is lowered adiabatically, 
the collision rate $ \Gamma $ increases by
\begin{equation}
\frac{\Gamma_{final}}{\Gamma_{initial}} = \frac{e^{1/2}}{2} \left[ 
\frac{ \mu_{m} B_{0 initial} }{k_B T_{initial} } \right]^{1/2} 
\label{eq:harmtolincollisions}
\end{equation}
We can apply 
estimates for adiabatic compression in
a harmonic trap to linear potentials by noting that $ B' $ 
confines a cloud 
at the same temperature and density as a harmonic trap with curvature 
$ \mu_{m} B_{equiv}'' =\mu_{m} B'/r = \mu_{m}^{2} {B'}^{2} / k_B T $, where r 
characterizes the size of the cloud.
If we use this equivalent curvature to characterize the compressed 
trap, we get the same result as eq.\ 
(\ref{eq:harmtolincollisions})
--- apart from the factor 
$ {e^{1/2}}/{2} $.  This emphasizes that lowering the bias 
field beyond a value $ k_B T / \mu_{m} $ does not compress the cloud 
any further.

\subsubsection{Figure of merit for magnetic traps}

The major goal when designing magnetic traps for BEC experiments is to get the 
highest collision rate after compression.  
Therefore, we regard this as the figure of merit of a trap.
In a 3D harmonic oscillator, the collision rate increases due to 
adiabatic compression in proportion to the geometric mean of the 
three curvatures:  $ ({B''}_{radial}^2 {B''}_z )^{(1/3)} $.

In a Ioffe-Pritchard trap, after full compression to a final temperature 
$ T $, 
the equivalent radial curvature is $  \mu_{m} {B'}^{2}/k_{B} T $. 
Knowing that the final 
temperature scales with the geometric mean of the trapping 
frequencies, we find that the collision rate will depend on $ B'  
B''^{1/4} $, implying that it is much more important to have strong field 
gradients than curvatures.

Let's compare this result to a TOP trap, assuming that the TOP trap is 
compressed until the trap depth due to the circle of death is  $ 5 k_B T $, 
a typical barrier for evaporative cooling.  In this case,
the effective radial curvature is

\begin{equation}
    {B''}_{TOP,r} = \frac{\mu_{m} {B'}^2}{40 k_{B} T}
\end{equation}
This implies that the TOP trap corresponds to an IP trap with a
$ \sqrt{40} $ smaller value of $ B' $.  Along the axial direction, the effective 
curvature $ B'' $ of a TOP trap is typically 100 times larger than 
that of an 
IP trap.  The figure of merit, $ B'  {B''}^{1/4} $, 
is slightly higher for 
an IP trap, but this does not take into account the fact that one can usually
obtain higher gradients in spherical 
quadrupole traps than in IP traps.

If we wish to compare the confinement for condensates after cooling, 
the effective radial 
curvatures of TOP traps and IP traps are $ B'^{2}/2 B_{0} $ and $ B'^{2}/ 
B_{0} $ respectively.  The maximum confinement depends on how small $ B_{0} $
can be, as determined by the adiabatic condition which 
ensures stability of the trap (eq.~\ref{eq:majorana}). Usually $ B_{0} $ is smaller 
in IP traps 
than in TOP traps, because the use of time-dependent 
fields in the TOP trap requires larger Larmor frequencies to prevent 
non-adiabatic spin flips.
However, experiments have not yet explored 
the lower limits for $ B_{0} $ in either type of trap (see 
\cite{suku97} for a theoretical treatment).

The bottom line is that both traps work very well.   
The TOP trap might have advantages in studying 
vortices due to the built-in rotation, and does not require 
careful bias-field cancellation. The 
advantages of the IP trap are the variable aspect ratio and the use of 
only dc fields.
Several groups have now built magnetic traps using room temperature 
electromagnets which provide sufficient confinement for evaporation 
to BEC.  Even tighter confinement, and therefore faster evaporation,
could be achieved with permanent magnets or superconducting magnets,
but at the price of less flexibility.

\subsection{Evaporative cooling\label{sec:evapcool}}

Gaseous Bose-Einstein condensates have so far only been obtained by 
evaporative cooling.  Evaporative cooling is done by continuously 
removing the high-energy tail of the thermal distribution from the 
trap.  The evaporated atoms carry away more than the average energy, 
which means that the temperature of the remaining atoms decreases.  
The high energy tail must be constantly repopulated by collisions, thus 
maintaining thermal equilibrium and sustaining the cooling process.
Evaporative cooling is a common 
phenomenon in daily life --- it's how hot water cools down in a 
bathtub or in a cup of coffee. 
Evaporative cooling of trapped atoms 
was developed at MIT as a method for cooling atomic hydrogen 
which had been pre-cooled by cryogenic methods 
\cite{hess86,masu88,doyl91stic}.  The first suggestion by Hess 
\cite{hess86} was soon followed by an experimental demonstration 
\cite{masu88}.  Evaporative cooling has been reviewed in 
\cite{walr96,kett96evap}, and we only briefly summarize the basic 
aspects here.


The essential condition for evaporative cooling is a long lifetime of the 
atomic sample compared to the collisional thermalization time.  
Trapped atom clouds are extremely dilute (about ten orders of 
magnitude less dense than a solid or a liquid) and collisional 
thermalization can take seconds.
A major step was
taken in May 1994 when the MIT and JILA groups
combined laser cooling with evaporative cooling, extending the 
applicability of evaporative cooling to alkali atoms 
\cite{davi94icap,petr94icap}.

In these experiments, the evaporation of atoms was controlled by radio 
frequency 
radiation (rf induced evaporation).  
This technique was proposed by Pritchard \cite{prit89} and 
Walraven \cite{hijm89} and first demonstrated by our group when 
spatial truncation of magnetically trapped atoms was observed 
\cite{kett93evap}.  Increases in phase-space density were reported by 
the MIT and Boulder groups at IQEC in May 1994 and at ICAP-XIV 
\cite{davi94icap,petr94icap} and published after further progress had 
been achieved \cite{davi95evap,petr95}.  Other early work on 
evaporation of alkali atoms was done at Rice  
\cite{brad95bec} and 
Stanford \cite{adam95}.

In rf-induced evaporation, the rf radiation flips the atomic spin.  
As a result, 
the attractive trapping force turns into a repulsive force and expels 
the atoms from the trap (fig.~\ref{figclover}).  
This scheme is energy-selective because the 
resonance frequency is proportional to the magnetic field, and 
therefore to the potential energy of the atoms.  In the case of 
transitions between magnetic sublevels $m_{F}$, the resonance 
condition for the magnetic field strength $B$ is $|g| \mu_B B = \hbar 
\omega_{rf}$, where $g$ is the Land{\'e} g-factor and $\mu_B$ the Bohr 
magnetron.  Since the trapping potential is given by $m_{F} g \mu_B 
[B(r)-B(0)]$, only atoms which have a total energy $E > \hbar 
|m_{F}|(\omega_{rf}-\omega_0)$ will evaporate ($\omega_0$ is the rf 
frequency which induces spinflips at the bottom of the trap).  Rf induced 
evaporation has several advantages over other evaporation methods  
\cite{kett96evap}.  First, the evaporation process can be 
completely separated from the design of the magnetic trapping 
potential.  In particular, there is no need to 
weaken the trapping potential in order to lower its depth.  
This makes it easier to reach runaway evaporation where the decrease 
in temperature leads to a net increase in density and collision rate 
despite the loss of atoms.
Furthermore, atoms evaporate from the whole surface 
where the rf resonance condition is fulfilled.  This make the 
evaporation  three-dimensional in velocity 
space, and therefore very efficient.

Let us briefly summarize the major 
developments in the three years since our earlier review 
\cite{kett96evap}.  First, evaporative 
cooling has become a standard technique of atomic physics.  The 
complete list of experiments in 1995 comprised only six groups --- 
since then, thirteen more groups have implemented evaporative cooling and 
achieved BEC, and there are more to come!  Second, 
the evaporation process has been simulated in more detail 
\cite{wu96sim,holl96traj,wu97,sack97evap,arim98}.
In particular, Monte Carlo simulations 
can account for realistic experimental situations \cite{arim98} providing 
insight on how to optimize the cooling cycle.  A careful study at 
Amsterdam demonstrated that saddle-point evaporation is 
one-dimensional and was the bottleneck for achieving BEC in hydrogen 
\cite{pink98}.
This problem was finally overcome when rf evaporation was implemented 
\cite{frie98}.  
Several groups reported 
evaporation in TOP traps by the circle of death 
\cite{ande98pra,phil98var,han98}.
This scheme should be one- or 
two-dimensional depending on the orientation of the circle of death, 
but no striking detrimental effects due to that have been reported.

Sympathetic cooling between rubidium atoms in different hyperfine 
states was observed \cite{myat97}.
Combining sympathetic cooling with cryogenic 
buffer gas loading \cite{kim97} should vastly 
extend the number of species 
which can be cooled.  Finally,
cooling by adiabatic deformation of the trapping potential 
has been achieved \cite{pink97,stam98rev}.  The relation of this 
reversible, adiabatic scheme to irreversible evaporative cooling 
is discussed in section \ref{revers}.

\subsubsection{Trap loss and heating} Evaporative cooling requires a favorable ratio of 
elastic to inelastic collisions.  In the case of alkali atoms, as shown in 
fig.~\ref{fig:colvstemp}, 
the dominant loss mechanism over a large range of temperatures and 
densities is background gas 
collisions.  Therefore, experimentalists have made major efforts to reconcile the loading and 
trapping techniques with an extreme UHV environment.  The standard technique is to use a 
combination of an ion pump with a titanium sublimation pump, and to implement differential 
pumping to isolate the trapped atoms from the high gas load of the source of atoms.  Our atom 
beam source (the atomic beam oven) is separated from the magnetic trap by about one meter, and 
an additional 10 cm-long narrow tube ensures good differential pumping.  After loading the MOT 
for about 2 seconds, we close a shutter near the atomic beam oven, 
isolating our sample in a 
perfect UHV environment.  More compact experiments have to achieve the separation between a 
region of high vapor pressure and UHV with less space, making differential pumping more 
difficult to implement.

Several groups have reported that a background pressure in the low $10^{-11} $ mbar range is 
necessary to obtain magnetic trap lifetimes of one minute.  The background collision rate per 
trapped atom is $n_{bg} \sigma_{bg} v_{bg} $, where $n_{bg}$ and $v_{bg}$ 
denote the density and thermal velocity of the background gas 
molecules.
The collision cross section, $ \sigma_{bg} $, is typically $10^{3} 
\un{\AA}^{2}$, much larger than the size of the molecules.
This reflects the major contribution of the long-range van-der-Waals potential 
and small-angle scattering to the total cross-section.
Background collisions 
are discussed in detail in refs.~\cite{myat97thes, corn98var}.

Another possible source of heating is magnetic field 
fluctuations which modulate the trapping potential. The simple 
model of ref.~\cite{sava97} predicts that the characteristic time $t_{heat}$ for the exponential 
growth of the energy of the trapped atoms is given by

\begin{equation}
	\frac{1}{t_{heat}} = \pi^{2} \nu_{trap}^{2} S( 2 \nu_{trap} ) 
\end{equation}	
where $\nu_{trap}$ is the trapping frequency and $S(\nu )$ is the power spectrum of the 
fractional magnetic field noise.  An estimate for $S(\nu )$ is given by 
$\epsilon^{2}/ \Delta \nu $ where $\epsilon$ denotes the rms fractional fluctuations of the magnetic 
field in a bandwidth of $ \Delta \nu $.  Assuming an inductance
limited bandwidth of 1 kHz 
and a trapping frequency of 100 Hz, we obtain
\begin{equation}
t_{heat} = \left[ \frac{10^{-2} \mbox{sec}}{\epsilon^{2}} \right]
\end{equation}
Current fluctuations of 1\% result in $t_{heat}=100 $ seconds.  
Therefore, commercial power supplies 
with a current stability of $10^{-3}$ to $10^{-4}$ are fully adequate for achieving BEC.  
Stability requirements are more severe at higher trapping frequencies as in optical traps.

Another source of heating is mechanical noise, but it seems not to be very critical.  As a 
precaution we usually don't use any mechanical pumps close to the experiment, but there were 
occasions where these pumps were running without affecting Bose condensation.  However, a 
fan-cooled, high-power rf amplifier which was on the experimental table prevented us from 
getting BEC until we isolated it with a foam pad. So although BEC 
is usually forgiving, it is still wise to consider sources of mechanical 
vibration as potential problems.

\subsection{Manipulation of Bose-Einstein condensates}

Having achieved BEC, one would like to study the properties of condensates in different shapes 
and symmetries, and to explore their dynamic behavior.  Therefore, we need tools to shape and 
excite the condensate.  The extremely low temperature of the condensate and the high sticking 
probability of atoms on cold surfaces forbid direct-contact manipulation 
of condensates, limiting us to  
various forms of electromagnetic fields.

\subsubsection{Magnetic fields}

\begin{figure}[htbf]
\epsfxsize=60mm
\centerline{\epsfbox[0 0 113 128]{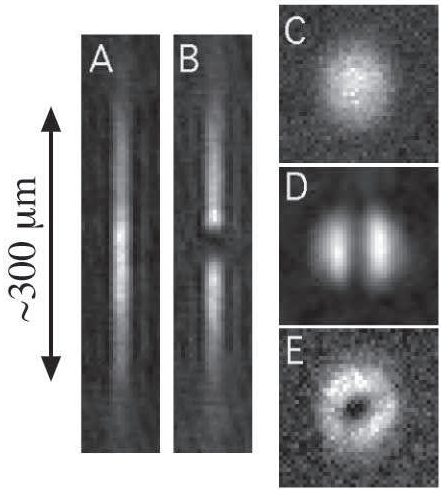}}
\caption[Getting a Bose-Einstein condensate into shape!]
{Getting a Bose-Einstein condensate into shape!  In-situ 
phase-contrast images show the variety 
of condensate shapes which can be realized with magnetic traps and far-off resonant blue-detuned 
laser beams. 
In a cloverleaf trap, the trap shape is varied from elongated (A,B) 
to spherical (C-E) by decreasing the radial confinement.
The light-shift potential of a focused, blue detuned laser can be 
used to repel atoms.
Depending on the intensity and shape of the laser beam, this 
results either in two condensates with adjustable separation (B,D) or in a 
toroidal (``doughnut'' shape) condensate (E).  
The same scale is used in all images.
Figure taken from ref.~\cite{mies98rev}.}
\label{fig:varioustraps}
\end{figure}

Magnetic fields have been used to adiabatically expand the condensate 
\cite{ande95,andr96,andr97prop} and to change its aspect ratio.
By varying the field parameters $B_{0}, B'$ and $ B''$ of a Ioffe-Pritchard 
trap (sect.~\ref{sec:iptraps}) one can 
vary the aspect ratio of the condensate. In our experiments, we have varied the aspect ratio between 
more than 20 and unity (fig.\  \ref{fig:varioustraps}).

Magnetic fields can also be used to launch atoms.
The standard technique of accelerating molasses would severely heat 
up atomic samples
at sub-recoil temperatures.  Thus, we have used a pulsed magnetic field gradient to launch 
a Bose condensate into an atomic fountain \cite{town96tops,town97icap}.

\subsubsection{Optical dipole forces}
\label{sec:odforces}

The dipole force of focused off-resonant laser beams can be used
 for ``microsurgery'' on condensates.  
This is not possible with magnets or coils which are typically a few 
cm away from the trap center.
Optical dipole forces have been used to create deformed (non-parabolic) trapping 
potentials (sect.~\ref{revers}).  We have created toroidal  condensates 
as part of
an effort to see 
persistent currents (sect.~\ref{sec:vortices})  and realized double well potentials 
to make two separated 
condensates which subsequently interfered \cite{andr97int}.  Finally, infrared light has been 
used to achieve all-optical confinement of a condensate \cite{stam98odt}.  The combination of 
magnetic fields and far-off-resonant light is very versatile --- the magnetic field provides the 
general confinement, blue light is used to add blips and red light to add dips to the trapping 
potential.  So we have all the tools to shape, slice, kick, shake and stir condensates!

\subsubsection{Rf fields}
\label{rffields}

Radio-frequency radiation can be used to change the hyperfine state of trapped atoms.  This was 
used to switch atoms from a trapped to an untrapped hyperfine state, thus realizing an output 
coupler for an atom laser \cite{mewe97} (sect.~\ref{sec:outputcoupler}).  Rf methods were also used to transfer 
condensates between trapped states, either in an optical trap \cite{stam98odt} or in a magnetic 
trap using a two-photon rf transition \cite{hall98dyn,matt98,hall98phas}.

Another important aspect of rf fields is that they can be used to set the trap depth of the 
magnetic trap.  We introduced this method, referred to as ``rf shielding 
of the condensate,'' when we 
realized that the lifetime of the condensate is much longer when the trap depth is limited to a 
value on the order of microkelvin~\cite{mewe96bec}.  The tentative explanation is that this 
removes from the trap energetic atoms,
which are produced by inelastic collisions and collisions with 
the hot background gas in the UHV chamber,
before they interact with other atoms in the 
condensate.  However, a detailed study of the effects of ``rf shielding'' on heating and trap 
loss has not provided a consistent picture \cite{myat97thes,corn98var}.

\subsubsection{Bragg and Raman transitions}
Two-photon transitions induced by two intersecting laser beams can be used to manipulate 
condensates in various ways. 
The absorption and stimulated emission of photons transfers momentum 
to the condensate.
Bragg and Raman scattering have been used as a beam 
splitter for a condensate, and to accelerate condensed atoms up to 12 photon 
recoil momenta~\cite{kozu98,phil98var}. 

\subsubsection{Excitation of sound}
Sound is excited by perturbing an equilibrium situation creating an oscillatory response.  
All the processes discussed in this section can be used for that purpose, and many results are 
reviewed in section \ref{sound}.  Here we briefly summarize the many different ways to excite sound.

Modulation of the 
magnetic trapping field has been employed to excite the center-of-mass dipole oscillation (by 
translating the trap origin) or shape oscillations (by modulating the strength 
of the trap)~\cite{jin96coll,mewe96coll,jin97,stam98coll}.
Excitation of collective modes with angular momentum ($m\neq 0$) requires a 
modulation which breaks the axial symmetry of the 
trap~\cite{jin96coll,jin97}.

Our group has used a far-off resonant laser 
beam to to create excitations which could not be excited by modulating the trapping potential.  
This includes the antisymmetric dipole oscillation \cite{stam98coll} and 
localized sound pulses \cite{andr97prop}.

Rf transitions can excite collective excitations in various ways.  Sudden truncation of the 
wings of a cloud by rf evaporation leads to shape oscillations~\cite{mies98form}.  Sudden 
output coupling~\cite{mewe97} induces shape oscillations by creating a 
trapped cloud which is larger than in 
equilibrium due to reduced mean-field energy.  Transferring a 
condensate to a different hyperfine state can change either
the atomic magnetic moment or the 
mean field energy.  Both cases lead to oscillations.  This technique was used to determine the 
difference between the scattering lengths of two hyperfine states \cite{matt98}.

\subsection{Atoms for BEC \label{sec:atomsforbec} }

A review of experimental BEC would be incomplete without presenting 
the major actors, the atoms themselves.  The only indispensable 
property for BEC is that the atom be bosonic, which is not too 
restrictive: all stable 
elements 
with the exception of beryllium have at least one bosonic isotope.  
The choice of an atom for a BEC experiment is mainly determined by the 
cooling and trapping techniques.  Magnetic trapping requires atoms 
with a strong magnetic moment and therefore an unpaired electron.  
Laser cooling favors atoms with strong transitions in the visible or 
infrared region where commercial cw lasers are available. To date
BEC has been realized with hydrogen, lithium, rubidium and sodium, and  
experiments on cesium, potassium, metastable helium and neon are in 
progress.

A requirement for evaporative cooling is a favorable ratio between 
the elastic collision rate (which provides evaporative cooling) and 
the                           
inelastic collision rate (which leads to trap loss and heating).  
Collisional properties of the 
relevant atoms with respect to evaporative cooling are discussed in 
ref.~\cite{kett96evap}, and in the contributions of Dan Heinzen and Dan 
Kleppner to this volume.  Several review papers \cite{verh95, wein95, 
walk94} summarize the field of cold collisions.

\begin{figure}
    \begin{center}
   \epsfig{file=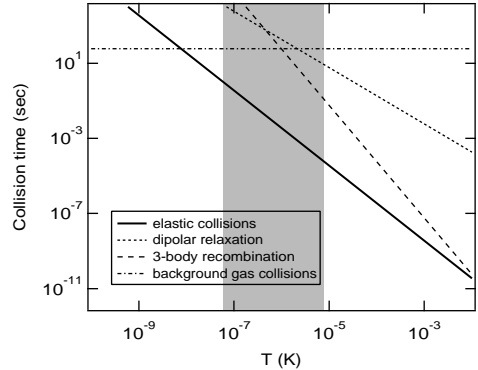,height=2in}
   \caption[Elastic and inelastic collision times for a sodium gas at 
   the BEC transition]{Mean collision time for several elastic and 
   inelastic processes in a sodium gas as a function of 
   temperature at the critical density for Bose-Einstein 
   condensation. 
   The ``BEC window,'' where 
   the lifetime of the sample exceeds 
   0.1 seconds and the rate of elastic collisions is faster than 1 Hz
   is shaded.
   This figure uses a scattering length of $ 50 \, a_{0} $ 
   and rate coefficients for
   two- and three-body inelastic collisions of $ 10^{-16} \mbox{cm}^{3} 
   \mbox{s}^{-1}$ and $ 6 \cdot 10^{-30} \mbox{cm}^{6} 
   \mbox{s}^{-1} $ respectively.}
 \label{fig:colvstemp}
 \end{center}
\end{figure}

Figure \ref{fig:colvstemp}  shows the situation for sodium.  The elastic and 
several 
inelastic rates are plotted vs.\  temperature (at the density of the 
BEC transition).  The region where the elastic rate is much larger 
than the inelastic rate spans most of temperature range displayed. 
Within this region, we 
have shaded the ``BEC window'' where the lifetime of the sample exceeds 
0.1 sec and where the rate of elastic collisions is faster than 1 Hz.  
The fact that 
it covers several orders of magnitude in temperature and density allows 
studies of BEC over a large range of parameters.  In our experiments, 
we have realized condensates with densities between $2 \times 10^{13} 
$ and $3 \times 10^{15} \, \mbox{cm}^{-3}$ and crossed the BEC 
transition at temperatures from 100 nK to 5 $\mu$K (see sect.~\ref{opticalsection}).  
BEC can be studied at 
lower and higher densities by expanding or compressing 
a condensate after it has been formed. 
Although 
this window is in agreement with early estimates of collisional 
properties \cite{prit86trap}, it is fortunate that 
this window exists at all.  We know now that 
collisional properties at ultralow temperature are more complicated 
than we first expected.  
Experiments suggest 
that there may be no window at all for magnetically trapped 
cesium (see J. Dalibard's contribution to this volume and 
ref.~\cite{guer98,arnd97}).  $^{85}$Rb has a very small elastic cross section at 
intermediate temperatures which makes it hard to reach runaway 
evaporative cooling \cite{corn98priv}. 
Differences found in trap loss and heating for $^{87}$Rb atoms in 
the upper and lower hyperfine state are not understood \cite{myat97thes}.  
So it is fortunate that sodium and 
$^{87}$Rb have an almost ideal combination of properties.


\section{Techniques to probe Bose-Einstein condensates}
\label{probesection}
Everything we know about gaseous Bose condensates has been obtained by 
optical diagnostics.  ``Contact probes'' cannot be used because the 
samples are much smaller ($ \sim 10^7 $ 
atoms) than even a 10 $\mu$m sized probe ( $ \sim 10^{13} $ 
atoms), which would cause the atoms to equilibrate with the probe 
rather than the opposite.  With the ``cooling power'' of 
rf evaporation we can cool at most $10^8 $ atoms per minute, and 
it would take several months just to cool the sensor tip.  
Fortunately, optical diagnostics are very versatile, and the ease with 
which light scattering methods are implemented for dilute atomic samples 
is a major advantage over condensed-matter systems.

The two most important techniques for observing Bose-Einstein 
condensates are in-situ and time-of flight imaging.  In both 
cases, one obtains an image which 
reflects the density distribution of the atoms either in a trapped 
state or 
in ballistic expansion.  In this section we discuss the basic
interactions between atoms and light and the physical principles of 
absorptive and dispersive imaging.
Finally, we refer the interested reader to 
Appendix~\ref{sec:imageprocessing} for a discussion of
techniques for processing absorption and phase-contrast images.

\subsection{Atom - light interactions}
\label{sec:atomlight}
The interaction of atoms with a beam of light involves three processes: 
spontaneous absorption of photons, re-emission of photons, 
and shifting the phase of the transmitted light.  These properties are used in
absorptive, fluorescence, and dispersive imaging methods, respectively.

The interaction can be described by the complex index of refraction of the atoms 
$ n_{ref} = \sqrt{1+4 \pi n \alpha} $,  where $\alpha$ is the atomic polarizability,
and $ n $ is the density of atoms.  
Assuming $n_{ref} - 1 \ll 1$, the index of refraction for a two-level system can 
written (in the rotating wave 
approximation):

\begin{equation}
n_{ref}= 1 + \frac{\sigma_{0} n \lambda}{4 \pi} \left[
\frac{i}{1+\delta^{2}} - \frac{\delta}{1 + \delta^{2}} \right]
\label{eq:indexofref}
\end{equation}
where $ \sigma_{0} $ is the resonant cross-section  
(for a two level atom  $ \sigma_{0} = 6 \pi \lambdabar^{2} $), 
and $\delta  \equiv \frac{\omega - \omega_0}{\Gamma/2}$ is the 
detuning in half linewidths.
If more than two levels are involved
(e.g. Zeeman sublevels, see  sect. \ref{sec:swirls}), 
several resonances contribute to the polarizability. 

Eq.~\ref{eq:indexofref} assumes the limit of a weak probe laser beam.
Saturation may be non-negligible for near-resonant absorption imaging.
In this case, one has to replace $\delta^{2}$ by $\delta^{2} 
(I/I_{\mbox{SAT}})$ in the expression for the Lorentzian lineshape.
$I_{\mbox{SAT}}$ is the saturation intensity for the transition, and 
the intensity $I$ depends on $x$, $y$ (beam profile) and $z$ (due to 
absorption).
 
Assuming that light enters and exits the cloud at the 
same $ (x,y) $ coordinate (the ``thin lens'' 
approximation applied to the atom cloud),
the atoms simply
attenuate and phase shift the probe light:

\begin{equation}
{E} = t {E_0} e^{ i  \phi } 
 \label{eq:eoftphi}
\end{equation}

The transmission coefficient $ t $ and phase shift $\phi$ depend on the 
product of the column density $ \tilde{n} = \int n \cdot d z $, and 
 $ \sigma_0 $ (eq.~\ref{eq:indexofref}).

\begin{eqnarray}
& t  = e^{ - \tilde{D} / 2 } = \exp{ \left( - \frac{\tilde{n} \sigma_0}{2} \frac{1}{1 + 
\delta^2} \label{eq:t}
\right)}\\
& \phi  = -\delta \frac{\tilde{D}}{2} = -\frac{ \tilde{n} \sigma_0}{2} \frac{\delta}{1 + \delta^2} 
\label{eq:phi}
\end{eqnarray}
where  $ \tilde{D} = \tilde{n} \sigma_{0} / (1 + \delta^{2}) $ is the 
off-resonance optical density. 
Section \ref{sec:qmscatter} discusses a quantum mechanical 
treatment of light scattering.

Most BEC experiments have relied 
 on spatially imaging the real or imaginary part of the index of 
 refraction.  We discuss various imaging methods in this section.  An 
 alternative and in many aspects complementary method is high 
 resolution spectroscopy, where information is obtained from line 
 shifts and line broadenings.  Such methods have so far only been applied 
 to hydrogen condensates (see Kleppner and Greytak's article in this 
 volume), but can be extended to alkalis in the 
 form of Doppler sensitive Raman scattering.

\subsection{Absorptive and dispersive methods} 
\label{absdispmethods}

Absorption imaging is done by illuminating the atoms with a laser beam and 
imaging the shadow cast by the atoms onto a CCD camera.
Because photosensors aren't sensitive to phase, the 
absorption image shows the spatial variation of $ t^2 $ (eq.~\ref{eq:t}).

To image a transparent object, information encoded in the phase shift 
of the light must be converted into intensity information which can be 
detected by a photosensor.  Several methods are commonly used in microscopy.  
The Schlieren method was introduced by A. Toepler in 
1864 and the phase-contrast method by Fritz Zernike in 1934 (earning him the 
1953 Nobel prize in 
physics). All dispersive methods rely on the ability to separate scattered and unscattered
components of the probe light
and manipulate them independently.  This is usually done by spatially 
filtering in
a Fourier plane of the imaging system 
where the unscattered probe light comes to a focus.

\begin{figure}
  \begin{center}
	 \epsfig{file=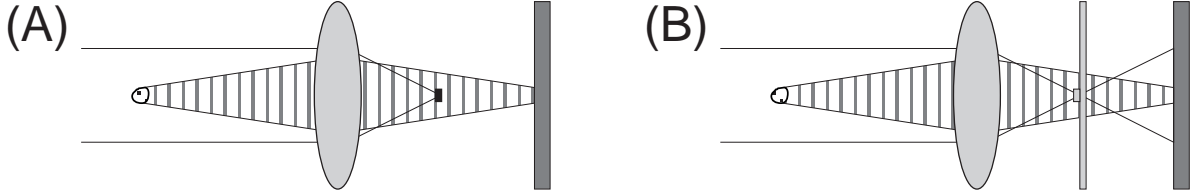,height=1 in}
     \caption[Dark-ground and phase-contrast imaging set-up]{Dark-ground (A) and phase-contrast (B)
imaging set-up.  Probe light from the left is dispersively scattered 
by the atoms.  In the Fourier plane of the lens, the unscattered 
light is filtered.  In dark-ground imaging (A), the unscattered light is blocked, 
forming a dark-ground image on the camera.  In phase-contrast imaging 
(B), the unscattered light is
shifted by a phase plate (consisting of an optical flat 
with a $\lambda / 4$ bump or dimple at the center), 
causing it to interfere with the
scattered light in the image plane.}
     \label{fig:dgsetup}  \label{fig:pcsetup}
  \end{center}
\end{figure}

{\it Dark-Ground Imaging:}
The simplest form of spatial filtering is to block the unscattered light 
by placing a small opaque object into the Fourier plane
(fig.\ \ref{fig:dgsetup}).

\begin{figure}
  \begin{center}
     \epsfig{file=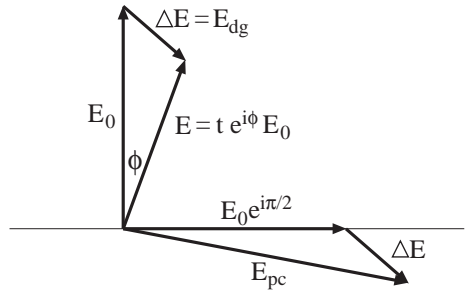,height=1.5in}
     \caption[Phasor diagram of dispersive imaging]{Phasor diagram of 
     dark-ground and phase-contrast imaging.  A ray of 
incident light with an electric field $ {E}_0 $ is scattered by the atoms, causing the light
to be attenuated and shifted in phase, resulting in the electric field $ 
{E} $.  The dark-ground
method images $ \Delta {E} = {E}_{dg} $, the difference between incident 
and scattered
electric fields.  Phase-contrast methods cause $ \Delta {E} $ and $ 
{E}_0 $ to interfere by
rotating the phase of $ {E}_0 $ by $90^{\circ}$, resulting in 
the field $ {E}_{pc} $.}
     \label{fig:phasor}
  \end{center}
\end{figure}

The probe light field after passing through the 
atoms (fig.\ \ref{fig:phasor}) can be separated into the 
scattered and unscattered radiation.

\begin{equation}
{E} = t {E}_0 e^{i \phi} = {E}_0 + \Delta {E}
\end{equation}
Blocking the unscattered light gives the dark-ground signal:

\begin{equation}
\langle I_{dg} \rangle   =   \frac{1}{2}
\left| {E} - {E}_0 \right|^2 =
 I_0 \left[1 + t^2 - 2 t \cos{ \phi} \right]\
     \label{eq:dgintensity}
\end{equation}
For small $ \phi $ the dark-ground signal is quadratic in $ \phi $.

{\it Phase-contrast imaging:}
Phase-contrast imaging can be regarded as a homodyne detection scheme 
in which the unscattered light acts as the local oscillator and 
interferes with the scattered radiation.  This is accomplished by 
shifting the phase of the unscattered light by $ \pm \pi / 2 $
in the Fourier 
plane of the imaging lens (fig.\ \ref{fig:pcsetup}).  This is done 
with a ``phase plate'' which is an optical flat with a small bump or dimple in the center. 
The result of this phase shift is shown in the phasor 
diagram in fig.~\ref{fig:phasor}.

The intensity of a point in the image plane is then

\begin{equation}
\langle I_{pc} \rangle  = \frac{1}{2}
\left| {E} + {E}_{0} \left( e^{ \pm i \frac{\pi}{2} } - 1 \right) 
\right|^2
 =  I_{0} \left[ t^2 + 2 - 2 \sqrt{2} t \cos{\left( \phi
      \pm \frac{\pi}{4} \right) } \right] \label{eq:pcintensity}
\end{equation}
where the $\pm$ sign corresponds to a phase plate which 
changes the phase of the unscattered light by $ \pm \pi/2 $.  
For small $ \phi $ one obtains

\begin{equation}
\langle I_{pc} \rangle \simeq I_{0} \left[ t^2 + 2 - 2 t \pm 2 t 
\phi
     \right]
\end{equation}
which is linear in $ \phi $.  This makes phase-contrast imaging 
superior to dark-ground imaging for small signals.

{\it Polarization-contrast imaging:}
Another dispersive imaging method takes advantage of the anisotropic 
polarizability of spin-polarized atoms \cite{sack97bec,brad97bec}.  It 
relies on different phase-shifts for orthogonal polarization 
components of the probe light.  A polarizer before
the camera can block the unscattered light, resulting in a 
dark-ground image.  Rotating the polarizer leads to 
interference of the two polarizations, resulting in a phase-contrast 
image.

In the case of high magnetic fields, the degeneracy between $ 
\sigma_{+} $, $ \sigma_{-} $, and $ \pi $ transitions is lifted, and 
it is possible to
choose a probe detuning such that only one polarization component is 
close to an atomic resonance and experiences a phase shift.  The 
other polarization component then serves as a uniform local oscillator.

This method has the advantage that it does not require spatial 
filtering in the Fourier plane, making it easier to correct 
for stray light (sect.~\ref{sec:pcprocessing}).  
However, this method is not universally 
applicable.  The standard imaging geometry for a Ioffe-Pritchard trap 
involves a probe beam propagating perpendicular to the axis of 
a low magnetic bias 
field.  For our typical detunings (much larger than the excited state 
hyperfine structure),
the phase shift is the same for both polarizations leading to zero 
signal.  However, the independence of the phase shift of polarization 
is advantageous for phase-contrast imaging, making it 
insensitive to the polarization of the probe beam.

{\it Imaging dilute clouds:} 
\label{sec:optimumimage}
Dispersive and absorptive methods can be discussed together by 
generalizing the previous treatment to a generic phase plate with a central 
spot which retards the unscattered light by an arbitrary phase $ 
\gamma $ and transmits a fraction $\tau^{2}$.  Absorption 
($ \tau = 1 $, $ \gamma=0 $), dark-ground ($ \tau = 0 $), and 
phase-contrast ($ \tau = 1 $, $ \gamma = \pm \frac{\pi}{2} $) imaging 
are obtained as special cases.

The intensity at a point in the image plane is:

\begin{equation} 
\left< I_{gen} \right> 
 = I_0  \left[ 1 + t^2 + \tau^2
     + 2 t \tau \cos{ \left( \phi - \gamma \right) } - 
        2 t \cos{\phi}  - 2 \tau \cos{\gamma} \right]
\end{equation}
If we expand this around $ \tilde{n} = 0 $ (i.e.\ $ t = 1 $, $ \phi = 0 $) and 
keep only the first order term we get

\begin{equation}
 \left< I_{gen} \right> 
= I_{0} \tau^{2} - I_0 \sigma_0 \tau 
     \left[ \frac{\delta}{1 + \delta^2} \sin{ \gamma } + 
     \frac{1}{1 + \delta^2} \cos{\gamma} \right]  \tilde{n}
     \label{eq:didn}
\end{equation}

This shows that for a given detuning $ \delta $, the signal is maximized 
when $ \tau = 1 $ and $ \tan{\gamma} = - \delta $.
The 
absolute maximum signal is obtained by choosing $ \delta = \gamma = 0 
$, the case of resonant absorption imaging.  However, the maximum 
phase-contrast signal ($ \gamma = \pm \pi / 2 $), obtained when $ \delta 
= 1 $, is only a factor of two lower. 
Dark-ground imaging is far from optimum at any detuning

High sensitivity is usually 
only an issue when imaging expanded clouds in time-of-flight, or for 
imaging the wings of the spatial distribution \cite{hau98}, because 
trapped condensates are generally optically dense.

{\it Imaging optically dense clouds:}
The typical resonant optical densities of our condensates (in the 
radial direction) is on the order of 300.  When such a dense cloud is 
probed with resonant light, the image is ``blacked out,'' making 
it extremely difficult to extract atomic column densities from the 
image.

Usually, an optimum detuning for absorption imaging gives about 50\% 
absorption.  However, for non-zero detuning, the atomic 
susceptibility has a finite real component which causes the 
atoms to refract light like a lens.  This does not affect the images 
as long as all of the refracted light is collected by the imaging 
system.  Otherwise, the refracted light appears as a false absorptive 
signal.  Since the magnitude of this effect varies across the cloud, it 
affects both relative and absolute density  measurements.

The refraction angle for a cloud of atoms with 
diameter $ d $ and a maximum phase shift $ \phi $ can be estimated as $ {2 \lambda 
\phi} / { \pi d } $.  As long as this phase shift is less than $ \pi / 2 $, 
the refraction angle is smaller than the diffraction angle $ \lambda / 
d $ due to the finite size of the object. For larger phase shifts, 
refraction leads to a noticeable additional divergence of the light after 
passing through the atoms. For a diffraction limited imaging system, 
diffraction of the smallest resolvable object just fills the solid angle 
extended by the lens.  Therefore, a dispersively dense object (i.e.\ 
phase shifts larger than $ \pi / 2$) will degrade the spatial 
resolution for 
quantitative absorption imaging.

For a cloud with a resonant optical density $ \tilde{D}_0 = \tilde{n} 
\sigma_{0} \gg 1 $, the highest 
contrast for absorption imaging is obtained with a detuning $ \delta = 
(\tilde{D}_0)^{1/2} $ at which the optical density drops  to unity.  
However, at this detuning, the 
cloud is dispersively dense with a phase shift of $ \phi \approx 
{(\tilde{D}_0)^{1/2}} / {2} $.  For large clouds, quantitative absorption imaging is 
still possible and has been applied \cite{davi95evap,hau98}, but for small 
clouds, refraction is a limiting factor.

These limits are avoided by larger detunings.  However, when the phase 
shift is reduced to $ \pi / 2 $ by choosing $ \delta = {\tilde{D}_0} /
{\pi} $, the optical density has dropped to $ \tilde{D} = {\pi^2} / {\tilde{D}_0} $.
For typical condensates of optical density $ \tilde{D}_0 = 300 $, the absorption
signal is far too small to be detected.  In 
contrast, dispersive imaging still features high signal levels.

\begin{figure}
  \begin{center}
     \epsfig{file=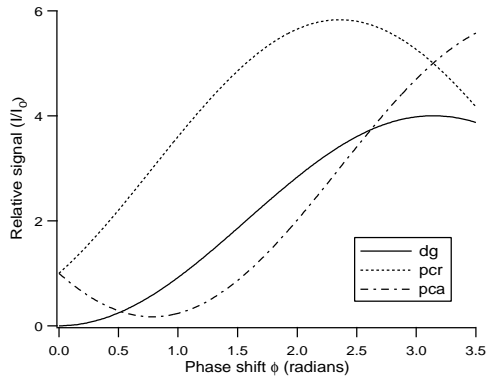,height=2in}
     \caption[Comparison of signal for dark-ground and phase-contrast 
     imaging]
     {Relative signal as a function of phase shift for
     dark-ground imaging (dg) and phase-contrast imaging with a phase plate 
     which either
     retards (pcr) or advances (pca) the phase of the unscattered light.
     For negative phase shifts (as would be generated with a blue 
     detuned probe), (pcr) and (pca) are swapped.}
     \label{fig:pcvsdg}
  \end{center}
\end{figure}

Equations \ref{eq:dgintensity} and \ref{eq:pcintensity} show that dark 
ground and phase-contrast signals are periodic in $ \phi $.  So unlike 
absorption methods, dispersive imaging does not saturate at large $ 
\phi $, and arbitrarily high phase shifts can be measured by counting 
fringes.  However, large phase shifting should be avoided since the interpretation of 
images is easier  if  the dispersive signal never ``rolls over.''  
For image analysis, an approximately linear relationship between 
signal and phase is desirable. Figure 
\ref{fig:pcvsdg} shows that a wide range of approximate linearity is 
achieved for phase-contrast imaging with a phase plate which retards 
the phase of the unscattered light ( $ \omega t \rightarrow \omega t + 
\pi/2 $) for red-detuning, or advances the 
phase ( $ \omega t \rightarrow \omega t - 
\pi/2 $) for blue-detuning of the probe light.
For this choice of phase plate, 
the maximum signal before rolling over is 5.8 times the 
incident light level. 
This gives phase-contrast imaging a larger dynamic range than 
absorption imaging.

\subsection{Quantum treatment of light scattering\label{sec:qmscatter}}
Further insight into the interaction between the probe light and the 
atoms is obtained from a quantum-mechanical treatment \cite{poli97}.  
For a single atom in state $ |i> $, the differential cross section for 
light scattering involves a sum over all possible final states $ |f>$:

\begin{equation}
\frac{ d \sigma_R }{d \Omega} = C \sum_f \left| \left< i \right| e^{i \mathbf{\Delta k} \cdot
\mathbf{r}} \left| f \right>
\right|^2 \label{eq:rayleigh}
\end{equation}
This is simply the differential cross section for Rayleigh scattering.  
We consider 
only the low intensity limit, i.e.\ the central peak of the Mollow 
triplet \cite{cohe92} and assume that Doppler shifts are small, i.e.\ 
$ | \mathbf{k_f} |= | \mathbf{k_i} | $, where 
$ | \mathbf{k_f} - \mathbf{k_i} | = | \mathbf{ \Delta k } | = 2 k \sin{( \theta/2 )} $ is 
the change of the 
photon's wavevector due to scattering by an angle $ \theta $.  
The atomic matrix elements and 
fundamental constants are contained in $ C $.

We now extend the treatment to $ N $ atoms in a trap, where each trap level $ 
|j> $ is 
initially populated with $ N_j $ atoms, still assuming that the cloud 
is optically dilute. The relevant matrix elements are those of the 
operator $ e^{i \mathbf{\Delta k} \cdot \mathbf{r} } $ with the symmetrized many-body eigenstates, 
characterized by the populations $ N_{j} $.  For bosonic atoms, one 
obtains \cite{poli97} 

\begin{equation}
\frac{ d \sigma }{d \Omega} 
 = C  \left| \sum_i N_i \left< i \right| e^{i \mathbf{\Delta k} \cdot \mathbf{r} } \left| i \right> 
\right|^2 +
C \sum_{  i \neq f } N_i (N_f + 1) \left| \left< i \right| e^{i 
\mathbf{\Delta k} \cdot \mathbf{r} } \left| f \right>
\right|^2  \label{eq:elasticandinelastic}
\end{equation}

The scattering is naturally divided into coherent and incoherent 
scattering.  The first term (the coherent part) represents elastic 
scattering in which the state of the atom is unchanged.  It is 
related to the Fourier transform of the density distribution:
\begin{equation} 
\sum_i N_i \left< i \right| e^{i \mathbf{\Delta k} \cdot \mathbf{r}} \left| i \right>
 = \int n(\mathbf{r}) e^{i \mathbf{\Delta k} \cdot \mathbf{r}} d 
\mathbf{r} \label{eq:fourier}
\end{equation}
For scattering angles smaller than the diffraction angle, $ \Delta k <
{1}/{d} $, the intensity of the scattered light is $N^{2} $ times 
the Rayleigh scattering of a single atom.  This term creates the 
real part of the index of refraction and generates the signal in 
dispersive imaging.

The second term in (\ref{eq:elasticandinelastic}) represents inelastic 
scattering.  We can separate the ``spontaneous part'' from the part 
due to bosonic stimulation:

\begin{equation}
\left( \frac{d \sigma}{d \Omega} \right)_{incoh} = C
\sum_{ i \neq f } N_i \left| \left< i \right| e^{i 
\Delta k r} \left| f \right> \right|^2 +
\sum_{ i \neq f } N_i N_f \left| \left< i \right| 
e^{i \Delta k r} \left| f \right> \right|^2  \label{eq:incoh}
\label{eq:spontbosesep}
\end{equation}

The first term is $ N $ times the single atom Rayleigh 
scattering (\ref{eq:rayleigh}), if we 
assume that the sample is much larger than an optical wavelength 
so that we can drop the condition $ { i \neq f } $. This term represents 
incoherent scattering  into a solid angle of $ 4 \pi$.  
Since photons scattered by a large angle are ``missing'' from the probe 
beam, it is this term which gives rise to the absorption signal.

The second term, describing Bose-enhanced scattering, is only non-vanishing 
near the phase transition.   Politzer distinguishes two contributions, 
scattering in and out of the condensate, and scattering between 
excited states \cite{poli97}.  This scattering is limited to angles 
smaller than the ratio of the optical wavelength to the thermal de 
Broglie wavelength.  Because this contribution is much weaker than the 
diffractive 
signal, it might be difficult to observe it. It is neglected in the 
quantitative analysis of images. 
The effects of quantum statistics can also be included in the 
index of refraction \cite{mori95}.  This treatment confirms 
 that non-classical light scattering effects are small,
 are
noticeable only near the phase transition, 
and vanish as $ T \rightarrow 0 $.

\subsection{Non-destructive imaging\label{sec:nondest} } 

It is obvious from the previous discussion that the different imaging 
techniques do not excite the atoms in a different way --- they only 
``collect'' different parts of the scattered electric field.  Whether 
the atom cloud is heated during the probing or not only depends on 
the intensity and duration of the probe light.

A discernible image with $30 \times 30$ pixels and 100 detected photons per 
pixel involves $10^{5} $ photons. Even if each photon would ``knock'' 
an atom out of the condensate, this would be ``non-perturbative'' if 
the sample had many millions of atoms.  Therefore, the large hydrogen Bose 
condensates could be observed non-destructively with absorption 
spectroscopy \cite{frie98}.

An important figure of merit of the different imaging techniques is 
the ratio of signal to heating.  In this regard, dispersive 
imaging often has a 
big advantage.  In absorption imaging, each 
absorbed photon (i.e.\ photons scattered at large angles) heats up the 
cloud by about one recoil energy.  The dispersive signal is based on the 
coherent term (\ref{eq:fourier})  which leaves the atom in its initial 
quantum 
state, and therefore does not contribute to the heating.  
The strength of the signal in dispersive scattering is best estimated 
from the number of photons collected in dark-ground imaging.  The 
total number of forward scattered photons (\ref{eq:fourier}) 
is larger than the 
total number of absorbed photons (\ref{eq:spontbosesep})
by a factor which is 
approximately $ N $ times the solid angle of the diffraction cone, $ 
\lambda^{2} / d^{2} $. Since
\begin{equation}
N \frac{\lambda^{2}}{d^{2}} \simeq 
\tilde{n} \lambda^{2} 
\end{equation}
this factor is about the resonant optical density $\tilde{D_{0}}$, identifying some 
common roots between forward scattering and superradiance \cite{alle75}.  
More quantitatively, the ratio of 
the signal in dark-ground imaging (\ref{eq:dgintensity}) 
to the number of Rayleigh scattered 
photons  ($1 - t^{2}$) in the 
limit of large detunings is $ \tilde{D}_0/4 $.
For typical values of $ \tilde{D}_{0} \simeq 300 $  this means that we can 
obtain two orders of magnitude more signal from dispersive imaging than from 
absorption imaging for the same amount of heating.  
With absorption methods, we could take 
only a single ``non-destructive'' image, but with dispersive 
imaging we could take real-time movies consisting of up to 100 
frames, more than the storage capacity of our 
camera.

The elastic scattering of photons at small angles is analogous to the 
M{\"{o}}ssbauer effect --- it is recoil-free scattering of 
photons where the momentum transfer to the photons is absorbed by the 
trap and not by the individual atoms.
This picture is valid when the pulse duration of the probe light is 
longer than the trapping period.  Rays of light passing through the atoms 
are bent in opposite directions on opposite sides of the atom cloud.  
If the pulse duration is longer than the trapping period, atoms have 
time to travel across the width of the cloud during the probe
and the recoil is averaged to 
zero.  For short pulses, atoms on opposite sides of the cloud 
experience opposite momentum transfer.  However, since the scattering 
angles are small, typically about 10 mrad, the recoil heating is $ 10^4 $ 
times lower than for a Rayleigh-scattered photon.  Therefore, even for 
short probe pulses, the dominant source of heating is large-angle 
Rayleigh scattering (as long as $ \tilde{D}_{0} \leq 10^{4}$). 

Since the relative figure of merit of dispersive vs.\ absorptive 
imaging is $ \tilde{D}_0/4 $, dispersive imaging is usually only applied to 
very dense clouds, either trapped or in the early phase of ballistic 
expansion.  Absorption imaging is used for time-of-flight 
imaging with sufficiently long expansion times so that the resonant optical 
density has dropped to values around unity.  There is no obvious 
advantage of dispersive imaging over absorptive imaging for dilute 
clouds.

\subsection{Other aspects of imaging}

\subsubsection{Fluorescence imaging}
So far we have not discussed fluorescence imaging.   
In fluorescence and absorption, one detects the same photons either as 
missing or as counted photons.  In comparing signal strengths, 
however, fluorescence imaging is inferior by a factor of typically 100, 
since only a small fraction of the scattered photons are collected by 
the imaging system.  Fluorescence methods usually have the advantage 
of being background-free, in contrast to absorption methods, where a 
small signal might be invisible due to fluctuations of the laser 
intensity.  However, in BEC experiments, one can usually choose a 
detuning where the absorptive or dispersive signal is comparable to 
the incident light intensity.

\subsubsection{Maximum light intensity in single-shot imaging}
Above we discussed heating by the probe pulse as the signal-limiting
factor in non-destructive imaging.  The situation is 
different if the goal is to take only one picture of a condensate. 
All groups reported their first observations of BEC using this 
``single-shot'' technique.
In this case, the figure of merit is not how much the condensate is heated 
during the probe pulse, but rather by how much is the image blurred 
due to recoil-induced motion.

If an atom in the sample scatters $ N_p $ photons in a time $ \Delta t $, it 
acquires a  velocity of $ v_{rec} N_{p} $ (where $ v_{rec}$ is the 
recoil velocity) along the direction of 
propagation of the probe beam due to absorption, and a random 
(approximately isotropic) velocity $  \sqrt{N_{p}} v_{rec} 
$. This random velocity $ v_{rms} $ leads to a random position of
$ r_{rms} = v_{rms} \Delta t  / \sqrt{3} =  
\sqrt{N_{p}/3} \, v_{rec} \Delta t 
$ \cite{joff93}.  A large number 
of photons $N_{p}$ per atom can be scattered if the pulse duration is 
short.  This strategy finds its limit when the atomic transition is 
saturated.  The acceptable amount of blurring depends on the required 
spatial resolution which is higher for a trapped cloud than for a 
cloud in ballistic expansion.  Typical numbers for sodium are that 100 
photons scattered in 50 $\mu$s give a final $v_{rms} = 30 
 $ cm/s   and $x_{rms} = 9 \, \mu$m.

\subsubsection{Optical pumping}
\label{tomographic_imaging}
The scattering of many photons per atom gives high signal levels 
and results in a very good signal-to-noise ratio, making the optical 
system less susceptible to stray light. However, it can only be 
applied on a cycling 
transition.  If the atoms are trapped in a different state, they can 
be transferred to the cycling transition by optical pumping.

Optical pumping was used by our group for two other reasons.  In the 
imaging of spinor condensates (sect.\ \ref{spinor})  atoms trapped in different 
hyperfine states were pumped into the same state for probing, allowing 
easy comparison of relative atom numbers without correcting for 
different transition matrix elements.
Spatially selective optical pumping can also be used to select only a part 
of the cloud for subsequent absorption imaging.  We used this method to 
get around the line-of-sight integration in normal absorption methods: 
a thin light sheet selected atoms by optical pumping in a tomographic 
way, and only these atoms absorbed the probe light which propagated 
perpendicular to the light sheet (fig.\ \ref{fig:catscan}).  
This method proved invaluable when 
imaging the fringes created by two interfering Bose condensates 
\cite{andr97int}.

\subsubsection{Imaging in inhomogeneous magnetic fields \label{sec:swirls}}
When the 
temperature of the cloud is larger than the natural linewidth, the 
spectroscopic nature of light scattering becomes important.
In this regime,
the inhomogeneous trapping potential can ``tune'' the atoms in and out of 
resonance.  This was the situation for the experiment on Lyman-$\alpha$
spectroscopy of hydrogen in Amsterdam, and also for our first 
evaporative cooling experiments \cite{davi95evap}.

\begin{figure}
    \begin{center}
\caption[Absorption images in a spherical quadrupole trap]{
Absorption images of clouds of sodium in a spherical quadrupole trap 
showing ``swirls'' and ``resonant shells'' (see text).
Image (A) 
    shows a cloud at $ \sim 80 \mu$K in a 
    quadrupole field with $ B' = 500 $ G/cm probed with linearly polarized 
    light detuned  
    -97 MHz from the F=1 to F'=2 transition. Image (B) is a 
    simulation of image (A).  Images (C), (D), (E), and (F) 
    show clouds at $ \sim 500 \mu$K in a 
    quadrupole field with $ B' = 250 $ G/cm at different detunings.  The experimental data is on 
    top, and the simulation below.  Image (C-F) were taken with -80, 
    -40, 0, and +80 MHz detuning respectively.
    The true rms size of the cloud is similar to the black area in (E).  
    The width of images (A) and (B) is 1.5 mm, and the width of 
    images (C-F) is 5 mm.  Figure taken from ref.~\cite{andr98thes}.
	(E-print: Separate figure)}
            \label{figswirls}
    \end{center}
\end{figure}

Figure \ref{figswirls} shows the variety of spectroscopic absorptive
shadows cast by identical 
clouds in a spherical quadrupole trap at different probe detunings. Some 
shadows are much larger than the rms size of the cloud.
These figures nicely show 
resonant shells where the atoms are Zeeman shifted into 
resonance with the probe light.

Temperature and density can be extracted by comparing experimental 
images to simulations.  These simulations solve for the propagation of 
the complex electric field through a medium characterized by a 
susceptibility tensor which depends on the local atom density and magnetic field 
direction.  Simulations done by Michael Andrews \cite{andr98thes}
were 
based on the method used by the Amsterdam group \cite{luit93lym} and explained 
all observed features (fig. \ref{figswirls}).

It was surprising that some spectroscopic images had a lower symmetry 
than the ellipsoidal cloud and exhibited
swirl-like structures.  These structures are due to the optical 
activity of the trapped cloud which rotates the polarization of the 
probe light.  
The symmetry of the cloud in the quadrupole trap would suggest that 
the intensity distribution of the transmitted light has an angular 
dependence of $a_{0} + a_{1} \cos{2 \phi} $  (see eq.\  (21) in 
ref. \cite{luit93lym}).  
However, there is a complex phase of the transmitted electric 
field which changes the angular dependence to $a_{0} + a_{1} \cos{(2 
\phi - \delta )} $ causing the swirls seen in fig.\ \ref{figswirls}.
This angle $ \delta $ was not 
considered in \cite{luit93lym} but did not affect their angle-averaged results.

\subsubsection{Experimental aspects}

The current generation of low-noise CCD cameras meets all the needs of current BEC experiments.  
With about 5 electrons rms noise per pixel, the camera is already shot-noise limited for more 
than about 50 detected photons per pixel.  Usually the detection sensitivity is limited by bad 
probe beam profiles and imperfect image normalization (see Appendix \ref{sec:imageprocessing})
rather than sensor 
performance.  Image-intensified cameras have nominally higher sensitivity for very small 
signals, but for more than $\sim$ 50 incident photons
per pixel, the higher quantum efficiency of CCD sensors (40\%) compared to 
image intensifiers (15\%) guarantees better signal-to-noise ratios.

The low noise of the camera usually requires a slow readout over several seconds.  Rapid 
sequencing of non-destructive images is achieved by storing the whole sequence on the chip, and 
then reading it out slowly.  For this purpose, the sensor is divided into an area for exposure 
and a much larger area for storage.  The division is done by a mask, either on the sensor, or 
for more flexibility, in an intermediate image plane of the optical system.  After exposure, 
the accumulated charge is shifted into the storage area.  All CCD chips should be able to 
perform this shift (since this is the way the image is finally read 
out), but some are much faster than others.  We have used
this method to take series of images with less than 1 ms time steps.

Our imaging system is based on standard achromats which are limited 
to an f-number of four.  We have found this to be a good compromise 
between resolution and flexibility.  Imaging systems with smaller f-numbers 
are possible, but
have smaller field of view and depth of field.
%

\section{Quantitative analysis of images}
\label{dataanalysis}
The purpose of imaging and image processing is to provide reliable 
density distributions of the atomic cloud, either trapped or in 
ballistic expansion.
All properties of condensates and thermal clouds are inferred from 
these density distributions.
This is done by comparing the measured distributions with the results 
of models of the atomic gas.

We present below several such models, and the means by which physical 
properties are determined.
We consider models appropriate for 
three regimes of temperature and density.
At high temperatures ($T \geq T_{c}$) and low temperatures (pure 
condensates) these density 
distributions are well-understood, experimentally verified, and 
reliable.  In the intermediate regime, clouds contain both a 
condensed and uncondensed fraction, and there are only approximate 
predictions for their distributions.

\subsection{Thermal clouds above the BEC transition temperature}

Let us first consider a trapped gas above the BEC 
transition temperature in a harmonic trapping potential:
\begin{equation}
	U({\bf r}) = \frac{1}{2} m (\omega_x^2 x^2 + \omega_y^2 y^2 + \omega_z^2 z^2).
\end{equation}
An ideal Bose gas in this 
potential at thermal equilibrium can be described either in 
the canonical (constant $N$) or grand-canonical ensembles, taking into 
account the discrete energy levels.  
However, at temperatures higher than the level spacing, $k_B T\gg 
\hbar \omega_{x,y,z}$, we can use a semi-classical 
approach \cite{gold81,huse82,bagn87ext} to obtain the density
\begin{equation}
n_{th}({\bf r}) = \frac{1}{\lambda_{dB}^3}  g_{3/2} ( z({\bf r}) )
\label{thermal_number}
\end{equation}
where $\lambda_{dB} = (2 \pi \hbar^{2} / m k_B T)^{1/2}$, 
$z({\bf r}) = \exp( (\mu - 
V({\bf r})) / k_B T)$, $\mu$ is the chemical potential, and $T$ is the 
temperature.  The Bose function, generally given by $g_j(z) = \sum_{i} 
z^i/i^j$, introduces the effects of quantum statistics on the density 
distribution; compared to a distribution of 
distinguishable particles, the density of a Bose gas is increased by 
$g_{3/2}(z) / z$.

We can also use this semi-classical approach to determine the 
distribution in time-of-flight.  When the trap is switched off, the 
trapped atoms fly ballistically from their position in the trap at 
their velocity at the time of the switch-off.  For an 
atom starting at point ${\bf r_0}$ to arrive at a point ${\bf r}$ after a time $t$ of 
free expansion, its  momentum must be ${\bf p}  = m ({\bf r} - {\bf r_0}) 
/ t$.  Integrating over all initial positions ${\bf r_0}$ we determine the 
density distribution as a function of the expansion time $t$
\begin{eqnarray}
	n_{tof}({\bf r}, t) & = & \frac{1}{h^{3}} \int d^{3} {\bf r_{0}} d^{3} 
	{\bf p} \, \frac{1}{e^{ - (\mu - H({\bf r_{0}},{\bf p})) / k_{B} 
	T} - 1} \delta^{3}({\bf r} - {\bf r_{0}} - \frac{{\bf p} t}{m}) \\
	& = & \frac{1}{\lambda_{dB}^{3}} \prod_{i=1}^{3} \left( \frac{1}{1 + 
	\omega_{i}^{2} t^{2}} \right) g_{3/2} \left( \exp \left[ \mu - 
	\frac{m}{2} \sum_{i=1}^{3} x_{i}^{2} \left( \frac{\omega_{i}^{2}}{1 + 
	\omega_{i}^{2} t^{2}} \right) \right] \right) 
\end{eqnarray}

At large times ($t \gg \omega_x^{-1},\omega_y^{-1},\omega_z^{-1}$),
neglecting collisions during the expansion (see 
sect.~\ref{compare_insitu_tof},
the density profile becomes
\begin{equation}
	n_{tof}({\bf r},t) = \frac{1}{\lambda_{dB}^3} 
	g_{3/2}( e^{ (\mu-\frac{m r^{2}}{2 t^{2}}) / 
k_{B} T })
\end{equation}
Thus, the thermal cloud expands isotropically once it becomes much 
larger than its original size.

\subsection{Bose-Einstein condensates at zero-temperature}
\label{condensate_density}

The density distribution of a zero-temperature 
Bose-Einstein condensate has been extensively discussed 
theoretically\cite{baym96,edwa95,dalf98}.  Due to the diluteness of the gas, 
effects of quantum depletion can be neglected (at least in the 
density distribution).
The many-body ground state is 
described by a single, complex order-parameter $\psi({\bf r})$.  This state 
and its dynamics are described by the 
Gross-Pitaevskii equation
\begin{equation}
i \hbar \frac{d \psi}{d t} = - \frac{\hbar^{2}}{2 m} \nabla^2 \psi + 
U({\bf r}) \psi + \tilde{U} |\psi|^2 \psi	
\label{gpe}
\end{equation}
The parameter  $\tilde{U} = 4 \pi \hbar^2 a / m$
describes the effect of two-body collisions, 
where $a$ is the s-wave scattering length and $|\psi|^{2}$ the density.  
The ground-state wavefunction is
$\psi({\bf r}, t) = \psi({\bf r}) e^{-i \mu t}$, where $\mu$ is the energy of the ground state, 
and is identified as the chemical potential.

\subsubsection{Ideal-gas limit}

This wavefunction is easily described in two limits.  In the limit of 
weak interactions ($n \tilde{U} \ll \hbar \omega_{x,y,z}$) one can neglect 
the interaction term.  Then, the condensate wavefunction is simply 
the ground state of the harmonic oscillator, which gives a density 
for $N$ particles of
\begin{equation}
	n_c({\bf r}) = \frac{N}{\pi^{3/2}} \prod_{i=1}^{3} \frac{1}{x_{i,0}} e^{- 
	x_{i}^{2} / x_{i,0}^{2}}
\end{equation}
Here the length scales $x_{i,0} = x_{i,HO} = \sqrt{\hbar / m 
\omega_i}$ are 
the rms-widths of the condensate wavefunction in each of the three 
dimensions (labeled by $i$).

In free expansion, a Gaussian wavefunction remains Gaussian except 
for a phase factor.  Thus, after a time $t$ of expansion, the 
condensate size is simply rescaled according to $x_{i,0}^2 = x_{i,HO}^2 + 
v_{i,HO}^2 t^2$, where $v_{i,HO} = \sqrt{\hbar \omega_{i} / m} $
is the rms velocity of the trapped condensate.  This length 
can also be written as $x_{i,HO}^2 (1 + \omega_i^2 t^2)$.  As with the 
expansion of the thermal cloud, the dependence on the initial size
becomes negligible 
for expansion times much longer than $\omega_i^{-1}$.

\subsubsection{Thomas-Fermi limit}
\label{bectf}
In the limit of strong interactions ($n \tilde{U} \gg \hbar 
\omega_{x,y,z}$), 
the determination of the trapped condensate wavefunction is 
simplified by neglecting the kinetic energy term in eq.\ (\ref{gpe}) which is now much 
smaller than the interaction term.  In this limit, known as the 
Thomas-Fermi limit, the density is given by
\begin{equation}
	n_{c} ({\bf r}) = \max \left( \frac{\mu - U({\bf r})}{\tilde{U}}, 0 \right)
\end{equation}
Thus, one can 
think of the condensate in the Thomas-Fermi limit as ``filling in'' the 
bottom of the trapping potential up to a ``height'' of the chemical 
potential $\mu$.  In a harmonic trap, the condensate has a parabolic 
density profile,
\begin{equation}
n_{c}({\bf r}) = \frac{15}{8 \pi} \frac{N}{ \prod x_{i,c,0}} \max
\left (1 - \sum_{i=1}^{3} \frac{x_{i}^2}{x_{i,c,0}^2} ,0 \right)
\end{equation}
characterized by the half-lengths of the trapped condensate 
$x_{i,c,0}$ where the density goes to zero (the subscript $c$ 
indicates the condensate, and $th$ the thermal cloud).  These are determined by 
the chemical potential and the trap parameters as $x_{i,c,0} = \sqrt{2 \mu 
/ m \omega_{i}^2}$.


Conveniently, it has been shown  that when such a 
condensate is released from the trap, it evolves simply as a 
rescaling of its parabolic shape~\cite{cast96,kaga96evol,dalf97} 
(sect.~\ref{tfexpansion}).
For release from a cigar-shaped trap with radial frequency 
$\omega_\rho$ and aspect ratio $\omega_\rho / \omega_z = \epsilon^{-1}$, 
the half-lengths of the condensate evolve according to the following 
equations (to lowest order in $\epsilon$):
\begin{eqnarray}
\rho_0(t) &= &\rho_0(0) \sqrt{1 +\tau^2}	\\
z_0(t) &= &\epsilon^{-1} \rho_0(0) \left( 1 + \epsilon^2[\tau \arctan \tau - \ln 
\sqrt{1 + \tau^2}] \right)
\end{eqnarray}
where $\tau = \omega_{\rho} t$.

This solution describes three stages in the time-of-flight expansion: 
(1) a radial acceleration as interaction energy is converted to 
kinetic energy ($\tau < 1$), (2) radial expansion with little 
axial expansion beyond the original axial size ( $1 < \tau < \epsilon^{-2}$), 
and (3) radial and axial expansion at an asymptotic aspect ratio of 
$z_0(t)/\rho_0(t)= \pi \epsilon^2/2$ \  ($\epsilon^{-2} < \tau$).
Note that the velocity of radial expansion for $t \gg \omega_\rho^{-1}$
satisfies $\frac{1}{2} m v_{\rho}^{2} = \mu$.
We typically use magnetic traps of 
frequencies $\omega_z ~= 2 \pi \times 20$ Hz and $\omega_\rho ~= 2 \pi 
\times 250$ Hz.  In this case, the 
mean-field energy is released in about 1 -- 2 ms as 
the cloud accelerates outward.   The axial expansion 
is only noticeable after about 100 ms, by which time the atoms have 
fallen about
5 cm.  We typically probe the cloud between 20 and 60 ms 
time-of-flight, and thus observe condensates that have not yet 
reached their asymptotic aspect ratios.

It is important to note
that the validity of both the ideal-gas and Thomas-Fermi limits is 
generally different for each axis of the harmonic oscillator.  For 
example, in the single-beam optical traps used at MIT, the 
cigar-shaped traps are highly asymmetric, with aspect ratios as large 
as 70 and radial trapping frequencies as high as several kHz.
A sodium condensate with a maximum density of $n_c(0) \simeq 1 \times 
10^{14} \, \mbox{cm}^{-3}$ 
has an interaction energy $\tilde{U} n_c(0) ~= h \times 1.4$ kHz.  In this case, the 
Thomas-Fermi approach would be valid along the long axis of the 
cloud, but not in the radial directions.
The proper description of condensates in this intermediate regime 
requires numerical solutions to the Gross-Pitaevskii equation 
~\cite{holl97int}. 

\subsection{Partly condensed clouds}

As we have noted above, the density distribution of a Bose gas in two 
limits (high and low temperatures) is well understood.  Between 
these two limits, these distributions are the subject of 
current theoretical and experimental 
scrutiny~\cite{huse82,gior97,gior97jltp,dodd98two,nara98semi}.
Accurate measurements of the density distributions are needed
to discriminate between the predictions of different 
finite-temperature theories.

These density distributions have one striking and useful 
feature: their bimodality. 
The assignment of the two components of 
the bimodal distribution to the condensate and non-condensate 
densities is an experimental advantage of dilute Bose-Einstein 
condensates over liquid helium, in which only indirect measurements 
of the condensate are made.

To glean properties of the mixed cloud from images, the observed 
density profiles are fit with a bimodal distribution.  The choice of 
distribution is somewhat arbitrary.  For our work, we chose a 
distribution that correctly described the cloud in the two limits 
described above: a Bose-enhanced Gaussian distribution at the 
transition temperature, and a parabolic density distribution at zero 
temperature.  Thus, we fit the observed density $n_{tot}({\bf r})$ to
\begin{equation}
n_{tot}({\bf r}) = n_{th} \, g_{3/2} \left( \prod_{i=1}^{3} e^{-x_{i}^{2} / 
x_{i,th,0}^{2}}\right) + n_c \, \max \left( 1 - 
\sum_{i=1}^{3} \frac{x_i^2}{x_{i,c,0}^2} , 0 \right)
\label{generic3d}
\end{equation}
We do not constrain these fitting parameters by the aspect ratios or 
maximum densities predicted by the ideal-gas or Thomas-Fermi models.
Thus, eq.~\ref{generic3d} is a rather generic parametrization of a 
bimodal distribution.

\subsection{Column densities}

So far we have discussed the predicted density distributions for 
various regimes.  However, with the exception of 
tomographic imaging (sect.~\ref{tomographic_imaging}),
all imaging techniques measure the column density of the cloud along the 
imaging axis.  Taking this axis to be the $y$-axis, and labeling 
the remaining coordinates $\rho$ and $z$, we obtain the column 
densities for the thermal cloud $\tilde{n}_{th}$ and the condensate $\tilde{n}_{c}$ as
\begin{eqnarray}
	\tilde{n}_{th}(\rho,z) & = & \frac{\tilde{n}_{th}(0)}{g_{2}(1)} \, 
g_2\left[e^{(1 - \rho^2/\rho_{th,0}^2 - 
z^2/z_{th,0}^2)} \right]
\nonumber \\
	\tilde{n}_c(\rho,z) & =& \tilde{n}_{c}(0) \max \left( 1 - 
	\frac{\rho^2}{\rho_{c,0}^{2}} - \frac{z^2}{z_{c,0}^{2}}  ,0 \right)^{3/2}	
\label{2dtot}
\end{eqnarray}
These column density distributions are used as fitting functions to 
the experimental images.
This fitting function 
describes all our observations very well.  In comparison, using a Gaussian 
distribution for the condensate gave inferior results.

\subsection{Extracting static quantities}
\label{derived_quantities}

\begin{itemize}

\item {\it Temperature:}  The temperature of the gas can be determined 
from the shape of the 
spatial wings of the distribution ascribed to the thermal cloud.  The 
density in the wings of the spatial distribution decays generally as 
$e^{-x_i^2/x_{i,th,0}^2}$, even in the case of Bose enhancement.  Using the 
results from above, we can determine the temperature $T$ of the cloud as
\begin{equation}
	k_{B} T = \frac{1}{2} m \left( \frac{\omega_{i}^{2}}{1 + \omega_{i}^{2} 
	t^{2}} x_{i,th,0}^{2} \right)
\end{equation}
where $t$ is the time-of-flight.

To obtain temperature measurements for mixed clouds that are 
independent of the specific choice of bimodal density distributions, 
it is necessary to limit the fit to regions of the image in which the 
condensate is clearly absent.  Further, to eliminate possible 
problems due to the Bose enhancement of the thermal cloud near the 
boundary of the condensate, it is advisable to further restrict the 
fit to the spatial wings.  From our experience, the systematic errors 
introduced by fitting too close to the condensate distribution are on 
the order of 20\% of the temperature.

\item {\it Chemical potential and peak density:}
The chemical potential $\mu$ is given by the size of the condensate.  
From a fit of 
the form given in eq.\ (\ref{2dtot}) we determine $\mu$ from in-situ measurements 
as
\begin{equation}
	\mu = \frac{1}{2} m \omega_{i}^{2} x_{i,c,0}^{2}
\end{equation}
where  $x_{i,c,0}$ is the half the length of the condensate in the $x_i$ direction.  
From condensates released from cigar-shaped clouds, we use the 
half-length of the cloud in the radial direction as
\begin{equation}
	\mu = \frac{1}{2} m \left( \frac{\omega_{\rho}^{2}}{1 + \omega_{\rho}^{2} 
	t^{2}} \rho_{c,0}^{2} \right)
	\label{mufromtof}
\end{equation}
Within the Thomas-Fermi approximation, this value of $\mu$ 
determines the maximum condensate density $n_c(0)$ of the trapped cloud 
as $n_c(0) = \mu m / 4 \pi \hbar^2 a$
where $a$ is the scattering length.

\item {\it Total number:}
The total number of atoms is determined by summing over the 
absorption or phase-contrast signal seen across the two-dimensional 
image of the cloud.  Thus, in resonant absorption  one obtains
\begin{equation}
N = \frac{A}{\sigma_{0}} \, \sum_{pixels} -\ln(\tilde{T}(x,y))	
\end{equation}
and in phase-contrast
\begin{equation}
N = \frac{A}{\sigma_{0}} \frac{2 ( \Gamma^{2} + 4 \Delta^{2} )}{2 \Delta \Gamma} 
\sum_{pixels} \phi(x,y)
\end{equation}
Here, $A$ is the imaged area per pixel, $\sigma_{0}$ is the resonant 
cross-section for light absorption, $\Gamma$ is the natural 
linewidth, $\Delta$ is the detuning from resonance, $\tilde{T}(x,y)$ is the 
transmission observed in absorption images, and $\phi(x,y)$ is the phase 
measured in phase-contrast images.  Note that $\sigma_{0}$ depends on the 
polarizations of the atomic cloud and the optical probe and
must be known to properly quantify the total number of 
atoms.
Sect.~\ref{sec:atomlight} discusses how to account for non-zero 
detuning and saturation.  Note that the chemical potential and temperature are based on 
length measurements which are 
independent of the total signal strength.

\item {\it Condensate number and condensate fraction:}
While $T$, $\mu$, and $N$ 
can be determined with few assumptions, measuring the condensate 
fraction or number is less straightforward.  Typically, one fits the 
total density distribution with a bimodal distribution, such as that 
given in eq.\ (\ref{2dtot}), and ascribes the total number from the central 
distribution to the condensate.  Thus, the condensate fraction 
depends critically on the assumed shape of the bimodal distribution, 
particularly in the central region where the non-condensed density 
distribution must be inferred.  Determinations of the condensate 
number are most reliable near the transition temperature, where the 
thermal cloud is clearly defined and can be extrapolated reliably 
into the central region.  For this reason, one can argue that instead 
of comparing the inferred condensate 
fractions with theory, one should rather directly compare the whole 
density distribution, which can be determined by these same 
theories.  Finally, let us note that the condensate fraction could in 
principle be 
measured by exploiting the coherence of the condensate, and not the 
bimodal distribution.
For example, if the
original condensate is split in two and then recombined in 
time-of-flight, the observed fringe contrast determines the 
condensate fraction (see related discussion in ref.\ \cite{dodd97coh}).

\item {\it Derived quantities:}
Other useful quantities can be derived from $T$, $\mu$, $N$ and 
the condensate number $N_{0}$ using relations among 
them.  To begin with, the Thomas-Fermi expression for the 
chemical potential of a harmonically confined condensate is
\begin{equation}
\mu^{5/2} = \frac{15 \hbar^2 m^{1/2}}{2^{5/2}} N_{0} \bar{\omega}^3 a	
\label{meanfieldvsnumbereq}
\end{equation}
where $\bar{\omega} = (\omega_{x} \omega_{y} \omega_{z})^{1/3}$ is the 
geometric mean of the trap frequencies.
This expression relates four quantities ($\mu$, $N$,$ \bar{\omega}$, 
$a$), 
and by knowing three of them, one can derive the fourth.  
For example, assuming a value for $a$, as derived from spectroscopic 
measurements, we have deduced the average trap frequency of an 
optical trap from single time-of-flight images of condensates.

Another useful relation is that for the number of atoms $N_c$ at the 
critical temperature $T_c$, neglecting the small effect of interactions 
(consider eq.\ (\ref{thermal_number}) for $\mu=0$, see 
\cite{bagn87ext}):
\begin{equation}
N_c = g_3(1)  \left(\frac{k_B T_c}{\hbar \bar{\omega}}\right)^{3}
\end{equation}
From images of clouds at the critical temperature, one can again 
derive one of three quantities ($N$, $T$, ${\bar{\omega}}$) by knowing the 
other two.  Note that this expression does not contain the scattering 
length.  Thus, the combined use of these two equations is a powerful 
tool to check for and eliminate systematic errors in number or length 
measurements.

Other useful quantities are the reduced temperature, 
 and the phase space density.  The reduced temperature 
is the ratio of the observed temperature $T$ to the BEC transition 
temperature $T_{c}^{(0)}$ for an ideal gas with the observed total number of atoms, i.e.
\begin{equation}
\frac{T}{T_{c}^{(0)}} = \frac{k_B T}{\hbar \bar{\omega}} \left(\frac{g_3(1)}{N}\right)^{1/3}	
\end{equation}
The maximum phase-space density of a non-condensed gas is given by ${\mathcal D} = 
n(0) \lambda_{dB}^3$, where $n(0)$ is the maximum density of the cloud, and 
$\lambda_{dB}$ the thermal deBroglie wavelength.  ${\mathcal D}$ is determined from 
the relations
\begin{eqnarray}
{\mathcal D} &= &g_{3/2}(z(0))	\\
N & = & g_3(z(0)) \left( \frac{k_B T}{\hbar \bar{\omega}}	\right)^{3}
\end{eqnarray}
Again, for the purpose of calibration, it is useful to compare clouds
above the transition temperature with those at the 
transition, where ${\mathcal D} = g_{3/2}(1)$.  Such calibration allowed for accurate 
measurements of the increase in phase-space density due to adiabatic 
deformation of a magnetic trapping potential by a focused infrared 
beam~\cite{stam98rev}.

\end{itemize}

\subsection{Extracting dynamic properties}

Aside from revealing equilibrium thermodynamics, in-situ and 
time-of-flight images also reveal the dynamic response of trapped 
Bose gases to external perturbations.  Such responses are probed 
stroboscopically by taking a time-series of images, each of which 
measures the state of the system at a fixed time after application of 
the perturbation.  In this regard, there is an important difference 
between destructive and
non-destructive imaging.  Using destructive imaging to probe 
the gas implies that for each data point in the accumulated 
time-series,
the experiment 
is cycled through the entire loading and cooling cycle.  Thus, such a 
time-series of data is susceptible to shot-to-shot fluctuations.
On the other hand, by using 
non-destructive imaging techniques, a rapid sequence of images can be 
taken on a single atomic sample, leading to much faster data taking, 
and eliminating the problem of controlling experimental conditions 
precisely from shot-to-shot.  For example, our observation of bosonic 
stimulation in the formation of a Bose condensate was only possible 
using rapid-sequence phase-contrast imaging (fig.~\ref{formfigure}) 
since the formation 
process was acutely sensitive to the conditions of the trapped clouds 
on the verge of condensation~\cite{mies98form}.

The collective excitations of trapped atomic 
gases have been a major focus of BEC research.  These excitations are 
analyzed by applying single- or bimodal fits to data 
from a time-series of probes after the perturbation.
Three types of collective excitation have been studied:
\begin{itemize}
	\item {\it Center-of-mass oscillations of the entire distribution:}  
	Trap frequencies are measured unambiguously by tracing the 
	center-of-mass motion of the cloud over time.  
We have used rapid sequences of 
phase-contrast images to measure frequencies of our magnetic trap to 
better than $10^{-3}$ single-shot precision, and to check for anharmonicities.
	\item  {\it Shape oscillations of the condensate and the 
thermal cloud:}  These excitations correspond to a periodic rescaling 
of the lengths and aspect ratios of the cloud.  The frequency and 
damping of such modes have been studied both in time-of-flight 
~\cite{mewe96coll,jin96coll,jin97,matt98}, and in-situ~\cite{stam98coll}.  
High-precision (better than $10^{-2}$) measurements were made using 
rapid-sequence imaging which provided a set of images before the 
perturbation, to determine the starting conditions, and three sets of 
images during the oscillation, covering periods as long as 0.5 s 
~\cite{stam98coll}.  To obtain as many as 28 non-destructive images, a 
light fluence as low as 10 photons/$\mu\mbox{m}^2$ was used.
\item {\it Sound propagation in a Bose-Einstein condensate:}
One-dimensional sound
propagation along the long axis of the cloud was analyzed using 
rapid-sequence phase-contrast imaging~\cite{andr97prop}.  The location 
of the sound pulse was determined by subtracting out the shape of an 
unperturbed condensate, and determining the maximum or minimum of the 
resulting density difference.  The axial length of the condensate
was used to determine the condensate density using eq.\
(\ref{mufromtof}).
\end{itemize}

\subsection{Comparison of time-of-flight and in-situ images}
\label{compare_insitu_tof}
As discussed above, detailed information about trapped Bose gases can 
be obtained from both in-situ and time-of-flight images.  Let us now 
compare the two methods, and find when each is best applied.

The onset of BEC is most striking for anisotropic traps in time-of-flight 
images.
The appearance of the condensate suddenly ``breaks the symmetry'' of 
the isotropic expansion above the transition temperature.
The expansion of a cigar-shaped condensate into a disk-shaped object 
is easy to observe even with a misaligned imaging system.
In contrast, in the Thomas-Fermi limit, the aspect ratios of the 
condensate and the thermal cloud are the same for in-situ imaging, and
observing the onset of BEC requires high-resolution imaging

The interpretation of time-of-flight images 
to determine temperature and chemical potential assumes a sudden, 
ballistic, and free expansion. 
A variety of effects 
may invalidate this assumption.  These include residual magnetic 
fields, such as a slow shut-off of the trap (compared to the 
trap oscillation periods), or stray curvature fields during the 
free-expansion.  Another impediment is that thermal clouds released 
from anisotropic traps can expand anisotropically due to collisions 
soon after the trap is switched off ~\cite{wu98exp1,wu98exp2}.

The limitations to temperature measurements on non-spherical partly-condensed 
clouds are different for in-situ and time-of-flight techniques.
Table ~\ref{sizeofclouds} compares the sizes of the thermal cloud and 
the condensate from cigar-shaped traps, both in the trap and in 
time-of-flight.
Focusing our attention on the Thomas-Fermi case, one finds that 
in the trap, the two components have equal sizes at $k_{B} T = \mu$.
Thus, in-situ temperature measurements based on the density 
distribution are limited to $k_{B} T \geq \mu$~\cite{stam98coll}.
On the other hand, in time-of-flight, the condensate and the thermal 
cloud can be distinguished along the $z$-axis down to $k_{B} T = 
\mu \times 4 \pi^{-2} \epsilon^{2}$ where $\epsilon = \omega_{z} / 
\omega_{\rho}$, stretching the limits of temperature measurements 
to much lower values.

\begin{table}[h]
\caption[Length scales for thermal clouds and condensates in trap and 
in time-of-flight]{Comparison of length scales for thermal clouds and 
condensates.  For the thermal cloud and ideal-gas condensates we 
give the rms radius, and for the Thomas-Fermi condensates the half-lengths.
Time-of-flight is long enough to ignore finite-size effects.
Results for a Thomas-Fermi condensate in time-of-flight assume 
$\omega_{z} / \omega_{\rho} \rightarrow 0$.}
\begin{center}
\begin{tabular}{|l||c|c||c|c|}
	\hline
	& \multicolumn{2}{c||}{In-situ} & \multicolumn{2}{c|}{Time-of-flight} 
	\\
	& $\rho$ & $z$ & $\rho$ & $z$ \\ \hline
 & & & &  \\
	Thermal cloud &
	$\sqrt{\frac{2 k_{B} T}{m}}\frac{1}{\omega_{\rho}}$ &
	$\sqrt{\frac{2 k_{B} T}{m}}\frac{1}{\omega_{z}}$ &
	$\sqrt{\frac{2 k_{B} T}{m}} t $ &
	$\sqrt{\frac{2 k_{B} T}{m}} t $ \\
& & & &  \\

Ideal-gas condensate &
	$\sqrt{\frac{\hbar}{m}}\frac{1}{\sqrt{\omega_{\rho}}}$ &
	$\sqrt{\frac{\hbar}{m}}\frac{1}{\sqrt{\omega_{z}}}$ &
	$\sqrt{\frac{\hbar}{m}} {\sqrt{\omega_{\rho}}}$ &
	$\sqrt{\frac{\hbar}{m}} {\sqrt{\omega_{z}}}$ \\
& & & &  \\
Thomas-Fermi condensate &
	$\sqrt{\frac{2 \mu}{m}}\frac{1}{\omega_{\rho}}$ &
	$\sqrt{\frac{2 \mu}{m}}\frac{1}{\omega_{z}}$ &
	$\sqrt{\frac{2 \mu}{m}} t $ &
	$\sqrt{\frac{2 \mu}{m}} \frac{\pi \omega_{z}}{2 \omega_{\rho}} t$ \\ 
	\hline
\end{tabular}
\end{center}\vskip-\lastskip
\label{sizeofclouds}
\end{table}

Another aspect to consider in choosing between time-of-flight and 
in-situ imaging for anisotropic clouds is the different column 
densities 
of the condensate and the thermal cloud.
Recall that in time-of-flight from a cigar-shaped potential, the 
thermal cloud expands isotropically once $t > \omega_{z}^{-1}$ while 
the condensate expands axially beyond its initial length only when 
$t > \omega_{\rho} / \omega_{z}^{2}$.
For times $1 < \omega_{z} t < \omega_{\rho} / \omega_{z}$,
if the imaging is done along a radial direction,
the condensate optical density 
drops linearly with expansion time, while the non-condensate optical 
density drops quadratically.  Thus, in-situ imaging is preferable 
for imaging the entire cloud faithfully, while long time-of-flight expansions 
allow for more reliable measurements of the condensate number.

There are also differences in measuring 
dynamic properties.
One goal is measuring oscillations at the limit of 
zero-amplitude.   It has been shown that the relative
amplitude of shape  oscillations observed in 
time-of-flight was much greater than what would be observed 
in-situ~\cite{cast96,dalf97}.  Thus, time-of-flight imaging allows for the observation of 
smaller amplitude oscillations.  On the other hand, this benefit of 
time-of-flight techniques may be offset by slower data taking and 
greater susceptibility to technical noise.  Furthermore, for more 
complicated oscillations it is difficult to interpret time-of-flight 
images.  For example, while sound pulses were easily observed and 
located in-situ, time-of-flight images of such condensates revealed a 
complicated pattern of striations~\cite{andr97prop}.


\section{Static properties} 
\label{static}

The previous sections discussed all the machinery --- laser 
cooling, magnetic trapping, evaporative cooling, tools to manipulate 
ultracold atoms, imaging and image processing, data analysis --- that comes 
together in pursuit of Bose-Einstein condensation.  At long last, with 
all parts assembled and working simultaneously, 
 we arrive at our goal, the condensate itself.  

\begin{figure}
\epsfxsize = 6cm
\centerline{\epsfbox[0 0 354 340]{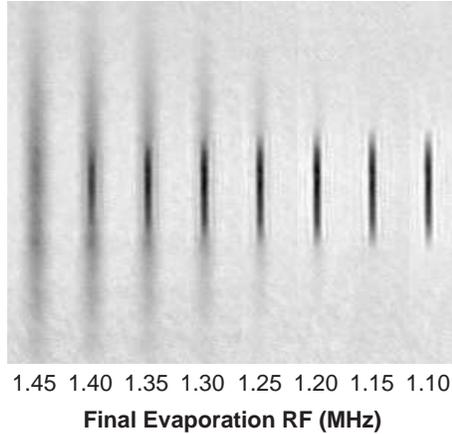}}
\caption[(Color) Phase contrast images of trapped Bose gases across the BEC 
phase transition]{(Color) Phase contrast images of trapped Bose gases across the BEC 
phase transition.
As the final rf used in evaporation cooling is lowered, the 
temperature is reduced (left to right).
Images show the onset of Bose condensation, the growth of the 
condensate fraction and contraction of the thermal wings, and finally 
a pure condensate with no discernible thermal fraction.
The axial and radial frequencies are about 17 and 230 Hz, respectively.
The pure condensate has a length of about 300 $\mu$m.}
\label{pctransition}
\end{figure}

\begin{figure}
\epsfxsize = 6cm
\centerline{\epsfbox[0 0 331 442]{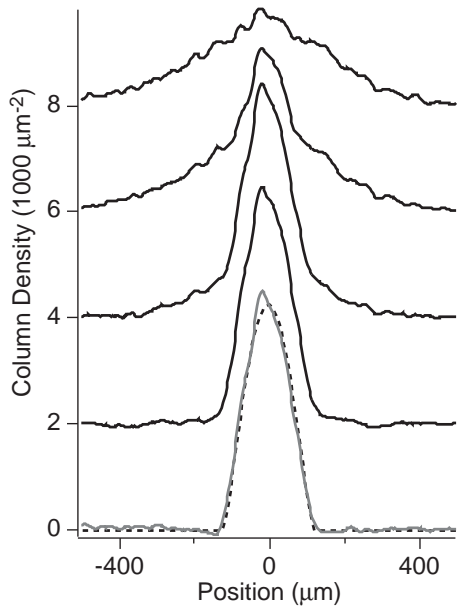}}
\caption[Axial column density profiles of Bose-Einstein condensed 
clouds from phase-contrast images]{Axial column density profiles of Bose-Einstein condensed 
clouds from phase-contrast images.
The bottom curve shows a parabolic fit to the data, according to eq.\ 
(\ref{2dtot}).}
\label{profiles}
\end{figure}

Fig.\ 
~\ref{pctransition} serves as a nice introduction to the gaseous system 
which is the subject of this book.  The final temperature is adjusted by the 
frequency at which the rf-evaporation is terminated (shown on the figure).  
At high values of the final rf, the temperature is high and the 
non-condensed cloud reaches far from the center of the harmonic trapping 
potential.  The gas lies above the BEC transition temperature, and the 
density profile is smooth.  As the rf is lowered, the temperature 
drops below the BEC phase transition, and a high-density core of atoms 
appears in the center of the distribution.  This is the Bose-Einstein 
condensate.  Lowering the rf further, the condensate number grows
and the thermal wings of the distribution become 
shorter.  Finally, the temperature drops 
to the point where only the central peak remains.  
  Each image shows an equilibrated gas 
 obtained in one complete trapping and cooling cycle.  
Axial profiles of several of these clouds are presented in fig.~\ref{profiles}.

Typically in our 
apparatus the transition temperature of 1 -- 2 $\mu$K is reached with 
50 -- 100 $\times 10^6$ atoms.  Further evaporation produces 
``pure'' condensates of 5 -- 20 $\times 10^6$ atoms.  These condensates are 
about 300 $\mu$m in length, have an aspect ratio of about 15, and a peak 
density of about 4 $\times 10^{14} \, \mbox{cm}^{-3}$.  These ``typical'' 
conditions can be and have been greatly modified in our various projects; 
condensates densities have ranged from $2 \times 10^{13}$ 
to $3 \times 10^{15} \, \mbox{cm}^{-3}$, and transition temperatures 
from 100 nK to 5 $\mu$K.  In other groups, condensate numbers have 
ranged from about $10^{2}$~\cite{ande98priv} to 10$^9$~\cite{frie98}.  This great 
variety allows for a thorough testing of the theoretical framework 
describing weakly interacting Bose gases.  Let us now present an overview 
of the recent experimental studies of these gases, beginning with a 
consideration of their equilibrium properties.  

\subsection{Zero temperature condensates} 

\subsubsection{Condensate density profile} 

As 
discussed in sect.\ \ref{condensate_density}, 
the condensate wavefunction at zero temperature is the solution of the 
non-linear Gross-Pitaevskii equation (eq.\  \ref{gpe}).  In the 
Thomas-Fermi limit ($\mu \gg \hbar \omega_i$), the condensate 
density is parabolic due to the parabolic shape of the confining potential.  
This is confirmed by in-situ images taken of condensates near 
zero-temperature  where no thermal cloud was visible.  For example, 
the bottom curve of fig.~\ref{profiles}, which gives the density 
profile of a pure condensate, is well fit by a function of the form given 
in eq. (\ref{2dtot}).  L.\  Hau and 
collaborators probed the boundary of a condensate in-situ with 
near-resonant absorption imaging~\cite{hau98} and
compared their data with the solution of the Gross-Pitaevskii equation for 
the known condensate number, trap frequencies, and scattering length.
The agreement was excellent.  

\subsubsection{Mean-field energy} 
These images of the condensate immediately show the 
 importance of  interactions in determining the shape (and 
the dynamics) of Bose condensates.  The condensate in figs.\  
~\ref{pctransition} and
~\ref{profiles} stretches for about 300 $\mu$m along the long axis of the 
magnetic trap.  In comparison, an ideal-gas Bose condensate would only have 
an rms length of $\sqrt{2 \hbar / m \omega_z} \simeq 7 \, \mu$m at the 
axial trapping frequency of 17 Hz.  The condensate swells to a much greater 
size due to self-repulsion.  

The interaction energy of a condensate, 
and its dependence on the condensate number, was studied by measuring 
 expansion energies in
time-of-flight imaging ~\cite{mewe96bec,holl97int}.
In the MIT study~\cite{mewe96bec}, the condensate number was varied
either by stopping the evaporation before 
a pure condensate was formed, or by
evaporating past that point. 
As all these condensates were in the Thomas-Fermi regime, their initial 
kinetic energy is negligible, and the mean-field energy (chemical 
potential) should scale as $N^{2/5}$ (eq.~\ref{meanfieldvsnumbereq}).  
The condensate numbers varied by a factor of 100 (between 
$\simeq 5 \times 10^{4}$  and $5 \times 10^{6}$), and the 
Thomas-Fermi prediction was nicely confirmed.  

Time-of-flight images from experiments at JILA 
were similarly analyzed~\cite{holl97int}.  Small condensates were 
studied between the ideal-gas and Thomas-Fermi limits.  The images were 
in quantitative agreement with a numerical model of the expansion, 
confirming the use of the scattering length, as determined from 
spectroscopic data, to describe mean-field interactions in a 
condensate.

\subsection{Bose condensates at 
non-zero temperatures: thermodynamics} 

\subsubsection{The BEC transition 
temperature} 

The predicted BEC transition temperature for a harmonically 
confined ideal gas of $N$ atoms is $k_{B} T_{c}^{(0)} = \hbar \bar{\omega} 
N^{1/3}/ g_3(1)$, where $\bar{\omega}$ is the geometric mean trapping 
frequency.  This relation reflects the condition $n \lambda_{dB}^{3} = 
g_{3/2}(1)$ which applies at the center of the trap~\cite{bagn87ext}.  The 
initial observations of Bose condensation at JILA~\cite{ande95} and 
MIT~\cite{davi95bec,mewe96bec} agreed with this ideal gas condition, 
confirming a 70-year old prediction~\cite{eins25qua2}.  

Subsequently, attention 
turned to changes in the transition temperature due to 
interactions and finite size effects.  For a weakly-interacting gas in a box, the 
transition temperature is predicted to shift upwards by an amount 
proportional to $\sqrt{n a^3}$~\cite{huan87} (see also recent work: 
~\cite{bijl96,grut98}).  This 
effect represents a lowering of the critical phase space density for 
Bose-Einstein condensation from its ideal gas value of $g_{3/2}(1)$.  
In an 
inhomogeneous potential, an additional shift arises due to 
interactions because the density at the center of the trap is lowered 
as the atoms repel one another.  This causes a downward shift in 
$T_{c}$~\cite{gior96}.
Finally, the critical temperature is also shifted downwards due to the finite 
number of atoms in the harmonic trap and the discretization of energy 
levels.
This can be accounted for by modifying the density of states in a 
harmonic oscillator to account for the zero-point energy
$\hbar (\omega_x + \omega_y + \omega_z)/2$~\cite{gros95lamb,kett96fin}.

Measurements of the 
transition temperature were performed at JILA~\cite{ensh96} with small 
trapped samples.  Condensate fractions were measured in time-of-flight, and 
compared with various predictions using the reduced temperature $T / 
T_c^{(0)}$.  Their results were consistent with the ideal-gas prediction.
The predicted temperature 
shifts were of the same 
magnitude as their measurement accuracy of 5\%.  Better precision can probably be 
obtained with larger samples -- encouraging progress was presented at 
Varenna by J.  Dalibard.  
Ultimately, measurements may be impaired by the uncertain 
interpretation of time-of-flight images (sect.\ 
\ref{compare_insitu_tof}) which can be obviated by in-situ imaging.

\subsubsection{Condensate fraction}
\label{sec:condfrac}

The first results on the condensate 
fraction $N_0/N$ versus temperature~\cite{mewe96bec,ensh96} clearly showed the sharp increase of 
the condensate fraction with decreasing temperature $T$, consistent with 
the theoretical prediction for an ideal Bose gas in an harmonic oscillator 
potential $N_0/N = 1- (T/T_c^{(0)})^3$~\cite{bagn87ext}.  For a 
homogeneous gas, the increase is softer according to 
$N_0/N = 1-(T/T_c^{(0)})^{3/2}$.  This dependence was observed in 
measurements of helium in Vycor which was shown to realize a dilute 
three-dimensional gas~\cite{croo83}.  

Interactions lower the condensate fraction.  This arises primarily 
because the ground state energy rises as particles accumulate in the 
Bose condensate.  This implies a rise in the chemical potential,
and therefore 
an increase in the number of particles in the thermal cloud.  The reduction of 
the condensate fraction has been evaluated numerically in several 
works~\cite{gior96,hutc97,dodd97} as well as analytically in a 
``semi-ideal'' approach, in which the thermal cloud is treated as an 
ideal gas in a combination of the external trapping potential and a 
mean-field repulsive potential due to the 
condensate~\cite{nara98semi,ming97}.  

The magnitude of this reduction is 
characterized by the parameter $\eta = \mu_{0}/k_{B} T_{c}^{{(0)}}$ where 
$\mu_{0}$ is the interaction energy if all trapped particles form a 
Bose condensate~\cite{gior97}.
$\eta$ is small for all current experiments with harmonic traps.
Further, any evidence of the reduction in the 
condensate fraction is lost if
$T_{c}^{(0)}$ is not well calibrated.  Indeed, no evidence of 
these interaction effects 
has been obtained for harmonically trapped samples.
Recently, clear evidence for the reduction of the condensate fraction 
 was obtained by using a non-harmonic trapping potential that 
accentuated the effect~\cite{stam98rev} (sect.\ \ref{revers}).

\subsubsection{Density profiles} 

\begin{figure}
\epsfxsize = 6cm
\centerline{\epsfbox[0 0 353 342]{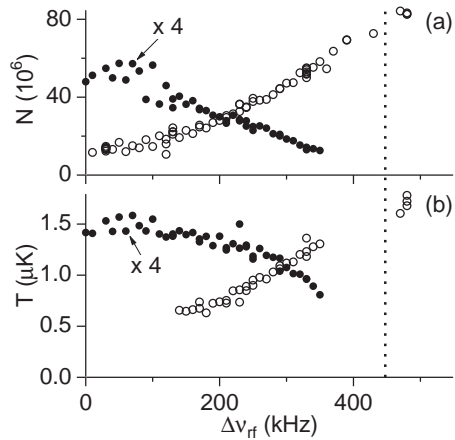}}
\caption[Equilibrium properties of trapped Bose gases above and below 
the phase transition]{Equilibrium properties of trapped Bose gases above and below 
the phase transition.
The total number $N$ (a, open circles)
was determined by summing over the observed column densities.
The approximate condensate number $N_0$ (a, closed circles, $\times 4$),
temperature $T$ (b, open circles) and chemical potential 
$\mu / k_B$
(b, closed circles, $\times 4$) were determined from fits.
$\Delta \nu_{\mbox{rf}}$ is the trap depth determined by the final 
rf frequency in the evaporative cooling cycle.
The dashed line indicates the observed transition temperature.
We estimate systematic errors in $N$ and $N_0$ of as much as 20\% due to
the calibration of our phase-contrast signal.
Figure taken from ref.~\cite{stam98rev}.}
\label{conditions}
\end{figure}

In-situ density profiles of partially condensed clouds were measured
at MIT~\cite{andr96,stam98coll}.
We examined a series of phase-contrast images of clouds at various temperatures
and from these determined
$T$, $\mu$, $N$, and an estimate for $N_{0}$ (fig.~\ref{conditions}).  
From these determinations and from time-of-flight measurements 
~\cite{mewe96bec} it can be seen that the Thomas-Fermi relation 
between the condensate energy and number remains valid at 
higher temperatures.  A more sensitive probe of non-zero temperature 
properties is obtained from precise measurements of collective 
excitation frequencies, discussed in sect.~\ref{sound}.

\subsubsection{Specific heat}

The derivative of the total release energy of 
the cloud with respect to temperature is related to the specific heat, but 
does not include the potential energy of the trap.  Theoretical 
calculations ~\cite{gior96}\ were in good agreement with the experimental 
results ~\cite{ensh96}\ and showed that the release energy and the total 
energy have a similar behavior as a function of temperature.  Both change 
their slope abruptly at the transition temperature.  

\subsection{Bose condensates with a negative scattering length}

The above discussion focused on Bose gases with repulsive 
interactions, for which the scattering length $a>0$.
Experiments have also been done using spin polarized $^{7}$Li gases, 
for which the scattering length is negative ~\cite{mcal95}.
In a homogeneous system, attractive interactions prevent the formation 
of a Bose condensate since a condensate lowers its 
energy by contracting to higher densities, leading eventually to 
extreme collisional losses.
In a trap, however, this collapse is forestalled by a kinetic 
energy barrier for condensates below a critical condensate number.
Experiments at Rice University~\cite{brad95bec,brad97bec} showed that, with 
a carefully characterized imaging system, condensates of less than 
1000 atoms could be discerned in in-situ images.
The observed condensate numbers were consistent with the 
predicted critical number for condensate collapse.

\newcommand{\mfp}{l_{\mbox{\scriptsize mfp}}}

\section{Sound and other dynamic properties}
\label{sound}

While Bose-Einstein condensates are produced and probed using the tools of
atomic physics, their connection to decades-old condensed-matter physics
is most evident in the study of sound.  Much of our understanding of the
nature of sound in quantum fluids comes from the context of
liquid helium (see discussions in refs.~\cite{nozi90} and \cite{grif93}). 
However, due to the strong interactions in liquid helium
(after all, it is a liquid) the fundamental connections of superfluidity
and sound propagation to Bose-Einstein condensation were not immediately
apparent.  Indeed, as studies of liquid helium progressed deeper into its
dynamic properties, two competing theoretical approaches emerged: the
empirical quantum hydrodynamic approach of the Landau school, and the
two-fluid picture of Tisza and London which emphasized the connection to
Bose condensation.

The convergence of these two approaches was suggested by the Bogoliubov
theory of weakly interacting Bose gases~\cite{bogo47}, which showed that Bose-Einstein
condensation could produce features of the excitation spectrum which
Landau's hydrodynamic theory postulated at its start.  The theory of
degenerate Bose gases was developed further in later years, providing
further insight into the nature of sound propagation at finite
temperature~\cite{lee59}, the interpretation of
experimental probes of liquid helium, and the nature of the BEC phase
transition.  Yet, these gases existed only in theoretical
papers, and none of these theories could be verified experimentally.

Now, with gaseous Bose-Einstein condensates produced in more than a dozen
laboratories around the world, researchers have turned to these
decades-old theories and begun testing their validity.
We leave the description of these theories to other chapters of
this book, and present this theoretical context only as it
pertains to recent experiments on magnetically trapped Bose condensed
gases.  By considering a hierarchy of length scales (shown in Table 
~\ref{soundlengths}),
we divide the
description of sound propagation into separate regimes, differentiating
between pulse propagation at short wavelengths and collective modes at
longer ones, and between collisionless and hydrodynamic behavior.
Experiments in the collisionless regime and zero-temperature have
confirmed the Bogoliubov mean-field description of Bose condensed gases,
while findings at non-zero temperature have challenged the theoretical
formalism for their description.  Most recently, experiments have pushed
toward the hydrodynamic limit, allowing a closer connection to the
phenomena of first and second sound in superfluid helium.

\begin{table}[ht]
\begin{center}
	\caption[The nature of collective excitations of the condensate and the 
	thermal cloud]{The nature of collective excitations of the condensate and the 
	thermal cloud.
	Various regimes are distinguished according to a 
	hierarchy of length scales: the reduced wavelength of the excitation 
	$\lambdabar_{ex}$, the healing length $\xi$, and the 
	mean-free path for collisions between quasi-particles $\mfp$. \label{soundlengths}}
\begin{tabular}{|c|c|c|c|}
	\hline
	Regime & Length scales & Condensate & Thermal cloud \\
	\hline
	collisionless & $\lambdabar_{ex} < \xi, \mfp $ & ballistic & ballistic \\
	collisionless & $\xi < \lambdabar_{ex} < \mfp $ & zero sound & ballistic \\
	hydrodynamic & $\xi < \mfp < \lambdabar_{ex}$ & second sound & first sound \\
	\hline
\end{tabular}
\end{center}\vskip-\lastskip
\end{table}

\subsection{Collisionless excitations in a homogeneous Bose gas}

The nature of collective excitations in a homogeneous Bose gas
depends on the hierarchy of
three length scales:

\begin{itemize}
	\item The reduced wavelength of the excitation $\lambdabar_{ex}$

	\item The healing length $\xi$ which is given by the condensate
density as
	$\xi = (8 \pi a n_0)^{-1/2}$.  Modifications of the condensate
	wavefunction on this length scale imply a kinetic energy which is
equal to
	the chemical potential.

	\item The mean-free path $\mfp$ for collisions between 
	quasi-particles, or more specifically, between the collective
	excitation and the other excitations which comprise the thermal
cloud.
	This length scale can be estimated by considering collisions among
free
	particles in a thermal cloud in the absence of the condensate.  A more
	exact determination requires careful consideration of the
modification of
	collisions due to the condensate (see chapter by K.\ Burnett in this 
	volume).
\end{itemize}

The condition $\lambdabar_{ex} \ll \mfp$ defines the collisionless 
regime (in the sense of collisions among quasi-particles),
which applies at low densities of the thermal cloud.  The excitations
in this regime were derived for zero temperature by Bogoliubov 
~\cite{bogo47}.
The
excitation energy $\epsilon_{k}$ at wavevector $k$ is given
by
\begin{eqnarray}
	\epsilon_{k} & = & \sqrt{ \frac{\hbar^{2} k^{2}}{2 m} \left(
\frac{\hbar^{2}
	k^{2}}{2 m} + 2 \mu \right) } \\
& = &  \frac{\hbar^2}{2 m} \sqrt{\frac{1}{\lambdabar_{ex}^2} 
\left( \frac{1}{\lambdabar_{ex}^2} + \frac{2}{\xi^2} \right)}
\end{eqnarray}
where $\mu = 4 \pi \hbar^{2} a n_{0}/m$ is the chemical potential as defined by the condensate
density $n_{0}$.
At long wavelengths ($\lambdabar_{ex} \gg \xi$) the excitation energy depends 
linearly on the wavevector, implying phonon-like excitations;
a packet of such excitations travels 
without spreading at the speed of {\it Bogoliubov (zero) sound} $c_0 = \sqrt{\mu / 
m}$.
At short wavelengths ($\lambdabar_{ex} \ll \xi$) the excitation energy is 
approximately
$\epsilon_{k} = \hbar^2 k^2 / 2 m + \mu$, i.e. the excitations are 
free-particle-like with a mean-field offset of $\mu$.  In an inhomogeneous 
trapping potential, this mean-field offset implies that the condensate 
repels the thermal cloud.
Collisionless excitations have been described at finite temperature as 
well~\cite{grif93}.  The distinction between Bogoliubov-sound phonons and free-particle 
excitation remains, while the condensate density $n_0(T)$ varies with 
temperature.

\subsection{Collisionless excitations in an inhomogeneous, trapped Bose 
gas}

The nature of collective excitations in an inhomogeneous, trapped Bose gas 
is influenced by the introduction of a new length scale: the length of 
the condensate $x_{i,c,0}$.  This divides the description of condensate 
excitation into three regimes:
\begin{itemize}

\item For excitations of wavelengths smaller than {\it all} dimensions of the 
condensate, $\lambdabar_{ex} \ll x_{i,c,0}$, the condensate can be treated as 
locally homogeneous, and the distinction between phonons and 
free-particles is as before. Indeed, it has been shown that 
finite-temperature thermodynamic properties such as condensate fractions 
and density profiles are well-described by a semi-classical approach 
using the Bogoliubov spectrum to describe localized 
excitations~\cite{gior97jltp,nara98semi}.
Excitations at these wavelengths have not been studied.

\item For longer wavelengths which approach the size of the condensate 
($\lambdabar_{ex} \simeq x_{i,c,0}$), the excitation spectrum becomes discretized, 
i.e. the collective modes of the system become standing sound waves at 
specific frequencies.  It is interesting to note that in the Thomas-Fermi 
limit, both the speed of Bogoliubov sound and the length of the condensate scale 
as $(a n_{0})^{1/2}$.  Thus, the frequency of the collective excitation 
$\omega \propto c_0/x_{i,c,0}$ is independent of the speed of sound.  Collective excitations in this 
regime have been studied over a wide range of temperatures, as we describe 
below.

\item For Bose condensates in anisotropic potentials there is an 
intermediate regime in which the wavelength of the excitation is larger 
than the size of the condensate in one or two dimensions, but smaller than 
the size of the condensate in the other directions.  In this case, the axial 
discretization of the collective modes is not apparent, and thus the 
pulses propagate as sound waves.  The connection between this 
phonon picture and the aforementioned discrete spectrum was laid out by 
Stringari ~\cite{stri98cigar}.
\end{itemize}

\subsection{Experiments on collective excitations near $T = 0$}

\begin{figure}[htbf]
\epsfxsize=70mm
\centerline{\epsfbox[77 200 584 532]{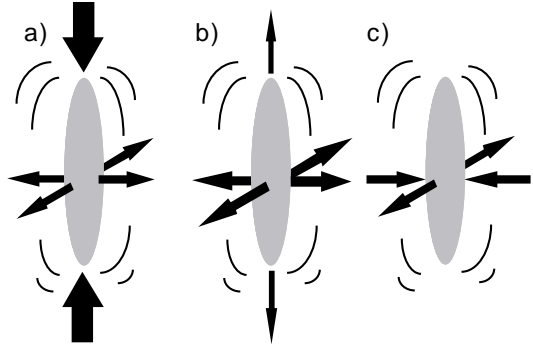}}
\caption[Shape of low-lying collective excitations]{Shape of low-lying collective excitations: a) slow $m=0$ 
quadrupolar oscillation (JILA, MIT), b) fast $m=0$ radial 
oscillation (MIT), c) $|m| = 2$ oscillation (JILA).}
\label{shapeofosc}
\end{figure}

Coming from the spectroscopy tradition of atomic physics, it was natural 
that
researchers focused early on the discrete excitation spectrum of a 
condensate.  First, low-lying excitations were studied at the limit of 
zero temperature.  Researchers at JILA studied two shape oscillations of 
condensates confined in a TOP trap~\cite{jin96coll}.  The first was a cylindrically 
symmetric $m=0$ quadrupole mode wherein the axial ($z$-axis) and radial 
lengths of the condensate oscillate out of phase (fig.~\ref{shapeofosc}a).  Here $m$ denotes the angular momentum of the 
excitation about the $z$-axis.  The second was the 
$m= 2$ quadrupole mode.  For this mode, the cylindrical 
symmetry was broken, and the lengths of the condensate in the two radial 
directions oscillated out of phase (fig.~\ref{shapeofosc}c).  
The condensates in these experiments 
were in a regime intermediate to the ideal-gas and Thomas-Fermi limits.  
Therefore, these low-lying oscillations were studied between the 
free-particle and the Bogoliubov-sound limits discussed above.
The measurements agreed well with the predictions of mean-field 
theory~\cite{edwa96coll}.

\begin{figure}[htbf]
\epsfxsize=70mm
\centerline{\epsfbox[0 0 307 102]{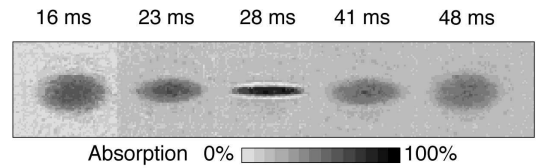}}
\caption[$m=0$ quadrupolar condensate oscillations viewed in 
time-of-flight]{$m=0$ quadrupolar condensate oscillations viewed in 
time-of-flight absorption imaging.  Oscillations in the aspect ratio of the expanding 
condensate are clearly visible.  The horizontal width of each cloud 
is 1.2 mm.  Figure taken from ref.~\cite{mewe96coll}.}
\label{tof_oscillations}
\end{figure}

\begin{figure}[htbf]
\epsfxsize=70mm
\centerline{\epsfbox[0 0 154 126]{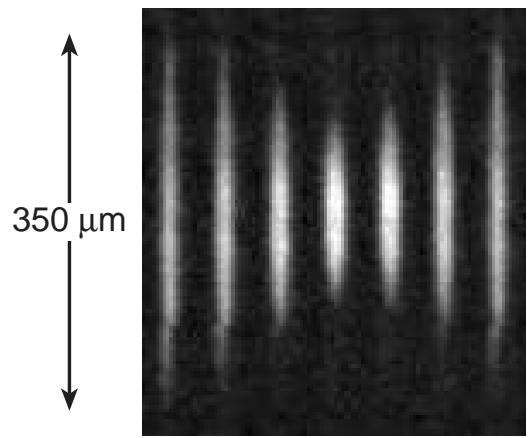}}
\caption[$m=0$ quadrupolar condensate oscillations viewed {\it 
in-situ}]{$m=0$ quadrupolar condensate oscillations viewed {\it 
in-situ}.  
Repeated phase-contrast images, taken at 5 ms intervals, show 
large-amplitude
oscillations of a low-temperature Bose-Einstein condensate.
Figure taken from ref.~\cite{stam98coll}.}
\label{pc_oscillations}
\end{figure}

We simultaneously studied oscillations near zero temperature of 
cigar-shaped condensates in a dc cloverleaf trap~\cite{mewe96coll}.
Collective modes were excited by sinusoidally varying the currents in 
the trapping 
coils.  Since this excitation scheme preserved the axial symmetry of the 
trap, we only expected to excite $m=0$ modes.  Fortunately, imperfections 
in the trapping coils also allowed us to excite center-of-mass 
oscillations of the condensate in the trap, providing for 
accurate measurements of the trap frequencies.  Again, two shape 
oscillations were excited.  The lower frequency mode was similar to the 
$m=0$ quadrupolar modes observed at JILA, with out-of-phase oscillations 
along the axial and radial directions.  The higher frequency mode was 
primarily a radial breathing mode (fig.~\ref{shapeofosc}b).
After locating the modes by a non-selective ``step'' excitation, we 
 used a five-cycle sinusoidal 
modulation of the trapping coils to resonantly excite the shape 
oscillations.
The subsequent free oscillations were clearly visible as periodic 
modulations of the aspect ratio in time-of-flight 
(fig.~\ref{tof_oscillations}) and in phase-contrast 
(fig.~\ref{pc_oscillations}) as observed later~\cite{stam98coll}.

Our experiments were performed with condensates which were well in the 
Thomas-Fermi limit.  The oscillations we observed were considered by 
Stringari, who provided the first analytical expression for their 
frequency and shape~\cite{stri96coll}.  The agreement between the predicted 
frequencies and the experimental results was quite good.  The fast 
oscillation at $\nu = 2.04(6) \cdot \nu_r$ agreed with the prediction of $2 
\cdot \nu_r$.  For the slow oscillation, we measured a frequency $\nu = 
1.556(14) \cdot \nu_z$ compared with the  prediction of $1.580 \cdot \nu_z$.
More recently, we improved our measurement to obtain a frequency of 
$1.569(4) \cdot \nu_{z}$ at the limit of low temperature ($\nu_{r}$ and 
$\nu_{z}$ are the radial and axial frequencies, resp.)~\cite{stam98coll}.
This close agreement constitutes a critical quantitative test of the mean-field 
description of excited states of a Bose condensate.

\subsection{Measurements of the speed of Bogoliubov sound}

The experiments described above studied the low-lying discrete 
oscillation modes of a trapped condensate.
In order to connect more closely with the continuous excitation 
spectrum of homogeneous system, we also studied density 
modulations at wavelengths  of 20 -- 30 $\mu$m that were smaller than the 
length of the 
condensate~\cite{andr97prop,andr98erra}.
For this, localized density perturbations were created using an 
off-resonant blue-detuned laser beam focused to the middle of the trap.
Positive perturbations 
were created by suddenly switching on the laser beam after the 
condensate had formed.  The repulsive optical dipole force expelled 
atoms from the center of the condensate, creating two density peaks 
which propagated symmetrically outward.
Alternatively, we formed a condensate in the presence of the laser 
light and then switched the laser off.
This created localized 
depletions of density which also propagated outward.

\begin{figure}[htbf]
\epsfxsize=70mm
\centerline{\epsfbox[0 0 395 381]{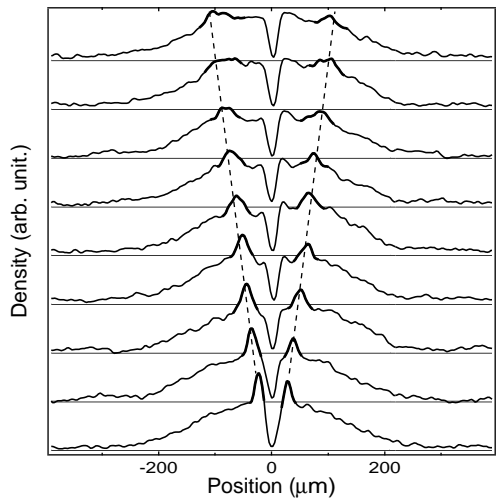}}
\caption[Observation of sound propagation in a Bose condensate]
{Observation of sound propagation in a Bose condensate.
A non-destructive phase-contrast image was taken every 1.3~ms.
Vertical profiles of the condensate density show two ``blips'' 
traveling out symmetrically from the center of the cloud.
Figure taken from ref.~\cite{andr97prop}.}
\label{soundprop}
\end{figure}

\begin{figure}[htbf]
\epsfxsize=70mm
\centerline{\epsfbox[0 0 429 266]{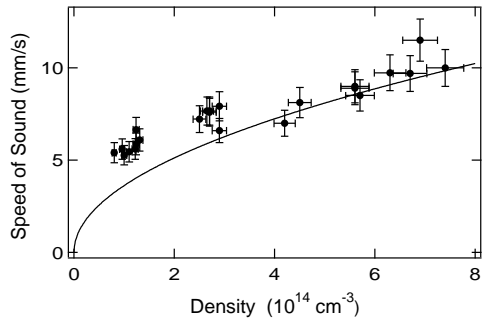}}
\caption[Speed of sound versus condensate peak density]{Speed of sound versus condensate peak density.
The solid line is the theoretical prediction with no adjustable parameter.
The error bars indicate only the statistical error.
Figure taken from ref.~\cite{andr98erra}, the erratum to the original 
experiment.}
\label{soundspeed}
\end{figure}

Fig.~\ref{soundprop} shows the propagation of density 
perturbations observed by sequential phase-contrast imaging of a 
single trapped cloud.
We observed one-dimensional axial propagation of sound at a constant 
velocity near the center of the cloud, where the axial density varies 
slowly.
The density dependence of the speed of sound was studied using 
adiabatically expanded or compressed condensates, yielding maximum 
condensate densities $n_{0}$ ranging from 1 to 5 $\times 10^{14}$ 
cm$^{-3}$ (fig.~\ref{soundspeed}).
The data were compared with the prediction of Bogoliubov theory, 
$c_s = (4 \pi \hbar^{2} a n / m^{2})^{1/2}$, where the variation of 
the condensate density across the radius of the cloud is accounted 
for by using $n = n_0 / 2$~\cite{zare97,kavo98,stri98cigar}.
The agreement between the data and this theory was good except
at low density where the assumption that the sound pulse is longer 
than the radial extent of the condensate began to break down.

It has been discussed that our scheme for producing propagating 
density depletions (cutting a condensate and then removing the ``knife'') 
should produce propagating dark solitons~\cite{morg97,rein97,jack98soli}.
Such solitons, familiar from other applications of non-linear 
propagation equations, are predicted to be 
stable in 
one-dimensional Bose-Einstein condensates (in relation to the 
Thomas-Fermi condition).
The condensates in our experiment were not one-dimensional, and 
the size of a soliton would be about a healing length ($\sim 0.2 \, 
\mu$m).
Thus, individual solitons are unlikely to be observed in our experiment.
Still, it is possible that the observed propagating ``dips'' formed the 
envelope of many unresolved solitons.

\subsection{Collective excitations at non-zero temperature}

While the above 
experiments were done with samples close to zero temperature,
the observed damping gave an indication of behavior at non-zero temperature.
The next set of experiments at JILA~\cite{jin97} and 
MIT~\cite{stam98coll} explored the excitations of a Bose gas as the 
temperature of the gas was varied.

The effects of non-zero temperature were 
three-fold: (1) shape oscillations of the thermal cloud were 
introduced, (2) the frequencies of the collective modes were influenced 
by interactions between the condensate and thermal cloud, and (3) the 
oscillations were increasingly damped at higher temperatures.

\subsubsection{Oscillations of the thermal cloud}

At non-zero temperature, a significant fraction of the gas is 
not condensed.  This thermal fraction can be 
clearly 
discerned in the bimodal density profiles observed in-situ and 
in time-of-flight.  Thus,
one can examine separately the response of both the thermal cloud
and of the condensate to modulations of the trapping potential.
Thus, in accordance with the two-fluid picture of partly condensed 
Bose gases, there are two ``collective excitations'' of the system 
for each type of shape oscillation: a condensate oscillation, and an oscillation of 
the thermal cloud.
The frequencies of the two oscillations are generally different.

In the JILA experiment~\cite{jin97}, the shape oscillation frequency of the 
thermal cloud was about twice the trapping frequency for both the 
$m=0$ and $m=2$ quadrupolar modes.  The thermal clouds in this 
experiment were in the collisionless regime.  The dynamics of a 
thermal cloud in this regime is not sound-like, but ballistic.
An oscillatory response comes only from the reflection of the free 
particles at the trap boundaries.
Such oscillations are persistent only for a harmonic 
trapping potential, in which free-particles of any velocity have the 
same oscillation period.  In contrast, for a collisionless thermal 
cloud in a box, in which the oscillation period scales inversely with 
the velocity, any collective response would quickly dephase.

In the MIT experiment~\cite{stam98coll}, the thermal cloud oscillated at a frequency 
which was not a multiple of the trapping frequency.  This response is 
due to the onset of hydrodynamic behavior in the thermal cloud which 
is discussed further in sect.~\ref{hydrosection}.

\subsubsection{Frequency shifts of condensate oscillations}

\begin{figure}
\epsfxsize = 7cm
\centerline{\epsfbox[0 0 499 441]{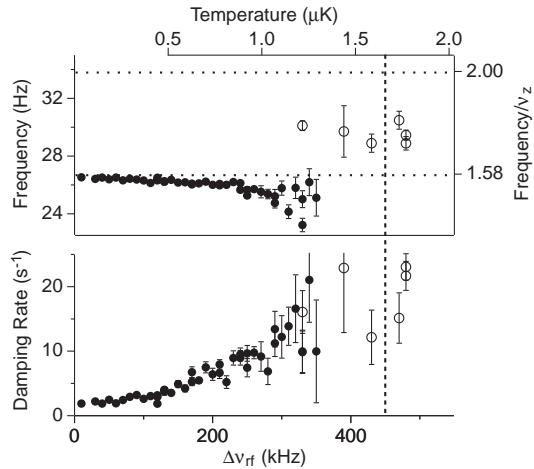}}
\caption[Temperature dependent frequencies and damping rates
of $m=0$ quadrupolar oscillations of the
thermal cloud and condensate]{
Temperature dependent frequency and damping rates
of $m=0$ quadrupolar collective oscillations of the
thermal cloud (open circles) and condensate (closed circles).
The free-particle limit of 2 $\nu_z$ and the
zero-temperature condensate oscillation limit of 1.580 $\nu_z$ are 
indicated.
The vertical dashed line marks the observed transition temperature.
The temperature axis is based on observations of clouds at equilibrium
for $T > 0.5 \, \mu\mbox{K}$.}
\label{nonzeroosc}
\end{figure}

Another effect of finite temperature is a shift in the frequency of 
the condensate excitations.
Such a frequency shift is expected due to the decrease of the 
condensate density as the temperature rises, shifting the excitations from
phonons to free-particles.
However, other effects were clearly evident.
For example, at JILA, the frequency of the $m=2$ quadrupolar mode was 
found to {\it decrease} at higher temperatures, rather than tending 
toward the higher frequency expected in the free-particle 
regime~\cite{jin97}.
In the MIT experiment, the condensates that were studied were always 
in the Thomas-Fermi limit.
Nevertheless, a distinct downward frequency shift of as much as 5\% 
was observed (fig.~\ref{nonzeroosc}) \cite{stam98coll}.

We also studied the effects of non-zero temperature on the propagation 
of Bogoliubov sound~\cite{andr97prop}.  When we varied the condensate density, the speed of 
Bogoliubov sound scaled as $\sqrt{n_{0}(T)}$ where $n_{0}$ is the 
temperature dependent condensate density.  This agrees with the 
expected behavior from finite temperature theories for homogeneous 
gases ~\cite{gay85,grif93}.
Due to a lack of precision, we could not detect the
$\sim$ 5\% shift seen in later 
experiments~\cite{stam98coll}.

The behavior seen in these experiments is not yet fully understood, 
and constitutes a major challenge to the theories developed to 
describe non-zero temperature behavior of Bose gases (see discussion 
in several chapters of this book).  A 
variety of explanations have been proposed:

\begin{itemize}

	\item The simplest extension of the Bogoliubov theory to finite 
	temperatures uses an approximation introduced by Popov which 
	neglects the correlations and motion of the thermal 
	cloud~\cite{popo87}.
	In this approach, the condensate oscillates in an effective potential 
	which is the sum of the trapping potential and the mean-field 
	potential due to the thermal cloud (recall, the condensate repels 
	and is repelled by the thermal cloud).  The failings of this approach were 
	evident already before the recent experiments on Bose-Einstein 
	condensation since it violated the Pines-Hugenholtz theorem which 
	requires a gapless excitation spectrum (see recent discussion in 
	~\cite{grif96cons,hutc98}).
	Applications of this 
	formalism to the experiments at JILA~\cite{dodd97,hutc97} have failed 
	to explain the results adequately.
	
	\item In response to the failure of the Popov approach, a gapless 
	formulation of the finite-temperature Bogoliubov theory has been 
	considered for the inhomogeneous condensates under study.
	This theory goes beyond the Popov approximation by accounting for 
	the influence of the condensate on 
	collisions between particles in the thermal cloud, giving rise to an 
	``anomalous density'' term~\cite{hutc98} (see also chapter by K. Burnett).
	This approach has had limited success in explaining the results of 
	the JILA experiment: it explains the observed downward 
	frequency trend of the $m=2$ oscillation, but not 
	the upward trend exhibited by the $m=0$ oscillation. 

	\item Finally, in recent work, attempts have been made to include the motion of 
	the thermal cloud and understand its influence on the observed 
	condensate excitations~\cite{olsh98coll,bijl98coll}.  Preliminary application of these ideas to 
	the JILA experiment have been promising, while their application to 
	the MIT data has not been done.

\end{itemize}

\subsubsection{Damping of condensate oscillations}

Early experiments on collective excitations had already noted the 
presence of damping, in spite of attempts to study oscillations at the 
limit of zero temperature.
In the variable-temperature experiments at JILA and MIT, the damping 
rate was found to increase dramatically with temperature, by as much 
as ten-fold 
near the BEC phase transition temperature.
The damping rates were found to vary even at temperatures where no 
thermal cloud was discernible by imaging.
In this sense, damping rates may become sensitive ``thermometers'' of 
condensates at extremely low temperatures.

The dependence of the damping rate on temperature is another way in 
which the presence of the thermal cloud influences the motion of the 
condensate.
The apparent mechanism for this damping is Landau damping, in which a 
quasi-particle disappears in a collision with a thermal excitation, 
promoting it to 
higher energy~\cite{liu97,pita97,gior98,fedi98}.
Recent treatments based on Landau damping have been quite successful 
in explaining the damping which was observed 
experimentally~\cite{fedi98}.
These treatments were discussed at Varenna by P. Fedichev.

Landau damping depends on the presence of thermal 
excitations, and thus is absent at zero-temperature.
Nevertheless, collective excitations at zero-temperature can decay 
away.
One mechanism, Beliaev damping, corresponds to the decay 
of a high-energy collective excitation into two lower energy ones.
For trapped Bose condensates and low-lying modes, this decay mechanism 
is not available since there are no modes into which to decay~\cite{pita97}.
Collective excitations are also subject to zero temperature dephasing 
due to the atom-number uncertainty in the condensate and to amplitude 
dependence of the mode.
Both can lead to an apparent damping~\cite{pita97phen,kukl97,grah97}.
Further, it has been shown that for higher amplitude oscillations, the 
non-linear mixing between modes can lead to ergodic behavior which 
results in an irreversible damping and an effective heating of the 
cloud~\cite{sina98}.
This provides incentive for future experiments to continue studying 
collective excitations closer to the zero-temperature limit, perhaps 
using more controlled evaporative cooling (such as provided in 
optical traps) to get there.

\subsection{First and second sound in a Bose gas}
\label{hydrosection}

So far, we have discussed excitations of a Bose condensed gas in the 
collisionless regime, where the wavelength of the excitation is much 
smaller than the mean-free path $\lambdabar_{ex} \ll \mfp$.  This regime 
applies at zero temperature and at low densities of the thermal cloud.
At higher densities of the normal component, when $\lambdabar_{ex} \gg 
\mfp$, collective excitations become hydrodynamic in nature, and one 
expects two phonon-like excitations which are the in-phase and 
out-of-phase oscillations of two hydrodynamic fluids (the normal fraction and the 
superfluid).
The presence of two hydrodynamic modes is similar to the case of bulk 
superfluid $^{4}$He, where they are known as {\it first} and {\it second sound}.
Superfluid $^{4}$He has a small coefficient of thermal expansion.
Thus the two eigenmodes decouple into density modulations (first 
sound) and temperature modulations (second sound), with both fluids 
participating equally in 
both modes.
In contrast, a gas has a large coefficient of thermal expansion.
This results in the oscillations of 
each fluid being nearly uncoupled.
The in-phase oscillation, which is analogous to first sound, involves 
mainly the thermal cloud.
The out-of-phase oscillation, which is analogous to second sound, is 
confined mainly to the condensate~\cite{lee59,grif93,grif97,shen98}.
Let us note another difference: the observation of second sound
was dramatic evidence for the presence of two fluids in 
superfluid $^{4}$He, whereas in trapped Bose gases, the visible separation 
between the normal and the superfluid components already confirms the 
two-fluid approach.

\begin{figure}
\epsfxsize = 7cm
\centerline{\epsfbox[0 0 472 246]{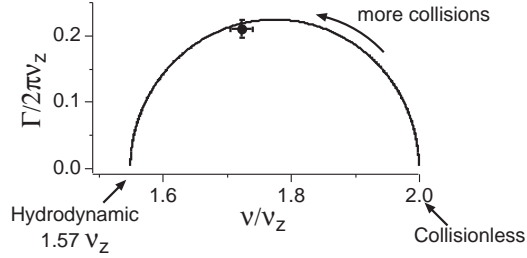}}
\caption[Frequency and damping rate of excitations of the thermal 
cloud between the collisionless and hydrodynamic limits]{Frequency 
$\nu$ and damping rate $\Gamma$ of excitations of the thermal 
cloud between the collisionless and hydrodynamic limits.
Line shows interpolation
according to eq.\ (\ref{hydroequation}).
The experimental data point shows that thermal clouds above the BEC transition temperature
were probed in an intermediate regime.}
\label{hydrodynamic}
\end{figure}

In the finite-temperature experiment at MIT, the hydrodynamic regime 
was approached for the first time~\cite{stam98coll}.
The onset of hydrodynamic behavior was indicated by the oscillations 
of the thermal cloud.
In this experiment, the thermal cloud oscillated at a frequency of
about 1.75 $\nu_{z}$ with a damping rate of about 20 s$^{-1}$
both above and below the BEC transition 
temperature (fig.~\ref{nonzeroosc}).
The observed frequency $\nu$ is between the predicted collisionless 
limit of
$\nu = 2 \cdot \nu_z$ and the hydrodynamic limit of
$\nu = 1.55 \cdot \nu_z$~\cite{grif96}.
These measurements indicate that the thermal cloud was in a density regime 
intermediate to the two limits.
In this intermediate regime, the complex angular frequency of the 
oscillation $\omega$ (the imaginary part of which gives the damping 
rate) is described by the interpolation~\cite{kavo98damp,kavo98rela}
\begin{equation}
	\omega^{2} = \omega_{C}^{2} + \frac{\omega_{H}^{2} - 
	\omega_{C}^{2}}{1 - i |\omega \tau|}
\label{hydroequation}
\end{equation}
where $\omega_{C}$ and $\omega_{H}$ are the angular frequencies of
the excitation in the 
collisionless and hydrodynamic limits, respectively, and $\tau$ is the 
rate of collisions with thermally excited quasi-particles.
The locus of points in the $\Gamma$ -- $\nu$ plane described by
the interpolation (\ref{hydroequation}) 
is shown in fig.~\ref{hydrodynamic} 
along with our measurement above the BEC phase transition.

One can also characterize the regime of our experiment by considering 
the length scales $\lambdabar_{ex}$ and $\mfp$.
One can estimate the collisional mean-free path
as $\mfp\simeq (n_T \sigma)^{-1} = 96 \, \mu\mbox{m} \times (T/\mu\mbox{K})^{-3/2}$
using the peak density of the thermal
cloud $n_T = 2.612 \, (m k_B T / 2 \pi \hbar^2)^{3/2}$,
and a collisional cross section  $\sigma = 8 \pi a^2$ with the 
scattering length $a = 2.75\, \mbox{nm}$~\cite{ties96}.
Around the transition temperature, we find the rms axial length of the 
thermal cloud to be $z_{th,0} \simeq 8 \, \mfp$.
This comparison of length scales,
the observed frequency shift away from $2 \cdot \nu_z$, and the
high damping rate all demonstrate that the collective behavior
of the thermal cloud is strongly affected by collisions.
Thus, the oscillations which we observe indicate the onset
of hydrodynamic excitations, i.e. first sound.
The hydrodynamic limit, characterized by low damping, would
only be reached for even larger clouds.

Similarly, a comparison 
between $\mfp$ and the axial length of the condensate $z_{c}$ at high 
temperatures ($z_c \simeq 4 \, 
\mfp$) indicates that hydrodynamic effects may be influencing the 
condensate oscillations as well.
Thus, these oscillations may constitute second sound in a Bose 
gas.
There are few theoretical predictions regarding the transition
from zero to second sound with which to compare our data.
In future experiments with larger condensates, the signature of this 
cross-over may appear in the damping rate of the oscillations, which
should decrease again at high-temperatures as one reaches the 
hydrodynamic limit.
Indeed, a recent analysis by Fedichev {\it et al.} (see chapter in 
this book) found
the damping rates we observed at high temperature
to be slightly lower than those 
expected based on a collisionless model of Landau damping, a 
tentative sign of the onset of hydrodynamic effects. 

Another collective excitation related to second 
sound is the anti-symmetric dipole oscillation, in which the 
centers-of-mass of the thermal cloud and the condensate oscillate 
out-of-phase.
This mode is analogous to second sound in liquid helium, where the 
superfluid and the normal fluid undergo out-of-phase oscillations of 
equal magnitude~\cite{zare97}.
We excited this mode using an off-resonant laser 
beam which was directed at the edge of the cloud,
where it overlapped only with the thermal cloud.
By tilting a motorized mirror, the laser beam was steered
toward and then away from the center of the 
cloud,
thereby pushing the thermal cloud in the axial direction
while not directly affecting the 
condensate.
The light was then turned off, and the cloud allowed to freely 
oscillate.

Over time, we observed the initially small oscillations of the 
condensate center-of-mass grow to an asymptotic oscillation with the 
center-of-mass of the entire cloud.
In the frame of this overall center-of-mass motion, the cloud 
underwent the anti-symmetric dipole oscillation, as we sought.
The downward frequency shift of this mode away from the trapping frequency 
(about 5\%) and its eventual dissipation are further signs of the 
interactions between the condensate and the thermal cloud~\cite{stam98coll}.
The description of these interactions requires a time-dependent 
treatment of the thermal cloud in contrast to the stationary treatment 
of most theoretical approaches.

\subsection{Challenges ahead}

The work over the last three years has elucidated all the basic 
aspects of sound in a Bose condensed cloud (Table ~\ref{soundoutlook}).  
Experiments have studied discrete 
standing-wave modes at zero and non-zero temperature, damping rates 
and frequency shifts, the propagation of sound pulses, and the onset 
of hydrodynamic behavior.
These experimental advances were accompanied by enormous progress in our 
theoretical understanding.
A comprehensive picture of collective excitations and sound 
propagation at zero temperature has been assembled.
At non-zero temperatures, progress has been made in understanding the 
coupled motion of the condensate and the thermal cloud.
Approaches to damping have been honed to the context of trapped Bose 
condensates with great success, and the hydrodynamic behavior of 
trapped gases above and below the BEC transition has been explored.

\begin{figure}[htbf]
\epsfxsize=70mm
\centerline{\epsfbox[0 0 105 90]{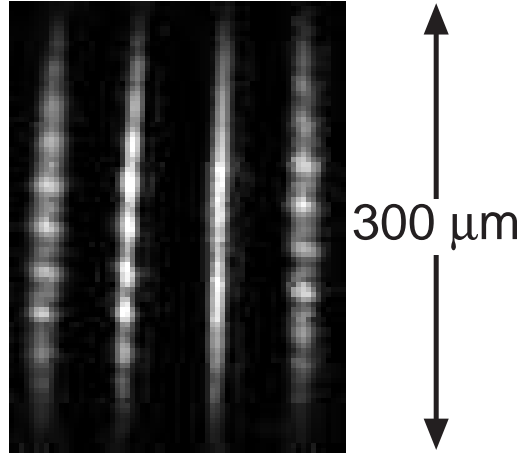}}
\caption[The shape of things to come: a high-lying (eight-node) axial 
excitation of a cigar shaped condensate]{The shape of things to 
come.  Four in-situ images at 1.3 ms intervals show a high-lying (eight-node) axial 
excitation of a cigar shaped condensate.  The oscillation frequency is 
about 250 Hz in a trap with axial frequency of about 17 Hz.}
\label{wildfigure}
\end{figure}

These advances point to many possibilities for further study.
For example, the demonstrated ability to measure collective 
excitation frequencies at the $10^{-3}$ level enables the search for 
small, but conceptually important effects, such as the small (about a 
percent) frequency shift 
due to quantum depletion~\cite{pita98}.
Only a few discretized collective modes have been studied closely, 
but the diverse ways which have been demonstrated to create 
excitations can be readily applied to study others.
Indeed, we created one such mode inadvertently: in the process of 
exciting radial center-of-mass oscillations to measure trap 
frequencies, we came upon a wild high-lying condensate excitation with 
as many as eight nodes along the trap axis (fig.~\ref{wildfigure}).
Studying these other modes may reveal the connection from discrete 
modes to continuous sound propagation~\cite{stri98cigar} or the 
possible chaotic propagation of localized excitations~\cite{flie98}.
Advancements in the preparation of ever colder samples may make 
accessible the study of dephasing and damping near zero temperature, 
while the creation of ever larger and denser condensates opens the 
door to studying the true hydrodynamic limit.

\begin{table}[ht]
\begin{center}
	\caption{Studies of sound in gaseous Bose-Einstein condensates.
	A comprehensive picture of the nature of sound has been assembled, 
	but many phenomena remain to be explored.}
\begin{tabular}{|l|l|l|}
	\multicolumn{3}{c}{$\leq 1998$} \\ \hline
	Experiment & Bogoliubov (zero) sound & $T = 0$ discrete modes  \\
	& & non-zero temperature \\
	& & damping   \\
	& & speed of sound \\
	& First and second sound & discrete modes \\
& 	& damping \\ \hline
	Theory & \multicolumn{2}{l|}{Landau damping} \\
	& \multicolumn{2}{l|}{zero, first, second sound} \\
	& \multicolumn{2}{l|}{discrete modes $\leftrightarrow$ sound} \\
	& \multicolumn{2}{l|}{hydrodynamic theory}\\  \hline
	\multicolumn{3}{c}{Experiments to come?} \\ \hline
	& $T = 0$ damping & Beliaev damping\\
	& & dephasing, collapse, revival\\
	& Higher lying modes & dynamic structure factor $S({\bf q},\omega)$\\
& 	Effects of quantum depletion &\\
	& Nonlinear effects & solitons, chaos, shock waves \\
& 	& mode coupling, frequency doubling \\
	& Hydrodynamics & \\ \hline
\label{soundoutlook}
\end{tabular}
\end{center}\vskip-\lastskip
\end{table}

\subsection{Other dynamic properties}
\label{dynamic2}
The previous discussion on the nature of sound in a Bose condensate 
focused on the linear response to external perturbations.
The study of higher amplitude motion is another strong test of the 
Gross-Pitaevskii equation.
In this section, we present these studies as well as summarizing 
other dynamical processes which include the formation and decay of 
the condensate.

\subsubsection{Free expansion and large amplitude oscillations of a Bose-Einstein condensate}
\label{tfexpansion}

The evolution of a Bose condensate during free expansion from a harmonic 
trap is described by the Gross-Pitaevskii equation (sect.~\ref{bectf}).
Careful study of this expansion is necessary, not only as a test of 
mean-field theory, but also to confirm the use of time-of-flight 
imaging to probe properties of trapped condensates.
Early studies of this expansion at MIT~\cite{davi95bec,mewe96bec} and 
JILA~\cite{ande95,holl97int} agreed well with theoretical
predictions~\cite{holl96exp,kaga96evol,cast96,nara96}.

\begin{figure}[htbf]
\epsfxsize=60mm
\centerline{\epsfbox[0 0 306 242]{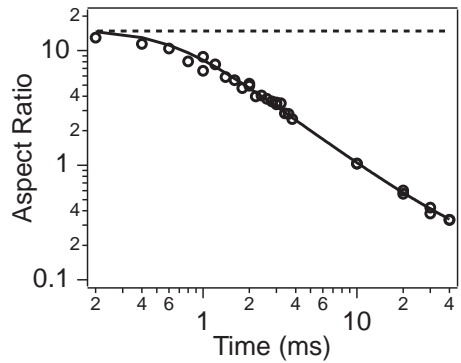}}
\caption[Aspect ratio of expanding Bose condensates]{Aspect ratio (axial to radial
widths) of expanding Bose condensates.
Data collected for times less than
10 ms were taken in 
phase-contrast imaging, and for later times
with absorption imaging.
The dashed line
indicates the ratio of trapping frequencies of 248(1) Hz radially
and 16.23(3) Hz axially.
The solid line gives the prediction of ref.\ ~\protect\cite{cast96}.}
\label{tof_figure}
\end{figure}

We recently probed this evolution in more detail, using both
phase-contrast (short
time-of-flight) and 
absorption (long time-of-flight) imaging~\cite{sten98odt}.
The
measured aspect ratios are presented
in
fig.~\ref{tof_figure}.
Two steps of the expansion are clearly shown: the initial radial 
acceleration in the first milliseconds, and the subsequent pure radial expansion.
The axial expansion would only become apparent at later times of 
expansion.
The axial and radial trapping frequencies were also determined, using 
in-situ imaging.
The theoretical prediction of ref.\ \cite{cast96}, which depends only 
on the measured trap frequencies, describes the data excellently,
confirming mean-field description of large-amplitude dynamics.

Other studies focused on large amplitude motion of trapped condensates.
To ensure that studies of collective excitations were done close 
enough to the limit of zero amplitude, where the excitation frequency 
reflects the quasi-particle spectrum, these frequencies were studied 
as a function of amplitude~\cite{jin96coll,mewe96coll,jin97,stam98coll}.  
For the low-lying $m=0$ quadrupolar 
condensate oscillation at low temperature, we observed a 1 Hz upward shift of 
the excitation frequency from its low amplitude limit as the relative amplitude 
of the oscillation rose to 50\%, in agreement 
with the theoretical analysis of Dalfovo \etal ~\cite{dalf97}.

Recently, the large amplitude dynamics of a two-component condensate 
were studied at JILA~\cite{hall98dyn}.  The 
observed sloshing and damping of the two-component system were treated 
theoretically as a manifestation of 
zero-temperature damping~\cite{sina98}.

\subsubsection{The search for vortices}
\label{sec:vortices}

Collective excitations, even those that possess angular momentum, 
describe single-particle excitations which can connect to the ground 
state by single-particle decays.
Vortices, on the other hand, describe collective motion involving the 
entire condensate.
The possible persistence of these rotating states relies on their 
metastability against single-particle decay --- although the energy of 
the condensate is lowered by placing it entirely in the non-rotating 
ground state, this many-particle relaxation is highly unlikely.
As the discussions at the Varenna summer school showed (see 
chapter by
A. Fetter), the search for quantized persistent vortices is a major 
aim of future BEC research.

\begin{figure}[htbf]
\epsfxsize=100mm
\centerline{\epsfbox[0 0 648 98]{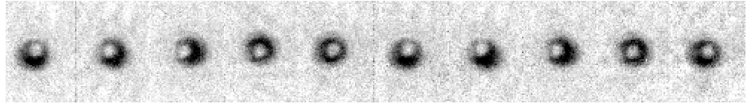}}
\caption[Picture of rotating condensates]{Phase-contrast images of a condensate 
being rotated at 1 Hz in a ``doughnut'' trap.  The condensate (about 
100 $\mu$m in diameter) remains 
pierced by an off-resonant laser beam during its motion.
Although angular momentum was clearly imparted to the cloud, we could 
not detect vortices.
Figure taken from ref.~\cite{andr98thes}.}
\label{rotating_condensate}
\end{figure}

Efforts in our laboratory have laid some of the ground work for this 
search.
While the stability of vortices in gaseous Bose condensates trapped
in a singly-connected harmonic trap is in doubt~\cite{rokh97}, 
vortices should be stable in a multiply-connected trap 
geometry~\cite{muel98,java98pers}.
Thus, we created a ``doughnut''-shaped container for Bose condensates by 
first forming a spherical harmonic magnetic trap by adiabatic 
decompression, and then shining a focused blue-detuned laser through 
the trap which repelled atoms from the trap center (fig.~\ref{fig:varioustraps}).
Then, we displaced the center of the 
magnetic trap using magnetic bias and gradient fields, and rotated it 
about the optical plug.
As shown in fig.~\ref{rotating_condensate}, this set the condensate 
in rotational motion about the optical plug imparting angular momentum 
to a condensate while maintaining the ``hole'' inside it.
We observed rotation after stopping the drive, and hoped that the 
system would relax into a state with a persistent current.
Nevertheless, when we probed for vortices by looking for a depletion in the 
center of the ballistically expanding condensate, 
no evidence for vortices was found
However, we 
cannot rule out their presence in the condensate due to a low 
signal-to-noise ratio in our detection scheme~\cite{andr98thes}.

\subsubsection{Collapse of a negative scattering length Bose condensate}

Condensates with negative scattering length have attractive interactions, and therefore 
show the new dynamic property of collapse when they exceed a certain size.  Formally, the 
collapse can be regarded as an unstable collective excitation having imaginary frequency.  
Manifestations of the collapse are the finite maximum number of atoms observed in lithium 
condensates~\cite{brad97bec} and fluctuations in the number of condensed 
atoms~\cite{hule98}.

\subsubsection{Formation and decay of the condensate}

Experiments have probed both the irreversible~\cite{mies98form} and 
the reversible~\cite{stam98rev} formation of the condensate.
An irreversible formation was observed after suddenly quenching a cold atom 
cloud below the BEC transition temperature in a magnetic trap.  
The intrinsic dynamics of condensate growth was in agreement with a model assuming 
bosonic stimulation (sect.~\ref{formationsection}).
Studies of the reversible formation of a condensate in a hybrid 
optical and magnetic trapping potential are discussed in 
sect.~\ref{revers}.

Refs.~\cite{burt97,stam98odt,sodi98deca} studied the decay of the 
condensate and explained it by three body recombination, which leads to molecule 
formation and loss of the atoms from the trap.
The difference in the rate constant for losses from a Bose condensate 
and from a thermal cloud observed by Burt \etal~\cite{burt97} reveals 
higher-order coherence of the condensate.
For typical experiments on sodium and rubidium, decay from dipolar 
relaxation (two-body collisions) is 
negligible, while it is the dominant decay mechanism for 
hydrogen~\cite{frie98}.
Another important finding was the {\it non-observation} of fast decays 
from a mixture of two hyperfine states of $^{87}$Rb in a magnetic 
trap~\cite{myat97}, which reveals a near-degeneracy of ground-state scattering 
lengths for that system~\cite{burk97,kokk97coll,juli97stab}
A dramatic increase of condensate decay was found in sodium near Feshbach resonances 
\cite{sten98stro} (sect.~\ref{fesh}).

	
\section{Coherence properties and the atom laser}
\label{coherencesection}

One fascinating aspect of Bose-Einstein condensation is the nature of coherence in a 
macroscopic quantum system.  Theoretically, the condensate should be described by a macroscopic 
wave function, behaving like a ``giant matter wave'' which is characterized by long-range 
order.  The first experiments on BEC focused on the energetics of Bose condensation, i.e.\ they 
showed that Bose condensates were spreading out with a very narrow momentum distribution when 
released from the atom trap \cite{ande95,davi95bec}.  In this section we discuss 
experimental studies which have directly probed the coherence properties of a Bose condensate, 
and discuss the concepts of the atom laser and bosonic stimulation.

\subsection{The atom laser and bosonic stimulation}

An atom laser is a device which generates a bright coherent beam of atoms through a 
stimulated process.  It does for atoms what an optical laser does for light.  The atom laser 
emits coherent matter waves whereas the optical laser emits coherent electromagnetic waves.  
Coherence means, for instance, that atom laser beams can interfere with each other.  The 
condition of high brightness requires many particles per mode or quantum state.  A thermal 
atomic beam has a population per mode of only $10^{-12} $, compared to 
a number $\gg 1$ for an atom 
laser.  The realization of an atom laser therefore required methods to enhance the mode 
occupation by a large factor.

Two different gain mechanisms have been discussed for an atom laser: 
evaporative cooling~\cite{wise96,holl96alas,wise97} and optical 
pumping~\cite{wise95atla,wise97,spre96,olsh96}.  In both cases, 
atoms are fed into a selected energy level of a matter wave resonator, and above a certain 
threshold, a macroscopic population is achieved.  Generally, the system is not in thermal 
equilibrium, and ``lasing'' can occur in any mode of the resonator.  Bose-Einstein condensation 
is a special case, where a macroscopic population builds up in the ground state of the system 
through elastic collisions which bring the system into thermal equilibrium.

\begin{figure}
\epsfxsize = 7cm
\centerline{\epsfbox[0 0 731 488]{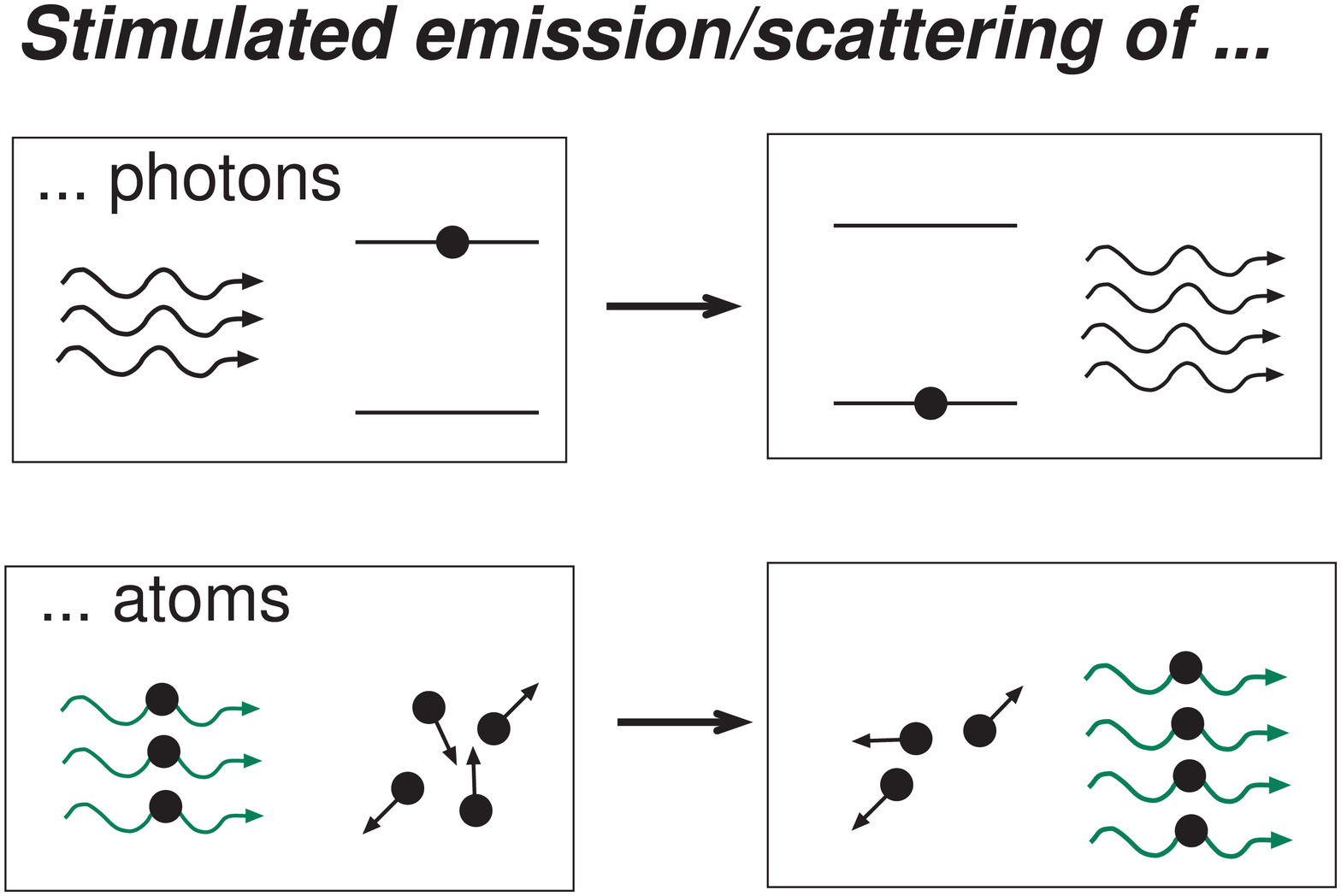}}	
\caption[Bosonic stimulation in optical and atom lasers]{Bosonic stimulation is responsible 
	for the amplification of light in the laser (upper part) and for ``matter wave 
	amplification''.  The lower part shows how a coherent matter wave is amplified by 
	stimulation of elastic collisions enhancing the scattering rate into the state already 
	occupied by the coherent wave.  This process is crucial in the formation of a Bose-Einstein 
	condensate.}
\label{Boseschematic}
\end{figure}

Laser light is created by stimulated emission of photons, a light amplification process.  
Similarly, an atom laser beam is created by stimulated amplification of 
matter waves (fig.~\ref{Boseschematic}).  The conservation of the number of atoms is not in conflict with matter 
wave amplification.  In an atom laser, atoms are taken out of a reservoir and transformed into 
a coherent matter wave.  Similarly, in an optical laser, energy is taken from a reservoir and 
transformed into coherent electromagnetic radiation.  An atom laser is only possible for 
bosonic atoms.  In a normal gas, atoms scatter among a myriad of possible quantum states.  When 
the critical temperature for Bose-Einstein condensation is reached, however, the atoms scatter 
predominantly into the lowest energy state of the system.  This abrupt process is closely 
analogous to the threshold for operating an optical laser.  The presence of a Bose-Einstein 
condensate causes stimulated scattering into the ground state, enhancing the scattering rate 
into the condensate by a factor of $N_{0}+1$ where $N_{0}$ is the number of condensed atoms.  Thermal 
equilibrium is reached when the number of collisions into and out of the condensate are equal 
(detailed balance).  Further evaporative cooling creates a cloud which is not in thermal 
equilibrium and relaxes towards colder temperatures.  This results in growth of the condensate.


There is some ongoing discussion of what defines a laser, even in the case of the optical laser 
\cite{klep97}; e.g. it has been suggested that stimulated emission is not necessary to obtain 
laser radiation \cite{wise97}.  In our discussion, we don't attempt to distinguish between 
defining features and desirable features of a laser.

\subsection{Derivation of Bose-Einstein statistics from bosonic stimulation}

In most textbooks, the Bose-Einstein distribution function is derived using maximization of 
entropy and counting statistics (see, e.g. \cite{grif95qm}) or using a grand-canonical ensemble 
which is based on similar assumptions \cite{huan87}.  However, bosonic stimulation is as 
fundamental as Bose-Einstein statistics: one can derive the Bose-Einstein equilibrium 
distribution just by assuming detailed balance and bosonic stimulation \cite{toda91}.

We assume a gas described by the populations $n_i$ in quantum state $i$ and some weak 
interaction, which scatters particles in states 1 and 2 into states 3 and 
4.  In thermal equilibrium, the two 
rates
\begin{eqnarray}
	W_{1,2 \rightarrow 3,4} &=& |M_{12,34}|^2 n_1 n_2 (n_3+1)(n_4+1) \\
	W_{3,4 \rightarrow 1,2} &=& |M_{34,12}|^2 n_3 n_4 (n_1+1)(n_2+1)
\end{eqnarray}
are equal due to detailed balance.  Furthermore, absolute values of the matrix elements 
$M_{12,34}$ and $M_{34,12}$ are equal.  Therefore, we get:
\begin{equation}
	\frac{n_{1}}{n_{1}+1}\frac{n_{2}}{n_{2}+1} = \frac{n_{3}}{n_{3}+1}\frac{n_{4}}{n_{4}+1}
	\label{detbalance}
\end{equation}

If we assume that the only conserved quantity is energy, we expect the populations $n_i$ only 
to depend on the energy $E_i$, or
\begin{equation}
	\frac{n_{i}}{n_{i}+1} = f(E_i)
	\label{nifuncEi}
\end{equation}
Eq.\  (\ref{detbalance}) yields $ f(E_1) f(E_2) = f(E_3) f(E_4) $ for all pairs (1,2) and (3,4) 
which are coupled by some weak interaction.  Assuming that the system does not have any 
conserved quantities besides energy, all pairs (1,2) are coupled to pairs (3,4) if 
$E_1+E_2=E_3+E_4$.
Therefore, the function $f$ obeys the functional relationship that $f(E_1) f(E_2)$ only depends 
on $E_1+E_2$ which is fulfilled by the exponential function
\begin{equation}
	f(E)= e^{-\beta(E-\mu)}
	\label{expfunct}
\end{equation}
where $\beta$ and $\mu$ are constants.  Eqs.\ (\ref{nifuncEi}) and (\ref{expfunct}) imply the 
Bose-Einstein distribution $n_i=1/(\exp( \beta(E_i-\mu) - 1)$.

The very existence of the condensate is already indirect evidence for bosonic stimulation.  If 
we have a condensate of $N_0 $ particles in equilibrium with a thermal distribution there is 
continuous scattering in and out of the condensate. The rate of scattering out of the 
condensate is proportional to $N_0 $.  Thus the condensate is only stable 
if there is bosonic 
stimulation enhancing the rate at which collisions between thermal atoms scatter into the 
condensate.

\subsection{Formation of the condensate in a magnetic trap}
\label{formationsection}

So far, we have not been able to study the passage of coherent atoms through an active atomic 
medium and observe single-pass gain and amplification.  However, evidence for gain 
(i.e.\ bosonic 
stimulation and growth) was obtained when the formation of the condensate was studied.

The dynamics of condensate formation involves non-equilibrium dynamics of an interacting 
many-body system and is not yet fully understood.  Initial
predictions for the time scale of 
condensation varied between infinite \cite{levi77} and extremely short \cite{stoo91}.  The 
early prediction for infinite time was based on a Boltzmann equation; in this framework the 
condensate fraction cannot grow from zero \cite{levi77,tikh90,stoo95}.  Thus, a separate 
process of nucleation had to be introduced with Boltzmann equations only describing the 
dynamics before \cite{snok89,wu97} and after \cite{ecke84,svis91,semi95} the nucleation.  Stoof 
suggested that a condensate is nucleated in a very short coherent stage 
\cite{stoo91,stoo95,stoo97} and then grows according to a kinetic equation.  Kagan and 
collaborators discussed the formation of a quasi-condensate which, in contrast to a condensate, 
has phase fluctuations, which die out in a time scale which increases with the size of the 
system \cite{kaga92,kaga94,kaga97,kaga95}.  In the thermodynamic limit, it would take infinite 
time to establish off-diagonal long-range order.  Recently, a fully quantum mechanical kinetic 
theory for a Bose gas has been formulated and was used to model the formation process of the 
condensate \cite{gard97, gard98}.

\begin{figure}[hhhh]
\epsfxsize=85mm
 \centerline{\epsfbox[0 0 362 228]{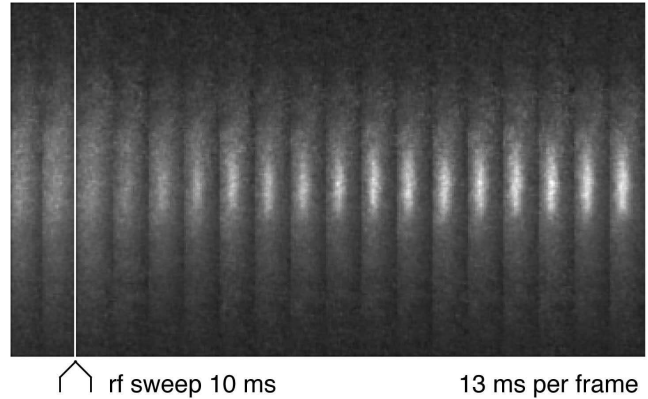}}
 \caption[(Color) The formation of a Bose-Einstein condensate]
{(Color) Observing how a Bose-Einstein condensate 
forms. Shown is a sequence of 18 phase-contrast images taken {\it in situ} of the same 
condensate.  The first two frames show a thermal cloud at a temperature above the transition 
temperature.  The following 16 frames were taken after the cloud was quenched below the BEC 
transition, and show the growth of a condensate at the center of the cloud at 13~ms intervals.  
The length of the images is 630 $\rm \mu m$.
Figure from ref.~\cite{mies98form}.}
\label{formfigure}
\end{figure}

The experimental realization of BEC certainly proved that condensates form within a finite 
time.  Likewise the observation of high contrast interference between two condensates 
demonstrated that condensates develop long-range coherence in a finite time \cite{andr97int}.  
However, a determination of the intrinsic time scales was not possible in this early work 
because the cooling was so slow that the system stayed close to thermal equilibrium.  
In order to observe the {\it intrinsic} relaxation towards condensation we suddenly cooled a 
cloud below the BEC transition temperature by truncating the wings of the spatial 
distribution using rf-induced spinflips.  The intrinsic dynamics of condensate growth 
was observed subsequently in a completely isolated system (fig.~\ref{formfigure}).

\begin{figure}[htbf]
\epsfxsize=85mm
 \centerline{\epsfbox[35 74 507 365]{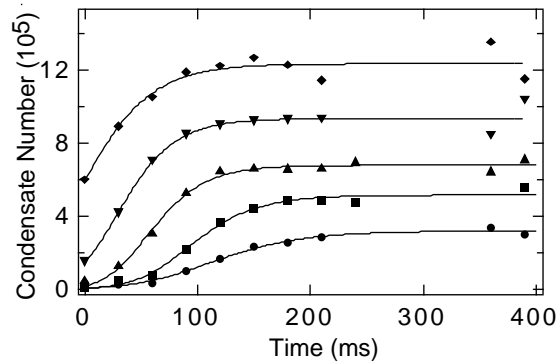}}
\caption[Growth of the condensate fraction towards equilibrium]{Growth of the condensate 
fraction towards equilibrium.  Shown is the number of condensate atoms versus the time after 
the end of a fast rf sweep which created a non-equilibrium situation.  The results were 
obtained by averaging about ten traces with similar equilibrium values for the number of 
condensate atoms.  The solid line are the predictions of a model which includes bosonic 
stimulation.
Figure from ref.~\cite{mies98form}.}
\label{growthfigure}
\end{figure}

Figure \ref{growthfigure} shows the equilibration of the number of condensate atoms 
for various initial conditions.  When no condensate was present right after the rf 
sweep, the growth started slowly and sped up after 50 to 100 ms.  In contrast, if there was 
already a substantial condensate fraction at the beginning of the rf sweep, the rapid growth of 
the condensate commenced immediately.  This ``avalanche'' behavior is characteristic of 
systems with gain and is distinctly different from an exponential relaxation into thermal 
equilibrium.  All our results were in agreement with a model based on bosonic stimulation 
\cite{mies98form}.

The model included a rate determining parameter $\gamma$.  
Our results for $\gamma$ were larger 
than the initial theoretical prediction~\cite{gard97}
by a factor which ranged between 4 and 30.  A more 
refined theoretical treatment gave reasonable agreement \cite{gard98}.  However, the theory predicts an increase of 
the rate $\gamma$ with both temperature $T$ and the number of condensate atoms in equilibrium 
whereas the observed trend is in the opposite direction.  A tentative explanation 
is saturation of the stimulated growth rate of the condensate due to local depletion of the 
thermal cloud.  This depletion might not be a depletion in density, but rather in the pair 
correlation function: when two non-condensed atoms approach each other, 
the presence of the condensate
stimulates them to collide, leading to a depletion of atom pairs with a small separation.  Such 
a process would be analogous to gain saturation in the optical laser.  Further insight into the 
dynamics of condensate formation is obtained from Monte-Carlo simulations which include a 
realistic treatment of the experimental steps \cite{arim98}.

\subsection{Interference between two condensates}

An intriguing property of a Bose condensate is the existence of a macroscopic wavefunction, 
i.e.~the existence of a common phase for the whole cloud.  This coherence 
is relevant for quantum 
fluids because the gradient of the phase is proportional to the 
superfluid velocity.  It also determines the properties of atom lasers based on BEC. In 
superconductors and liquid helium, the existence of coherence and of a macroscopic wavefunction 
is impressively demonstrated through the Josephson effect 
\cite{jose62,feyn64,pere97}.  In the dilute atomic gases, we were 
able to demonstrate the coherence even more directly by interfering two Bose condensates.

In the theory of BEC a breaking of symmetry naturally allows a condensate to be described by a 
coherent state~\cite{huan87}: the boson field is classical with a well-defined macroscopic 
wavefunction and a well-defined (but arbitrary) phase.  This assumption of a phase is 
convenient for the interpretation of interference experiments.  A coherent state is made up of 
a linear combination of number states which implies that there is no definite value for the 
number of atoms in the condensate.  In principle, however, one can measure this number and 
attribute the wavefunction to the appropriate number state.  Recent theoretical studies have 
shown that this apparent conflict is only superficial: in measuring the phase of a condensate 
(e.g. by observing the interference between two condensates), the detection process itself 
causes the condensate to evolve from a number state into a coherent state with a definite 
phase, without violation of number 
conservation~\cite{java96phas,cira96,cast97,nara96,wong96int}.  Thus for the purposes of 
describing an interference experiment one may simply adopt the picture of a macroscopic 
wavefunction with an arbitrary but well-defined phase.

The interference pattern of overlapping independent condensates is analogous to the observation 
of a Josephson current when two superfluids are brought into a weak contact \cite{ande86}, as 
recently observed with superfluid $^{3}$He \cite{pere97}.  Interference of two condensates is 
also analogous to the interference of two independent laser beams \cite{pfle67}, where each 
measurement shows an interference signal, but the phase is random for each experimental 
realization.  This differs from the interference of a single 
beam which is split and then recombined, for which the pattern 
depends only on geometry.

\begin{figure}
    \begin{center}
     \epsfig{file=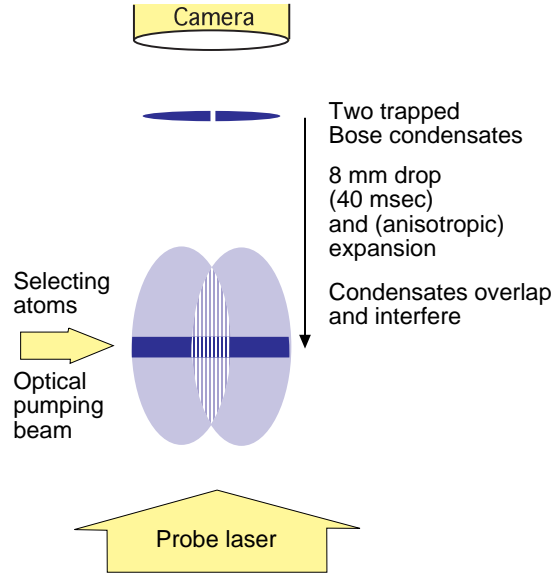,height=3in}
	\caption[Schematic for the observation of interference of two condensates]{ Schematic setup 
	for the observation of the interference of two Bose condensates, created in a double well 
	potential.  The two condensates were separated by a laser beam which exerted a repulsive 
	force on the atoms.  After switching off the trap, the condensates were accelerated by 
	gravity, expanded ballistically, and overlapped.  In the overlap region, a high-contrast 
	interference pattern was observed by using absorption imaging.  An additional laser beam 
	selected a thin layer of atoms by optically pumping them into the initial state for the 
	absorption probe.  This tomographic method prevented blurring of the interference pattern 
	due to integration along the probe laser beam.}
     \label{fig:catscan}
	 \end{center}
\end{figure}

The first step toward observing such interference was the production of two independent 
condensates by evaporatively cooling atoms in a double-well potential created by splitting a 
magnetic trap in half with a far-off resonant laser beam.  
After the trap was switched off, the falling atom clouds expanded 
ballistically and overlapped (fig.~\ref{fig:catscan}).  The 
interference pattern was observed using tomographic absorption imaging \cite{andr97int}.  
As shown in fig.~\ref{nice_interference}, the interference pattern consisted of straight lines with a spacing 
of about 15 $\mu$m, a huge length for matter waves; the matter wavelength of atoms at room 
temperature is only 0.5 \AA, less than the size of an atom.

The straight line pattern is expected for interference between two pulsed matter waves, which 
are initially localized wave packets with the corresponding momentum 
distribution.  When the pattern is 
observed after a time delay $t$, the two interfering components have a relative velocity $v= 
d/t$ where $d$ is the initial separation of the two condensates.  The interference pattern is 
determined by the de Broglie wavelength associated with this relative velocity.  This argument 
is independent of the position of observation, and therefore the fringe spacing is uniform 
throughout the overlap region.

\begin{figure}
\epsfxsize=85mm
\caption[Interference pattern of two expanding condensates]{Interference pattern of two 
	expanding condensates observed after 40 ms time of flight.  The width of the absorption 
	image is 1.1 mm.  The interference fringes have a spacing of 15 $\mu$m and are strong 
	evidence for the long-range coherence of Bose-Einstein condensates.
	Figure from ref.~\cite{andr97int}. (E-print: Separate figure)}
\label{nice_interference}
\end{figure}

\begin{figure}
\epsfxsize=85mm
\centerline{\epsfbox[ 0 0 334 276]{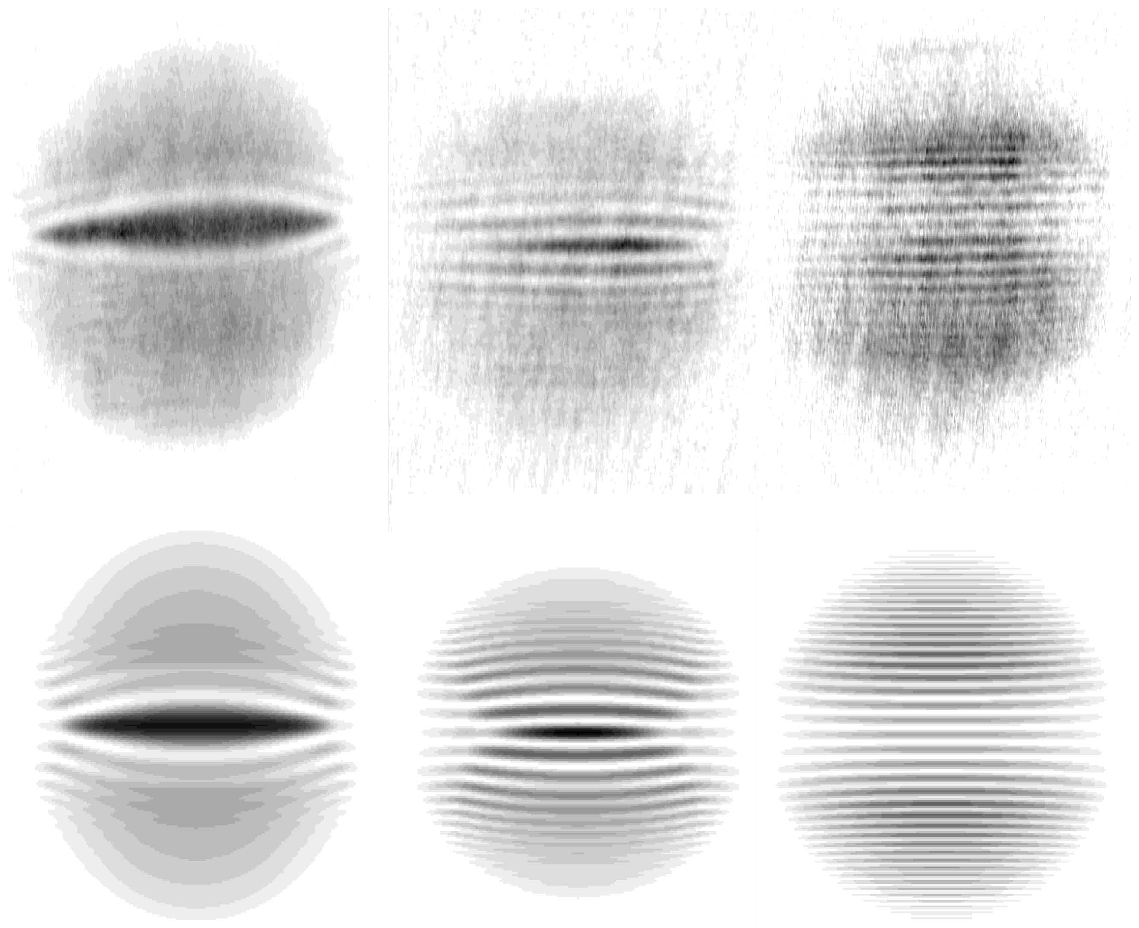}}
	\caption[Interference patters for three different barrier heights of the double-well 
	potential]{Interference pattern for two interfering Bose-Einstein condensates for three 
	values of the height of the barrier of the double-well potential.  At low values (left and 
	middle) the two condensates are not fully separated resulting in a strong central maximum.  
	The bottom row are simulations using the non-linear Schr\"odinger equation \cite{rohr97, 
	wall98}.  The field of view for each image is about 0.4 mm in height and 1.2 mm in width.}
\label{interference_vs_height}
\end{figure}

The contrast of the matter wave interference was estimated from the observed fringe contrast to 
be between 50 and 100\%.  A more precise value could not be given because finer details of the 
probing and imaging of the interference pattern were not quantitatively accounted for.  
Theoretical simulations of the interference showed that the results were consistent with 
numerical solutions of the non-linear Schr\"odinger equation which assumed 100\% coherence 
\cite{rohr97} (see refs.~\cite{nara96, host96, wall97phas} for related work).  When the condensates 
were not fully separated by the dipole force potential, the interference
fringes became curved, and a strong central maximum developed (fig.~\ref{interference_vs_height}).  The Garching group reproduced all these features using 
numerical simulations based on the Gross-Pitaevskii equation \cite{rohr97, wall98}.  These 
simulations of dynamics of Bose condensates were among the first which were sensitive to 
the {\it phase and coherence} of the condensate.  Many other dynamical aspects like collective 
excitations and ballistic expansion have classical analogues.

\subsection{Condensate interferometry}

In the experiment described above, two independent Bose-Einstein condensates were produced by 
evaporative cooling in a double-well potential thus creating two condensates which had never 
interacted with each other (neglecting tunneling and interactions mediated by high-energy 
thermal atoms).  Alternatively, we could first produce the condensate, and then ``cut'' it into 
two pieces.  This would impose a definite initial phase relation between the two condensates.  
Allowing this split condensate to overlap and interfere is akin to interference from a double 
slit illuminated with a single laser beam.  This constitutes a ``separated beam'' atom 
interferometry experiment with condensates where a condensate is first split into two parts and 
afterwards recombined.  In this case, one would expect a fringe pattern with a reproducible 
phase.  However, when we performed this experiment~\cite{mies98rev}, we could not 
read out a reproducible phase because the geometry involving the camera, the condensate and the 
``cutting'' laser beam was not mechanically stable to within 10 $\mu$m.

However, it is worthwhile to discuss how the 
pattern might have evolved in a more stable experiment.  For short delay times, we would expect 
the two condensates to interfere with a predictable and reproducible phase.  After long delay times, the 
condensates would have ``forgotten'' how they were prepared, and we would get the interference 
pattern of two independent condensates, each with high contrast, but with a phase which 
randomly varies from shot to shot.  In between, one would see the continuous transition from a 
well-determined phase to a random phase, i.e.\ the spread of the phase around its mean value 
will increase from (ideally) zero to $2\pi $ (see e.g. \cite{java97spli}).

The Boulder group was able to do such an experiment with an ``internal state'' condensate 
interferometer.  They observed the evolution of an interference pattern from 
predictable to random phase \cite{hall98phas}.  Further experiments which probe coherence 
properties of a condensate were performed at NIST, when multiple pulses of atoms were coupled 
out of the condensate using Raman transitions and interference was observed \cite{helm98}.  The Yale 
group created multiple condensates in a one-dimensional optical lattice \cite{ande98atla} and observed the coherence through temporal oscillation at the Josephson 
frequency.

In the absence of technical noise, the loss of coherence by phase 
diffusion (how the condensate ``forgets'' its phase) is determined by 
fluctuations $\Delta N =N_2 - N_1$ in the difference of the number of atoms $N_{1}$, $N_{2}$ in 
each condensates.  Preparing the condensates with a well-defined phase leads to fluctuations 
in the complementary variable $\Delta N$.  After preparation, the two condensates 
evolve with phase factors $\exp(i \mu_i t/\hbar)$ where $\mu_i$, $i$=1,2 are the chemical 
potentials of the two condensates which depend on $N_i$.  In the Thomas-Fermi approximation, $d 
\mu_{i}/ dN_{i} = (2/5) \mu_{i}/N_{i}$, and the rate of phase diffusion becomes 
\cite{lewe96phas, cast97, wrig96col1, molm98}
\begin{equation}
	\gamma_\phi = \frac{d \mu}{d N} \frac{\Delta N}{\hbar} = \frac{2}{5} 
	\frac{\mu}{\hbar} \frac{\Delta N}{N}
\end{equation}
If a condensate is split symmetrically into two parts, $\Delta N \approx 
\sqrt{N}$~\cite{java97spli}.  Typical 
phase diffusion rates are a few Hz.

\subsection{Higher order coherence}

The interference experiment above measures directly the first-order coherence.  It provided 
evidence for long-range correlations extending over the whole sample, and for the existence (or 
creation) of a relative phase between two condensates.  A full characterization of coherence 
requires knowledge of higher-order coherences.

Let us first discuss how higher-order 
coherences would manifest themselves if an interference experiment has shown 100\% contrast.  
In this case, the condensates are 100\% spatially coherent, i.e.\ each condensate is a ``perfect 
wave.''  Many repetitions of the experiment would show pulse-to-pulse fluctuations both in the 
phase and in the number of condensed atoms.  The number fluctuations contain further 
information on the atom statistics.  This is similar to the comparison of a pulsed thermal 
source which is passed through a single mode-filter and a pulsed single mode laser beam.  In 
both cases, the pulses are 100\% coherent (single mode!), but they differ in the shot-to-shot 
fluctuations.  The thermal statistics are characterized by an exponential distribution for 
which the most probable value is zero, whereas the laser-like distribution is a much narrower 
Poisson distribution \cite{loud83,kuo91}.  The number fluctuations in creating Bose 
condensates or coupling out pulses of a Bose condensate were small (typically 10\%, probably 
determined by the reproducibility of loading the atom trap) and therefore ``laser-like'' rather 
than ``thermal-like.''  

What we have just discussed was the question whether a stream of condensate pulses (each 100\% 
spatially coherent) would show \textit{total} number fluctuations which 
follow a thermal or a laser-like distribution.  We can now drop the assumption that each 
condensate is 100\% spatially coherent, and raise the question whether the distribution would 
be smooth (laser-like) or show fluctuations when we record the \textit{local} density 
vs.  time.  Those fluctuations are expressed by the second-order coherence function $g^{(2)} 
(r)$ at $r=0$ ($r$ is the interparticle separation) through
\begin{equation}
	\langle n^2 \rangle = \langle n \rangle^{2} + \langle (\Delta n)^2 
	\rangle = \langle n \rangle^2 g^{(2)}(0).
\end{equation}

The absence of density fluctuations is expressed by $g^{(2)}(0) =1$, whereas thermal 
fluctuations are characterized by $g^{(2)}(0) =2$.  The latter was recently observed for 
laser-cooled neon atoms \cite{yasu96}.  Elastic and inelastic two-body collisions are 
proportional to $n^2$, and can therefore be used to determine $g^{(2)} (0)$.  To state it more 
simply: binary collisions are proportional to the probability that two atoms are simultaneously 
at the same position, and this normalized probability is given by the second order correlation 
function.

Elastic collisions are responsible for the mean-field energy $U_{int}$ of a cloud, which can be 
expressed as \cite{kett97} 

\begin{equation}
	U_{int} = \frac{2 \pi \hbar^{2} a}{m}  \int d^{3} {\bf r}\,  g^{(2)}(0)|_{\bf 
	r} \, n({\bf r})^2
\end{equation}
Quantitative measurements of the mean-field energy done at MIT \cite{mewe96bec,cast96} and at 
Boulder \cite{holl97int} are all consistent with the prediction of $g^{(2)}(0) =1$ for a pure 
condensate.  Although they have error bars of 20\% to 50\% \cite{kett97}, they are 
inconsistent with $g^{(2)}(0) =2$ and provide evidence for the suppression of local density 
fluctuations in a Bose condensate.
 
The value of two for $g^{(2)}(0)$ for a thermal cloud can be traced back to an exchange term in 
the interaction matrix element.  This exchange term arises whenever the system occupies 
different quantum states which have spatial overlap \cite{walr96}.  $g^{(2)}(0) =1$ implies 
that the system can be described \textit{locally} by a single wavefunction, but it does not 
rule out the population of numerous non-overlapping quantum states.  Therefore, measurements of 
mean field energy and collisions probe only short-range correlations in a Bose condensate and 
cannot distinguish between quasi-condensates, which lack long-range correlations, and ``true'' 
condensates \cite{kaga95}.

The third-order coherence function $g^{(3)} (0)$ is proportional to $n^3$ and could be 
determined by monitoring the trap loss of atoms which, at high density, is mainly due to 
three-body recombination~\cite{burt97}.  Due to the counting statistics for bosons, one has 
$g^{(n)}(0) =1$ for a condensate and $g^{(n)}(0) =n!$ for a thermal cloud.  Therefore, one 
would expect that at the same density, a thermal cloud would decay due to three-body 
recombination six times faster than a condensate \cite{kaga85}.  Burt 
{\it et al.\ }  compared the trap 
loss of a $^{87}$Rb condensate to that of a thermal cloud, and obtained $7.4 \pm 2.6$ for the 
ratio of the $g^{(3)}(0)$ values in good agreement with the predicted value \cite{burt97}, 
clearly demonstrating the third-order coherence of a Bose condensate.

All experimental studies done so far are consistent with the standard assumption that a Bose 
condensate is coherent to first and higher order and can be characterized by a macroscopic 
wavefunction, but further studies are worthwhile and should lead to a more rigorous 
characterization of the coherence properties of condensates.

\subsection{Output couplers for an atom laser}
\label{sec:outputcoupler}

A laser requires a cavity (resonator), an active medium, and an output coupler.  Various 
``cavities'' for atoms have been realized, but the most important ones are magnetic traps 
(sect.~\ref{magnetictraps}) and optical dipole traps (see sect.~\ref{opticalsection}).  The 
major difference between these and an optical Fabry-Perot cavity is that traps or ``matter wave 
cavities'' usually operate in the lowest mode, similar to the maser.

\begin{figure}
\epsfxsize=50mm
\centerline{\epsfbox[0 0 294 598]{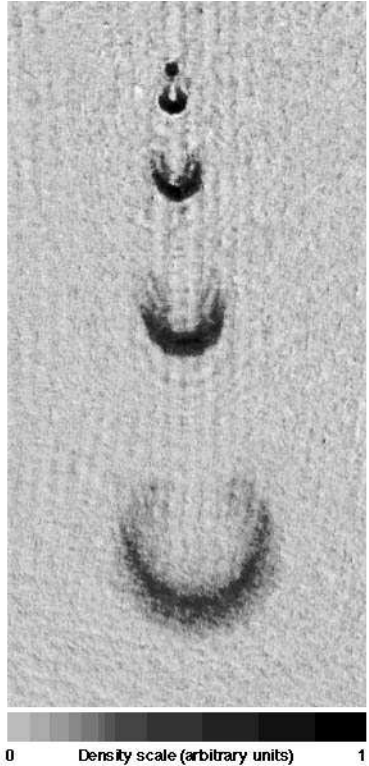}}
\caption[The MIT atom laser operating at 200~Hz]{The MIT atom laser operating at 200~Hz.  
Pulses of coherent sodium atoms are coupled out from a Bose-Einstein condensate confined in a 
magnetic trap (field of view 2.5 $\times$ 5.0~mm).  Every 5~ms, a short rf pulse transferred 
a fraction of these atoms into an unconfined quantum state.  These atoms were 
accelerated downward by gravity and spread out due to repulsive interactions.  The atom pulses 
were observed by absorption imaging.  Each pulse contained between $10^5$ and $10^6$ atoms.}
\label{mitatomlaser}
\end{figure}

The purpose of the output coupler is to coherently extract atoms out of the cavity.  A simple 
way to accomplish this is to switch off the trap and release the atoms.  This is analogous to 
cavity dumping for an optical laser and extracts all the stored atoms into a single pulse.  A 
more controlled way to extract the atoms requires a coupling between confined quantum states 
and a propagating mode.  Such a ``beam splitter'' for atoms was realized using a short rf pulse 
which rotated the spin of the trapped atoms by a variable angle.
The inhomogeneous magnetic 
trapping field then separated the atoms into trapped and out-coupled components 
\cite{mewe97}.  By using a series 
of rf pulses, a sequence of coherent atom pulses was formed (fig.~\ref{mitatomlaser}).  
The crescent shape of the propagating pulses can be qualitatively explained as the result of 
the forces of gravity and of the mean field.  In the absence of gravity, one would expect a 
hollow shell or a loop propagating mainly in the radial direction; the center is depleted by 
the repulsion of the trapped condensate.  However, due to gravity, atoms are not coupled out upward, 
resulting in a crescent instead of a loop.  This was reproduced in a numerical 
simulation using the Gross-Pitaevskii equation \cite{jack98flow} (see also
theoretical discussion in ref.~\cite{zhan98grav}).

A variable output coupler was realized in two different ways~\cite{mewe97}.  Resonant rf 
radiation was applied for a variable pulse duration causing full Rabi oscillations between trapped 
and untrapped states. Alternatively, the rf was swept through resonance.  A slow sweep rotated the spins of the atoms 
in an adiabatic way, whereas a fast sweep coupled only a small fraction 
of the atoms into untrapped states.  The rf output coupler acts on atoms in the same way that a partially silvered mirror acts on 
light; atoms in a coherent state are split into two coherent states.  If applied to atoms in a 
number state, the output is a superposition of number states entangled 
with trapped states~\cite{mewe97thes}.

The realization of a simple outcoupling mechanism for trapped condensates, using rf-pulses 
\cite{mewe97}, and the proof that the outcoupled pulses are coherent \cite{andr97int, 
mies98rev} demonstrated the potential of BEC to generate coherent atomic beams and realized a 
rudimentary pulsed atom laser.  Since then, several groups have devised different output 
coupling schemes.  The Munich group had much better magnetic field control and could operate 
the rf output coupler continuously~\cite{bloc98}.  The NIST group used Raman scattering to transfer two 
photon-recoil momenta to some of the atoms and ``propel'' them out of the condensate, thus 
realizing a directional output coupler.  At a high repetition rate, quasi-continuous 
outcoupling was achieved \cite{hagl98}.  The Yale group observed a self-pulsing atomic beam tunneling 
out of an optical lattice which has some analogies to a mode-locked laser~\cite{ande98atla}.



\section{Optically confined Bose-Einstein condensates}
\label{opticalsection}

In previous sections, we discussed how magnetic traps were essential
for producing Bose-Einstein condensates by providing quiet and deep
trapping potentials and allowing for forced evaporative cooling.
They have also provided well characterized trapping potentials that
have allowed the many recent experiments on the properties of
Bose-Einstein condensates.

However, magnetic traps have
 limitations.
Only weak-field
seeking hyperfine states can be confined, which leaves open the 
heating and loss from dipolar relaxation to strong-field seeking 
states.  
Having access only to weak-field seeking states
imposes limitations on studies of atomic properties such as collisional
resonances
which depend crucially on
the hyperfine state, or for
examinations of multi-component condensates.
Magnetic traps cannot trap atoms
in the $m_F=0$ state, which are
preferable for atomic clocks and other
precision experiments, thus
limiting the use of trapped condensates for
metrology.
Finally, the long range inhomogeneous magnetic fields interfere with 
the use of atom laser pulses which are output from magnetically 
trapped condensates 
(even $m_{F} = 0$ pulses are affected by quadratic Zeeman shifts)~\cite{mewe97}.

These limitations have led us to
the development of an
all-optical trap for Bose-Einstein condensates. In
the following,
we summarize the basic properties of this trap (sect.\ ~\ref{odt}), and then
illustrate various  new possibilities for experiments
on
Bose-Einstein condensation:
\begin{itemize}

        \item Optical traps allow precise spatial (micrometer) and
temporal
(microsecond) manipulation of Bose-Einstein condensates.
This
should allow the realization of box traps, atom guides, and
optical
lattices for condensates.  We have used the spatial
resolution afforded by optical
traps to create a new ``dimple'' trap in
which BEC was created
adiabatically and reversibly (sect.\  \ref{revers}).
    \item
Optical traps have a new external degree of freedom: they can be
operated
at arbitrary external magnetic fields.  We have used this
feature for the
observation of Feshbach resonances in strong-field
seeking states of
sodium which cannot be confined magnetically
(sect.\  \ref{fesh}).

\item Optical traps offer a new internal degree of freedom: the
orientation
of the atom's magnetic moment.  This  resulted in the
generation of spinor
condensates, condensates which populate all
three hyperfine states of the
$F=1$ state of sodium  and
possess a three-component
vectorial order parameter (sect.~\ref{spinor}) .
\end{itemize}

\subsection{Optical confinement of a Bose-Einstein condensate}
\label{odt}
One of the long-standing goals of optical cooling and trapping
techniques has been the creation, storage, and probing
of Bose-Einstein condensates by purely optical means.
This pursuit has provided one motivation for the
development of sub-recoil cooling
techniques~\cite{aspe88,kase92}, various
optical dipole traps~\cite{chu86,phil92,mill93,take95cs,kuga97}, 
and
Raman~\cite{lee96,kuhn96,lee98}
and evaporative~\cite{adam95} cooling in such traps.
However, to date, optical cooling schemes have not reached the
phase-space density necessary for BEC, the closest approach being a
factor of 300 short~\cite{lee96,lee98}.
One major difficulty is that reaching a phase-space density of unity
requires both low temperatures and high densities.
Using the recoil temperature as a benchmark, a gas will Bose condense
only at densities higher than one atom per cubic wavelength,
coinciding with the regime where high optical densities wreak havoc with
near-resonant laser cooling schemes.
In contrast, evaporative cooling --- which even in optical dipole 
traps
relies on atom-atom interactions rather than atom-light
interactions~\cite{adam95} ---
becomes {\it more efficient} at high
densities.

We have taken a different approach toward controlling BEC by 
purely
optical means: we first brought a gas to BEC by
evaporative cooling in a magnetic trap, and then transferred it to an
optical trap~\cite{stam98odt}.
We chose the parameters of the optical trap based on two criteria:
the trap had to be deep enough to confine Bose condensates, and
have sufficiently low heating rates
to permit studies of the condensates for long
periods of time.

The confining potential of an optical trap is due to the AC Stark
shift and given by
\begin{equation}
	U({\bf r})  =  - \frac{\hbar \omega_{R}^{2}({\bf r})}{4} \left( 
	\frac{1}{\omega_{0}-\omega_{L}} + \frac{1}{\omega_{0}+\omega_{L}}\right)
	\simeq  \frac{\hbar \omega_{R}^{2}({\bf r})}{4 \Delta}
	\label{ac_stark_near_resonance}
\end{equation}
where $\omega_{0}$ is the resonant frequency, 
$\omega_{L}$ the laser frequency and $\Delta = \omega_{L} -
\omega_{0}$ its detuning.
The Rabi frequency $\omega_{R}({\bf r})$ is position dependent, and 
conveniently defined through
$2 \omega_{R}^{2}({\bf r})/ \Gamma^{2} = I({\bf r}) / I_{\mbox{{\scriptsize SAT}}}$
where the intensity is scaled by the saturation intensity
$I_{\mbox{{\scriptsize SAT}}} = \hbar \omega_{0}^{3} \Gamma / 12 \pi 
c^{2}$ which is 6 mW/cm$^{2}$ for sodium (assuming an oscillator 
strength of unity).

Besides the conservative potential $U({\bf r})$, the laser light can
also induce heating via Rayleigh scattering and power or position
jitter of the beam.
The rate of Rayleigh scattering is given by
\begin{equation}
	\gamma_{R} = \frac{\alpha^{3}}{3}
	\frac{\omega_{L}^{3} \Gamma^{2}}{4 \omega_{A} 
	\omega_{0}} \left( 
	\frac{1}{\omega_{0}-\omega_{L}} + \frac{1}{\omega_{0}+\omega_{L}}\right)^{2}
	\, \frac{I}{I_{\mbox{\scriptsize SAT}}}
\end{equation}
where $\omega_{A}$ is twice the Rydberg frequency, and $\alpha$ the 
fine-structure constant.

Neglecting counter-rotating terms ($|\Delta| \ll \omega_{0}$),
the potential depth scales as $I / \Delta$ while the
scattering rate scales as $I / \Delta^{2}$.
While near-resonant light can provide large trap depths it can
also cause severe heating.
Thus, as indicated in Table \ref{odt_table},
optical dipole traps have tended toward larger detunings which 
has necessitated the use of high laser powers
to provide sufficient confinement for laser-cooled atoms.

\begin{table}[htbf]
\caption{Optical dipole traps: a light sampler.}
\begin{center}
\begin{tabular}{|l|c|c|c|c|c|} \hline
 & Detuning & Power & Depth  & Number & Density
 \\
 & $\Delta / \Gamma$ &  (Watts) &  (mK) & of atoms & (cm$^{-3}$) \\ \hline
Chu (`86) \cite{chu86} & $10^{4}$ & 0.2 & 5 & 500 & $8 \times 
10^{11}$
\\
Heinzen (`93) \cite{mill93} & $2 \times 10^{5}$ & 0.8 & 6 & 
$10^{4}$ & $2
\times 10^{12}$
\\
Chu {\it et al.} (`95) \cite{adam95} & $2 \times 10^{7}$ & 8 & 0.9 &
$5 \times 10^{5}$ & $4 \times 10^{12}$
\\ \hline
optically trapped BEC \cite{stam98odt} & $7 \times 10^{7}$ &
$\simeq 0.005$ & $\simeq 0.005$ & $10^{7}$ & $10^{14}$ -- $10^{15}$ 
\\ \hline
\end{tabular}
\end{center}\vskip-\lastskip
\label{odt_table}
\end{table}

In contrast, in our approach we use evaporative cooling as a
precursor for optical trapping and reduce the temperature of trapped
atoms by a factor of 100.
Thus, even at large detunings, only milliwatts of laser power are needed as compared with 
several
watts
used to directly trap laser-cooled atoms, making for traps which are 
easier
to handle and trapped atoms which are longer-lived.
Furthermore, since the
cloud shrinks while being cooled in the magnetic trap, the
transfer efficiency into the
small trapping volume of an optical dipole trap is increased.
Thus, magnetic trapping and evaporative cooling
are an ideal way to ``funnel'' atoms into an optical trap,
which might be useful for a host of applications such as
photoassociation spectroscopy and cavity QED.

The optical trap was formed by focusing a single near-infrared laser
beam into the center of the magnetic trap along the
axial direction.
Similar single-beam setups have been used in the
past~\cite{chu86,phil92,mill93,take95cs}.
The light intensity distribution at the optical focus, which gives
the trapping potential via eq.\ (\ref{ac_stark_near_resonance}), is
given by
\begin{equation}
	I(\rho, z) = \frac{2 P}{\pi w_{0}^{2} (1 + (z/z_{R})^{2})}
\exp \left({-\frac{2 \rho^{2}}
	{w_{0}^{2} (1 + (z/z_{R})^{2})}}\right)
\end{equation}
where $P$ is the laser power, $w_{0}$ is the $1/e^{2}$ beam waist
radius,
$z_{R}$ is the Rayleigh range, and
$\rho$ and $z$ are the distance from the focus along the radial
and axial directions, respectively.
The length scales $w_{0}$ and $z_{R}$ are related as $z_{R} = \pi
w_{0}^{2} / \lambda$ where $\lambda$ is the wavelength of the
trapping laser.
The bottom of the trapping potential can be approximated as a harmonic
oscillator with frequencies $\nu_{\rho}$ and $\nu_{z}$.
At the wavelength of $\lambda$ = 985 nm which we chose, the trap
depth $U$, the radial frequency $\nu_{\rho}$ and the aspect ratio
$\nu_{\rho} / \nu_{z}$ are
\begin{eqnarray}
	U/k_{B} & \simeq & 1\, \mu\mbox{K}\, \times \, \frac{P}{\mbox{mW}} \,
	\times \left( \frac{6 \, \mu\mbox{m}}{w_{0}}\right)^{2} \\
	\nu_{\rho}& \simeq & 1\, \mbox{kHz}\, \times \, \left(
\frac{U}{k_{B} \,
	\mu\mbox{K}}\right)^{1/2} \,
	\times \frac{6 \, \mu\mbox{m}}{w_{0}} \\
	\frac{\nu_{\rho}}{\nu_{z}} & \simeq & 28 \,
	\times \frac{6 \, \mu\mbox{m}}{w_{0}}
\end{eqnarray}
The benchmark $w_{0} = 6 \, \mu$m in the above equations is the beam 
radius obtained with an $f / \# = 10$ imaging lens.
Finally,
the small spontaneous
scattering rate from the far-detuned trapping beam ($5 \times 10^{-3}
{\rm s}^{-1}$ per $\mu{\rm K}$
trap
depth) has not limited our experiments, therefore there is little
incentive to reduce this rate even further by using far-infrared~\cite{take95cs}
or blue-detuned traps~\cite{kuga97,lee96}.

Condensates were transferred into the optical trap by holding them in
a steady magnetic trap while
ramping up the infrared laser power, and then suddenly switching
off the magnetic trap.
Nearly complete transfers of the atoms were observed, with
condensate numbers in the optical trap as high as $10^7$.
For various experiments, we have produced optical traps
with beam radii ranging between 6 and 20 $\mu{\rm m}$, and
with infrared powers ranging from 5 to 50 mW.

Condensates in the
optical trap are unusually pure, i.e.\ the condensate
fraction is very high.
This is because the optical trap
has a small volume, forcing the number of
uncondensed atoms to be very
low.  An estimate for this number can be obtained
by assuming that
the temperature $T$ of the cloud is about 1/10th of the trap
depth (the optical trap has ``built-in'' evaporative cooling due to 
the
limited trap depth).  The number of non-condensed atoms is 
approximately
$(k_{B} T/\hbar \bar{\omega})^3$, which for typical conditions is a 
few
times $10^4$ atoms, in quantitative agreement with our measurements. 
With
a condensate of about 5 million atoms, this estimate indicates a
condensate fraction greater than  99\%.  ``Pure'' condensates in 
magnetic
traps usually have a larger non-condensed fraction due to the
smaller trapping frequencies and the technical difficulty of
adjusting a small trap depth using rf-induced evaporation, since the 
trap
depth is sensitive to stray magnetic fields.
In the future, condensate
``purification'' by optical trapping may allow studies of
various predicted zero-temperature
phenomena, such as collapses and revivals of collective
excitations~\cite{grah97}, small shifts of excitation frequencies due to 
quantum
depletion which might otherwise be masked by finite-temperature
effects~\cite{pita98}, or generally for high-coherence atom lasers.

Our optically trapped condensates were long-lived.
The observed losses were dominated by three-body recombination losses, 
with loss
rates per atom ranging from 4
${\rm s}^{-1}$  at a peak
density $n_0=3\times
10^{15}\,{\rm cm}^{-3}$ to less than 1/10 s$^{-1}$
at
$n_0=3\times 10^{14}\,{\rm cm}^{-3}$.

From these lifetime measurements, we can estimate an upper limit for
the heating caused by beam jitter and
spontaneous scattering in the optical trap.
An analysis of the observed number losses indicated a density independent
loss rate of $0.03(2)
\, \mbox{s}^{-1}$. Assuming a trap depth of about 5 $\mu$K, this implies a heating
rate of about 150 nK/s, which is larger than the 25 nK/s heating rate one
would expect from spontaneous scattering alone.
Another estimate of this
heating rate was obtained by shining the infrared light onto a sample 
of
gas in the magnetic trap, and monitoring the
temperature of the gas in this combined potential (fig.\ \ref{acfigure})
\cite{stam98rev}.
Accounting for the fact that atoms spent only about 10\% of the time
in the optical potential, we arrive again at a heating rate of 
about 100 nK/s.

The optical dipole force exerted by a tightly focused laser beam 
easily exceeds the magnetic forces of typical magnetic traps.
Therefore, optically trapped condensates can be easily compressed to
extremely high densities, as high as $3 \times 10^{15} {\rm cm}^{-3}$ 
in some of our work.
At such densities, the three-body recombination rate is
extremely high, but transient experiments on such
dense samples are still possible.
At the low-density end, we have produced condensates with
several million atoms at densities of about $1 \times 10^{13} \, {\rm
cm}^{{-3}}$ in a decompressed magnetic trap with a mean frequency
$\bar{\omega} = 2 \pi \times 7 \, {\rm Hz}$.
This wide range of possible condensate densities is impressive, 
especially
considering that a few years ago there were many discussions whether
the ``BEC window'' existed at all (sect.\ \ref{sec:atomsforbec}).

Besides the new scientific possibilities discussed in the
following paragraphs, the optical trap may become an important tool
to manipulate and transport condensates.
It may serve as an ``optical tweezers'' to move condensates
 into optical or microwave cavities or close
to surfaces.

\subsection{Reversible formation of a Bose-Einstein condensate}
\label{revers}

In an ordinary cryostat, the
experimenter can raise and lower the
temperature of the sample reversibly.
In contrast,
evaporative cooling is irreversible due to the loss of the
evaporated
atoms.  Even if the temperature is raised again by 
heating of the sample, the number of atoms has already been
reduced
during the cooling stage.  This reflects the fact that the
trapped
atoms are both the sample and the ``working fluid'' of the
refrigerator.

If the strength of the trapping potential is changed
adiabatically,
one can reach higher or lower temperatures reversibly --- but
the
density changes simultaneously in such a way that the phase
space
density is invariant.  Thus adiabatic changes in the strength of the
trapping
potential cannot be used to cross the BEC transition
line with an ideal gas (the case of an interacting gas is discussed in 
ref.\ ~\cite{houb97}).

However, imagine that one could contain a small portion of
a
trapped atomic sample in a vessel which is kept in thermal contact
with
the remaining atoms, and then use this vessel to compress the small
sample.
This
would raise the density of a small portion of the trapped gas,
but
because of the thermal contact with a larger reservoir of atoms,
the
temperature rise would be slight.  Furthermore, if such a
compression
were performed slowly (quasi-statically), the total entropy of the
gas
would be unchanged, and one could reversibly (by compressing
and
expanding the small vessel) change the maximum phase-space density.

This adiabatic
(isentropic) increase in phase-space density can
only occur in the
presence of collisions.  Without collisional
thermalization, an
adiabatic process would preserve the number of
atoms in each quantum state
(i.e. the Ehrenfest notion of
adiabaticity) and therefore would not change
the phase-space
density.  Thus, the maximal increase in phase-space
density will occur if changes in the trapping potential
are made
slowly with respect to the collisional (and motional)
equilibration
times of the trapped gas.

Adiabatic
changes of the ground state population
were first demonstrated by Pinkse and
collaborators \cite{pink97}.  For
the power-law potentials considered in
their paper, the maximum
increase of phase-space density is limited to a
factor of 20.
However, the ``small box in a large box'' scheme discussed above can
increase the phase-space density by an arbitrary
factor.
We demonstrated this using a
combination of magnetic
and optical forces~\cite{stam98rev}.
Atoms in the magnetic trap
acted as the
reservoir while a small fraction of the cloud was
compressed to higher
densities by adding a focused red-detuned laser
beam which created a narrow
potential well (a ``dimple'') in the center
of the magnetic trap.

The increase of the phase-space density was measured by
cooling clouds to various
temperatures above the phase transition
temperature
and determining the infrared
laser power required to
produce a small Bose condensate \cite{stam98rev}.
A factor of 50 increase in phase-space density was obtained.
Larger increases were
hindered by limitations in laser power
and by limits to the ramp-up time
set by the various
heating and loss processes in the deformed
trap.

Further deformation of the potential takes the
sample across the BEC
phase transition.  The observed condensate fractions
were
considerably smaller than those predicted for an ideal gas.
This
constitutes the first clear evidence for the effect of interactions
on the thermodynamics of Bose-Einstein condensed gases 
(sect.~\ref{sec:condfrac}).
Indeed, the ``dimple'' trap is ideally suited for observing such
effects.
The importance of
interactions for thermodynamic quantities depends on a dimensionless
parameter
$\eta$~\cite{gior96,gior97} which is the ratio of
the mean
field energy at zero temperature $\mu_{0}$ to the transition
temperature $T_c$.  For
a harmonic trapping potential with mean
trapping frequency
$\bar{\omega}$, $\mu_{0} \propto \bar{\omega}^{6/5}$ and $T_c \propto \bar{\omega}$,
leading to $\eta
\propto \bar{\omega}^{1/5}$.
This weak dependence makes harmonic traps ill suited for observing
thermodynamic interaction effects.
In contrast, the ``dimple''
trap is characterized by two trapping
frequencies which can be controlled independently:
one for the broad
magnetic trapping potential which
determines $T_{c}$, and one for the
bottom of the optical trapping
potential which confines the condensate and
determines its mean field
energy.  Thus, the parameter $\eta$ can be
greatly increased.

\begin{figure}[htbf]
\epsfxsize=60mm
\centerline{\epsfbox[0 0 375 373]{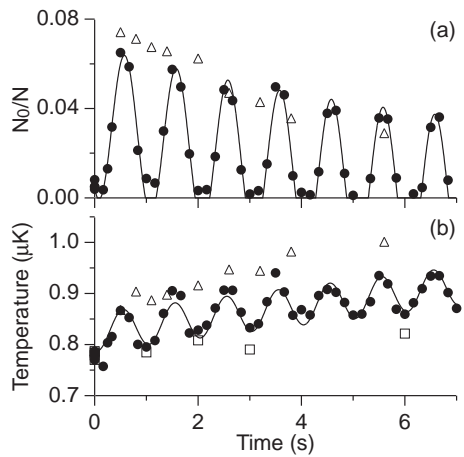}}
\caption[Adiabatic cycling through the phase transition]{Adiabatic cycling through the phase transition. Shown is the
condensate fraction (a) and the temperature (b) vs.\ time for the case
of a modulated infrared beam (closed circles), an infrared 
beam ramped up to a constant power
(open triangles), and no infrared light (open squares).
Rf shielding, discussed in sect.~\ref{rffields}, was used.
The solid lines are guides to the eye.
Figure taken from ref.~\cite{stam98rev}.}
\label{acfigure}
\end{figure}

Finally, since the condensate formation is adiabatic, it is
reversible.
This was demonstrated by preparing a magnetically trapped
cloud just above $T_{c}$, and then
sinusoidally modulating the power of the
infrared light.
Upon each oscillation, the trapped sample passed back and forth across 
the the BEC phase transition (fig.~\ref{acfigure}).
The peak of these oscillations
decreased slowly in time due to the slight heating of the optical trap.

This method of
creating condensates provides new opportunities for studying
condensate formation.
In the experiment described above, the
condensate fraction was found to lag about 70 ms behind the modulation
of the laser power, which is a measure for the formation time.  This
agrees with experimental \cite{mies98form} and theoretical \cite{gard97}
studies of the formation using a different method.  However, using the 
``dimple'' trap has the flexibility of a wide range of optical and 
magnetic
trap parameters
which
should help elucidate the dependence of the formation time on
temperature and density.

Since the optical trap can be switched on in microseconds,
it should be easier to
separate the formation time scale from other time scales
in the trapped gas.  Indeed, in one experiment, we switched on the
infrared light instantly,
and observed condensation on time scales
much
faster than the oscillation periods in the magnetic trap and
along the
weakly confining axis of the optical trap \cite{stam98rev,sten98odt}. 
Such
studies of
shock-condensation might give new insight into the
formation of
quasi-condensates and condensation into excited
states
\cite{kaga92,gard98quas}.

Also, deformation of the trapping
potential greatly extended the
range of accessible transition temperatures.  The
compression of the thermal cloud by the optical
potential resulted in
extremely high critical densities $n_c \approx 5 \times
10^{14} \,
{\rm cm}^{-3}$ and critical temperatures $T_c \approx 5
\,
\mu{\rm K}$.  At low densities, crossing the BEC transition in a
purely
magnetic trap is hindered by ineffectiveness of rf-evaporation
(caused by
the sagging of the cloud due to gravity~\cite{kett96evap}) and the
technical difficulties of
controlling very small trap depth with rf-
and magnetic fields.  However,
using our method of deforming the
potential, we could shine the infrared
laser into a decompressed
magnetic trap ($\bar{\omega} = 2\pi \times 7$ Hz) and cross
the transition at
$n_c \approx 2 \times 10^{12} \, {\rm cm}^{-3}$ and $T_c
\approx
100$ nK.

Finally, let us point out that this reversible method for increasing
the phase-space density has similarities to the irreversible method
of evaporative cooling.
Reversible cooling by adiabatic deformation of the
trapping potential
is achieved by a modification of the density of states.
When a
potential dimple is added to a broad harmonic oscillator
potential,
it lowers the energy of the ground state, but affects very few
of the
excited states of the system.  As the well-depth is increased,
the
ground state energy reaches the chemical potential, and BEC occurs.
An
alternative way of describing this process is to first compress
the whole
cloud by increasing the harmonic oscillator frequency to
the value at the
bottom of the dimple, and then to open up the
potential to the original
broad harmonic oscillator potential outside
a small central area.  The last
step increases the density of states
at high energy, and as a result,
elastic collisions preferably
populate these states.  This is very similar
to evaporative cooling
in which  high energy
states are coupled to the
continuum leading to a net collisional transfer of atoms to
these
states.  The major difference is that in evaporative cooling no
equilibrium is
established between the evaporated atoms and the
trapped
atoms, whereas in the case of the deformed potential, the
whole cloud stays
in thermal equilibrium.

\subsection{Observation of Feshbach resonances in a Bose-Einstein
condensate}
\label{fesh}
All the essential properties of Bose condensed systems --- the formation
and shape of the condensate, the nature of its collective excitations 
and
statistical fluctuations, the
dynamics of solitons and vortices --- are determined by the
strength of the atomic interactions. 
At low temperatures, these interactions are controlled solely by the 
s-wave scattering length(s) for elastic collisions between atoms (see 
chapters
by J.~Dalibard and D.~Heinzen).
But unlike many chemical properties which govern the behavior of 
condensed-matter systems, the scattering lengths are not immutable,
but can vary dramatically near a
zero-energy collisional resonance, either a shape
resonance~\cite{mari98} or, of current experimental relevance, a 
Feshbach resonance~\cite{fesh62}.

A Feshbach resonance occurs when a quasi-bound molecular state has
an energy equal to that of two colliding atoms.
Such a resonance strongly affects both elastic and inelastic 
collisions such
as dipolar relaxation~\cite{ties92bec,ties93} and three-body 
recombination.
Feshbach resonances have been studied in the past by varying the 
collisional energy to correspond to the fixed energy of a quasi-bound 
state~\cite{brya77}.
For ultracold atoms (in a Bose condensate), the near-zero collisional 
energy is fixed, thus one 
must bring the quasi-bound state energy down to zero.
Proposals have been made to do so with external
magnetic~\cite{ties92bec,ties93,moer95res,voge97,boes96},
optical~\cite{fedi96,bohn97}, rf \cite{moer96dres},
and electric fields~\cite{mari98}.

Recently, Feshbach resonances for ultracold atoms have been found in 
the presence of external magnetic fields.
Our group discovered several Feshbach resonances in sodium and
directly confirmed their dramatic effect on the interaction energy of a 
condensate~\cite{inou98},
demonstrating how minute 
changes in a magnetic field can bring about dramatic changes in a 
macroscopic system.
Observations in $^{85}$Rb were made with non-condensed atoms  via photoassociation 
spectroscopy, which probes the relative wavefunction between 
colliding atoms~\cite{cour98fesh,abee98}, and via measurements of the 
elastic collision 
cross section~\cite{robe98}.

Theoretical
calculations predicted Feshbach resonances for sodium at high fields and 
in hyperfine states that cannot be  magnetically
trapped \cite{moer95res,abee97priv}.
Thus, the optical trap was indispensable for studying Feshbach 
resonances with sodium condensates.

To locate the Feshbach resonances, we first placed the optically 
trapped condensate into the desired hyperfine state, and then ramped 
up the magnetic field to as high as 1200 G while repeatedly probing 
the atoms with phase-contrast imaging.
At specific values of the magnetic field, a rapid loss of atoms was 
observed.
This rapid loss was expected at a Feshbach resonance, 
due to either the collapse of a condensate 
as the scattering length turns negative, or to increased inelastic 
collision rates.
Furthermore, the losses were observed only for atoms in a specific 
hyperfine state, again indicating a collisional resonance.
This was used to locate three resonances near 853 G,
907 G, and 1195 G~\cite{inou98,sten98stro}.

\begin{figure}[htbf]
\epsfxsize=60mm
\centerline{\epsfbox[44 60 309 408]{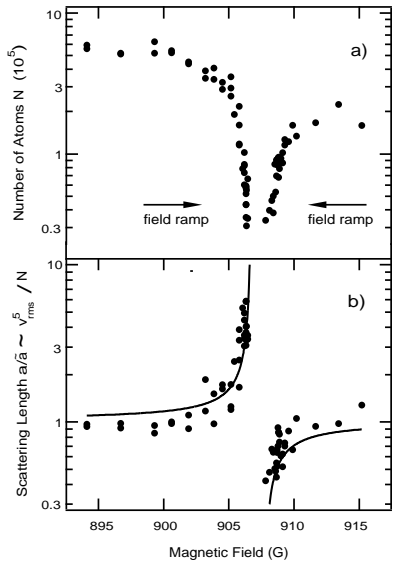}}
\caption[Observation of a $^{23}$Na Feshbach resonance at 907 G]{Observation of a $^{23}$Na Feshbach resonance at 907 G.
(a) Rapid number losses in sweeps across the resonance from the low- and 
high-field ends located the resonance.
(b) Measuring the scattering length based on time-of-flight images of 
expanding condensates confirmed its dispersive variation.
The solid line is a fit based on eq.\ (\ref{eq:a}).
Figure taken from ref.~\cite{inou98}.}
\label{feshfigure}
\end{figure}

To definitively identify the Feshbach resonances, we looked for the 
true ``smoking gun'' phenomenon: a drastic change in the interaction 
energy of a Bose condensate.
The scattering length was determined by measuring the condensate 
number and the width of the expanding condensate in time-of-flight 
images, making use of the Thomas-Fermi description of a condensate as 
detailed in sect.\  \ref{derived_quantities}.
These measurements are shown in fig.~\ref{feshfigure} as a 
function of the magnetic field for the resonance at 907 G.
Near a Feshbach resonance at a field $B_0$, the scattering length $a$ 
was predicted to vary dispersively as a function of magnetic field 
$B$ \cite{moer95res}:
\begin{equation}
a =\tilde a
\left( 1+\frac{\Delta B}{B_0-B}\right)
\label{eq:a}
\end{equation}
where $\tilde{a}$ is the scattering length away from the resonance, 
and
$\Delta B$ is the width of the resonance which is determined by 
the strength of coupling between the quasi-bound state and the 
free-particle state of the incoming atoms.
Our measurements verified this dispersive shape, and demonstrated a 
factor of 10 change in the scattering length by 
this collisional resonance.

Our observation confirms the theoretical predictions about ``tunability'' of 
the
scattering length with the prospect of ``designing'' atomic quantum 
gases
with novel properties.
For example, by setting
$a\approx 0$, one can create a condensate
with essentially non-interacting
atoms.
By setting $a<0$ one can make a once stable ($a>0$) condensate suddenly 
unstable, and observe the collapse of a macroscopic wavefunction in a 
controlled manner~\cite{ueda98,kaga98scat,sack98},
akin to the less controlled collapse of a $^7$Li 
condensate~\cite{brad97bec}.
Rapid variations of $a$ may also lead to novel forms of collective 
oscillations~\cite{kaga98scat}.
Sweeps across the resonance may pass condensed atoms adiabatically 
into molecular condensates~\cite{tomm98}.
Further, Feshbach resonances are also predicted to exist for 
collisions between atoms of unlike atomic species or hyperfine state.
Such resonances can be used to tune interspecies interactions, making 
multi-component Bose condensates overlap or phase-separate (sect.\  ~\ref{immiscibility}).
Finally, Feshbach
resonances may also be important in atom optics for modifying the
atomic
interactions in an atom laser, or more generally for controlling
non-linear atom-optical
coefficients.

However, our experiments suggest there may be severe limitations to 
the use of Feshbach resonances due to the concomitant increase in the 
rates of inelastic collisions and trap loss.
The rapid losses by which the Feshbach resonances were observed were 
exclusively the result of these increased rates, since the region of 
negative scattering length which would cause a collapse of the 
condensate due to {\it elastic} collisions was never accessed.
We ascribe these losses to three-body collisions, since dipolar 
relaxation, the mechanism for two-body decay, is prohibited for 
atoms in the lowest energy hyperfine state where two of these 
resonances occur.
The rate constant for three-body decay was found to increase on
both sides of the resonance, rising by more than three orders of 
magnitude~\cite{sten98stro}.
This behavior is not accounted for theoretically.
Whether the observed enhancement of the
three-body recombination rate is
generic for any Feshbach resonance or only
specific to sodium remains to be
seen.
To give enough time to study the dynamics of a sodium condensate at a 
Feshbach resonance, one would need lower density condensates using 
a large-volume optical trap.
In this regard, the $^{85}$Rb Feshbach resonance, which 
occurs for magnetically trappable atoms, may be better suited for 
such studies.

\subsection{Spinor Bose-Einstein condensates}
\label{spinor}
In a magnetic trap, the atomic spin adiabatically follows the 
direction of the magnetic field (sect.~\ref{magnetictraps}).
Thus, although alkali atoms have internal spin, their Bose-Einstein 
condensates are described by a scalar order
parameter similar to the spinless superfluid
$^4$He.
In contrast, an optical trap
confines atoms in all spin states.
This liberates the atomic spin as a new degree of freedom.
Thus,
optically trapped Bose-Einstein condensates are 
represented by a vector order parameter instead of a 
scalar, and are thus called spinor Bose 
condensates.
A variety of new phenomena are predicted for such condensates such as 
spin textures, spin waves, and coupling between atomic spin and 
superfluid flow~\cite{ho98,ohmi98,law98}.

A word about the name: by the term ``spinor'' we mean a $2 n + 1$ 
component wavefunction which describes the spin state of an $F = n$ spin 
particle.  
In the literature, some use the term ``spinor'' to denote the spin 
wavefunction of a spin-1/2 particle, and then construct 
higher-dimensional objects as a direct product of $2 n$ ``spinors''.
In that parlance, an $F=1$ particle can be
described by a subset spinor-tensors of rank 
2 with three independent components \cite{land77qm,petr69}.
Others distinguish between {\it spinors} which describe half-integer spin 
and {\it vectors} which describe integer spin, which behave differently 
under rotations of $2 \pi$~\cite{tink64}.

Two component condensates have been created 
recently by magnetically trapping two different hyperfine states of 
$^{87}$Rb and have
been used for a remarkable set of experiments~\cite{corn98jltp}.
Compared to these two-component condensates, spinor condensates have
several new features including the multi-component vector character of the
order parameter and the changed role of spin relaxation collisions which
allow for collisional population exchange among hyperfine states
without trap loss.
In contrast, in the $^{87}$Rb experiments trap loss due to spin
relaxation limits the lifetime.
In addition, as we discuss below, the properties of the spinor 
condensate can be conveniently manipulated with weak magnetic fields, 
allowing for studies of two-component mixtures (like the $^{87}$Rb 
experiments) which can be either miscible or immiscible (unlike the 
$^{87}$Rb experiments).

Having introduced this new quantum fluid, let us describe how we made 
it, and how we probed it.
We made spinor condensates starting with an optically trapped 
condensate in the $|F = 1, m_{F} = -1\rangle$ hyperfine state.
Then, rf-resonance techniques (Landau-Zener sweeps similar to those 
used for the atom laser output coupler~\cite{mewe97}) were used to 
distribute the condensate population among the $F=1$ magnetic 
sublevels.
To achieve an arbitrary hyperfine distribution, it was necessary to 
make these rf-transitions at large (15 -- 30 G) bias fields, 
separating the $|m_{F} = +1\rangle \rightarrow |m_{F} = 0\rangle$ and 
$|m_{F} = 0\rangle \rightarrow |m_{F} = -1\rangle$ transition 
frequencies due to the quadratic Zeeman shift.

\begin{figure}[htbf]
\epsfxsize=60mm
\centerline{\epsfbox[0 0 392 306]{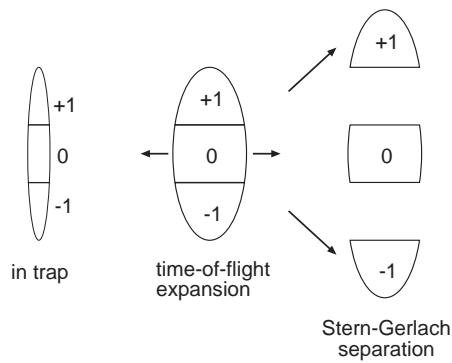}}
\caption[Probing spinor condensates]{Probing spinor condensates.
After release from the elongated optical trap, the trapped spinor 
condensate expands primarily radially while maintaining the axial 
hyperfine distribution.  A magnetic field gradient is then used to 
separate out the different components while preserving their shape.
A subsequent absorption probe reveals the spatial and hyperfine 
distributions in the trap.}
\label{spinorprobe}
\end{figure}

The spinor condensates were probed by time-of-flight imaging combined with a 
Stern-Gerlach spin separation (fig.\ ~\ref{spinorprobe}).
The optical trap was suddenly switched off, allowing the atoms to 
expand primarily radially from the highly anisotropic optical trap.
Then, after allowing about 5 ms for the interaction energy to be 
completely converted to kinetic energy, a magnetic field gradient was 
applied which separated the spin state populations without distorting 
them.
Finally, after 15 -- 30 ms, the atoms were optically pumped to 
the $|F = 2\rangle$ hyperfine manifold to give the same  
cross-section for all the atoms in the subsequent absorption probing 
on the $|F=2, m_{F} = 2\rangle$ 
to $|F'=3, m_{F'} = 3\rangle$ cycling transition.
This probing method determined both the
spatial and hyperfine distributions along the axis of the optical trap.

We have recently studied the equilibrium state of spinor
condensates in an optical trap~\cite{sten98spin}.
Spinor condensates with 
an overall spin projection 
$\langle F_{z} \rangle = 0$ along the magnetic field axis were prepared
in one of two ways: either the entire 
trapped condensate was placed in $|m_{F} = 0\rangle$,
or the condensate was placed in a 50-50 mixture of the 
$|m_{F}= +1\rangle$ and $|m_{F} = -1\rangle$ states.
The condensates were then allowed to equilibrate freely, changing 
their hyperfine distribution via spin relaxation wherein two
$m_F = 0$ atoms collide and produce a $m_F = +1$ and a $m_F = -1$
atom and visa versa.
From both starting conditions, the condensates equilibrated to the 
same ground-state distribution in which domains of all three 
hyperfine states were formed.

\begin{figure}[htbf]
\epsfxsize=60mm
\centerline{\epsfbox[0 0 529 379]{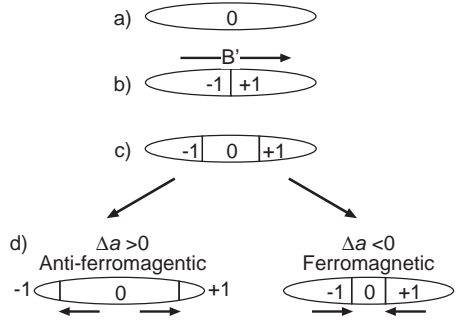}}
\caption[Constructing the ground-state $\langle F_{z}  \rangle = 0$
spinor in a trap]{Constructing the ground-state $\langle F_{z}\rangle = 0$ spinor in a trap.  
In each step (a -- d), another contribution to the Hamiltonian is 
considered (see text).}
\label{spinorground}
\end{figure}

To understand the nature of the ground state, let us consider the 
evolution of a pure $m_{F} = 0$ condensate as we slowly add the 
effects of magnetic fields and mean-field interactions (fig.\ 
~\ref{spinorground}).

\begin{itemize}
	\item[a)] {\it Linear Zeeman shift from homogeneous field.}
	In the presence of a slight magnetic field (in the $\mu$G range to 
	overcome nanokelvin scale interaction energies), the lowest energy state 
	is a pure $m_{F} = +1$ condensate.
	However, spin non-conserving collisions (dipolar relaxation)
	are negligible over the 
	lifetime of the condensate.
	Therefore, the total spin is regarded as a conserved quantity, 
	and the $m_{F}=0$ condensate is unchanged.
	
	\item[b)] {\it Linear Zeeman shift from field gradient.}
	A field gradient $B'$ makes it energetically favorable for two $m_{F} = 0$ 
	atoms to collide and, via spin relaxation, produce a $m_{F} = +1$ 
	atom on the high-field end of the cloud, and a $m_{F} = -1$ atom on 
	the low-field end.  The energy gain per atom is $\hat{p} B' z$ where $z$ 
	is the distance from the center of the cloud and $\hat{p} = 35\, 
	\mu\mbox{K}/\mbox{G}$.  Thus, the condensate is magnetically polarized 
	into two pure spin domains.
	
	\item[c)] {\it Quadratic Zeeman shift from homogeneous field.}
	The quadratic Zeeman shift in a field $B_{0}$
	causes the energy of a $m_{F} = 0$ atom to 
	be lower than the average energy of a $m_{F} = +1$ and $m_{F} = -1$ 
	atom by an amount $\hat{q} B_{0}^{2}$.
	For sodium, $\hat{q} = 20 \, \mbox{nK}/\mbox{G}^{2}$.
	This causes the formation of a $m_{F} = 0$ domain at the center of 
	the cloud with boundaries at $\hat{q} B_{0}^{2} = \hat{p} B' |z|$.
	
	\item[d)] {\it Spin-dependent mean-field interaction.}
	Rotational symmetry at zero-field simplifies greatly the treatment 
	of collisions among all the $F = 1$ manifold sublevels.
	The various scattering lengths depend solely on two scattering 
	lengths $a_{F_{tot} = 2}$ and $a_{F_{tot} = 0}$ where $F_{tot}$ is 
	the total spin of the colliding atoms.
	As a result, the spin-dependent mean-field interaction energy is 
	given simply by $c \langle {\bf F} \rangle^{2}$ with $c = n \cdot 4 
	\pi \hbar^{2} \Delta a / m$ where $\Delta a = (a_{F_{tot} = 2} - 
	a_{F_{tot}=0})/3$ and $n$ is the condensate density~\cite{ho98}.
	If $\Delta a > 0$, interactions favor domains of $|\langle {\bf F} 
	\rangle | = 0$ and thus the $m_{F} = 0$ domain is made larger.  This 
	is called anti-ferromagnetic coupling, and the global ground states 
	under such coupling are ``polar'' states.
	If $\Delta a < 0$, $|\langle {\bf F} \rangle| = 1$ domains are favored and 
	thus the $m_{F} = 0$ domain shrinks.  This is called ferromagnetic 
	coupling.
\end{itemize}

\begin{figure}[htbf]
\epsfxsize=80mm
\centerline{\epsfbox[0 0 586 468]{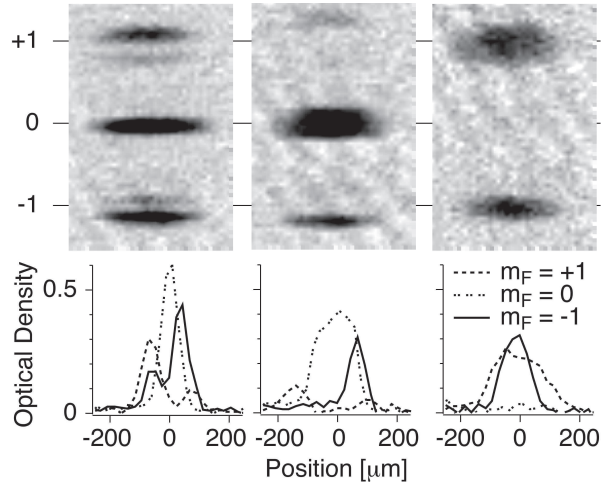}}
\caption[Ground state spin domains in $F=1$ spinor Bose-Einstein 
condensates]{Ground state spin domains in $F=1$ spinor Bose-Einstein 
condensates.  Time-of-flight images after Stern-Gerlach separation 
are shown, along with the indicated axial density profiles in the 
optical trap, for which the Stern-Gerlach separation was ``undone.''
Images at left and center show spin domains of all three components.
Image at right shows a miscible $m_{F} = \pm 1$ component condensate.
Conditions are: $B = 20 \, \mbox{mG}$, $B' = 11 \, \mbox{mG/cm}$ 
(left); $B = 100 \, \mbox{mG}$, $B' = 11 \, \mbox{mG/cm}$ (center);
$B = 20 \, \mbox{mG}$, $B' \simeq 0$, $\langle F_{z} \rangle > 0$ (right).}
\label{spinorcurves}
\end{figure}

These effects were summarized in a 
universal spin-domain diagram that describes bulk features 
(Thomas-Fermi approach) of both ferromagnetic and anti-ferromagnetic 
spinor condensates at arbitrary magnetic fields, field gradients, and 
average spin $\langle F_{z} \rangle$~\cite{sten98spin}.
This spin-diagram explained the ground-state spin structures which 
were
observed, as shown in fig.\ ~\ref{spinorcurves}.
By examining the nature of the ground state at various fields and 
field gradients, sodium spinor condensates were identified to be 
anti-ferromagnetic.
The scattering length difference $\Delta a$ was determined by 
measuring the size of the $m_{F} = 0$ domains, and
was found to agree fairly well with its predicted value \cite{burk98}.

At typical densities in our experiment ($\approx 3 \times 10^{14} \, 
\mbox{cm}^{-3}$), the anti-ferromagnetic 
mean-field energy which dictated the nature of the ground state was 
found to be just 2.5 nK.
This is remarkable: a 2.5 nK interaction energy is 
responsible for spin-domains and phase-separation in an optically 
trapped {\it gas} which is at a temperature of about 100 nK!
The thermal energy $k_{B} T$ is only available to the dilute thermal 
cloud, whereas the condensate finds the ground state with sub-nK 
precision.
Thus, BEC makes it possible to study nanokelvin-scale effects without 
producing such low temperatures.

\subsection{Miscibility and phase-separation of spinor condensate 
components}
\label{immiscibility}

These experimental results and more recent 
studies~\cite{mies98meta} showed clear evidence for the
miscibility of the $m_F=-1$ and $m_F =+1$
components and the immiscibility of $m_F =\pm 1$
and $m_F =0$ components.  We conclude this chapter with a brief review 
and discussion of miscible superfluids.

The study of multicomponent superfluid systems has been a tantalizing 
goal of low-temperature physics for decades.
The earliest discussion focussed on $^{4}$He -- $^{6}$He mixtures.
$^{6}$He is radioactive with a
half-life of 1 second.
An ambitious experiment by Guttman and Arnold~\cite{gutt53} in 1953 
sought evidence for the superfluid flow of $^{6}$He mixed with 
$^{4}$He to no avail.
Nevertheless, this pursuit touched off a series of theoretical works 
on two-component superfluid hydrodynamics~\cite[and others since]{khal57}.
In 1978, Colson and Fetter~\cite{cols78} considered such mixtures 
in the context of mean-field theories which apply directly to current 
experiments, and discussed the criterion for interactions between 
the superfluids to cause miscibility or phase-separation.
After progress in the stabilization of a spin-polarized atomic 
hydrogen gas, Siggia and Ruckenstein~\cite{sigg80} considered the use 
of different hyperfine states 
to achieve a mixture of superfluids.
Since the observation of gaseous Bose condensates, the interest in 
multicomponent condensates has been revived with a flurry of 
theoretical attention~\cite[for 
example]{ho96bin,esry97hf,gold97,law97}.
The first multicomponent condensates were 
created at JILA~\cite{myat97} by magnetically trapping 
different hyperfine states of $^{87}$Rb.
Finally, a miscible two-component condensate mixture was created with 
spinor condensates of sodium~\cite{sten98spin}, accomplishing a long 
standing goal.

The bulk miscibility or immiscibility of two-component condensate mixtures
is predicted by mean-field theory~\cite{cols78,ho96bin,esry97hf,ao98,gold97}.
The total mean-field interaction energy of such condensates is given 
by $2 \pi \hbar^{2}/ m \times  (n_{a}^{2} a_{a} + n_{b}^{2} a_{b} + 2 n_{a} n_{b}
a_{ab})$ where $m$ is the common atomic mass, $n_{a}$ and $n_{b}$ are the 
densities of each of the components, $a_{a}$ and $a_{b}$ are the 
same-species scattering lengths, and $a_{ab}$ is the scattering length 
for interspecies collisions.
Consider a two-component mixture in a box of volume $V$ 
with $N$ atoms in each component.
If the condensates overlap, their mean-field energy is
\begin{equation}
	E_{O} = \frac{2 \pi \hbar^{2}}{m} \, \frac{N^{2}}{V}
	\left(a_{a} + a_{b} + 2 a_{ab}\right)
\end{equation}
If they phase-separate, their energy is
\begin{equation}
	E_{S} = \frac{2 \pi \hbar^{2}}{m} \left( \frac{N^{2}}{V_{a}} 
	a_{a} + \frac{N^{2}}{V_{b}} a_{b}\right)
\end{equation}
The volumes $V_{a}$ and $V_{b}$ occupied by each of the separated 
condensates are determined by the condition of equal pressure:
\begin{equation}
	a_{a} \left( \frac{N}{V_{a}}\right)^{2} = a_{b} \left( 
	\frac{N}{V_{b}}\right)^{2}
\end{equation}
Comparing the energies $E_{O}$ and $E_{S}$ the condensates will 
phase-separate if $a_{ab} > \sqrt{a_{a} a_{b}}$, and will mix if 
$a_{ab} < \sqrt{a_{a} a_{b}}$.

In the $F = 1$ spinor system, as described above, the 
 scattering lengths are determined by $a_{F_{tot}=0}$ and
$a_{F_{tot}=2}$~\cite{ho98}.
Defining $\bar{a} = (2 a_{F_{tot}=2} + a_{F_{tot}=0}) / 3$
and $\Delta a = (a_{F_{tot}=2} - a_{F_{tot}=0}) / 3$, 
the scattering lengths for the
$m_{F} = 1 , 0$ two-component system (or equivalently the 
$m_{F} = -1, 0$ system) are given by
$a_{0} = \bar{a}$, and $a_{1} = a_{01} = \bar{a} + \Delta a$.
Since $\Delta a$ was measured to be positive~\cite{sten98spin}, the condition
$a_{01} > \sqrt{a_{0}a_{1}}$ applies and the components should phase-separate, 
as we have observed~\cite{sten98spin,mies98meta}.
Interestingly, this phase-separation should not occur in the 
non-condensed cloud because the same-species mean-field interaction 
energies are doubled due to exchange terms.

In the $m_{F} = +1, -1 $ two component system, the scattering 
lengths are $a_{1} = a_{-1} = \bar{a} + \Delta a$ and $a_{1,-1} = 
\bar{a} - \Delta a$.  Thus, $a_{1,-1} < \sqrt{a_{1} a_{-1}}$, and 
these two components should mix.
Indeed, as shown in fig.\ ~\ref{spinorcurves}c, an equilibrium spinor
condensate with $\langle F_{z} \rangle \neq 0$, small field gradient, 
and near-zero field consists of an overlapping mixture of atoms in 
the $m_{F} = \pm 1$ states.
This particular miscible two-component system has an important 
advantage.
If the trapping potential varies across a two-component condensate, 
the lowest energy state may be a phase-separated state if $a_{a} \neq 
a_{b}$ even though the condition $a_{ab} < \sqrt{a_{a} a_{b}}$ is 
fulfilled~\cite{pu98two}.  In this case, the atoms with the smaller 
scattering length concentrate near the trap center, making it 
harder to observe miscibility.
However, in the $m_{F} = \pm 1$ system, the two scattering lengths 
$a_{1}$ and $a_{-1}$ are equal by rotational symmetry near zero field, 
so the components mix completely even in a trapping potential.
%

\section{Conclusion}

The basic phenomena of Bose-Einstein condensation in gases was predicted more than 70 years 
ago.  The experimental realization required, first, the identification of an atomic system 
which would stay gaseous all the way to the BEC transition 
and not preempt BEC by forming molecules or 
clusters, and second, the development of cooling and trapping techniques 
to reach this regime.
After this had been accomplished, several studies of BEC were able to confirm 
theories which had been formulated decades ago, but had never been experimentally tested.  
However, the work has already progressed far
beyond the confirmation of old theories, as was 
impressively demonstrated during the Varenna summer school.  BEC in atom traps has several new 
qualitative features which are not included in the treatment of the weakly interacting dilute 
Bose gas as it was developed mainly in the `50s:

\begin{itemize}
\item \textit{The inhomogeneous trapping potential.}  It leads to a partial spatial separation between 
condensate and thermal cloud --- this changes dynamical properties in a profound 
way, and
enables experimentalists to directly ``see'' the condensate.

\item \textit{Mesoscopic physics.} The finite size of the sample introduced a new length scale.  
Several groups have calculated finite-size
corrections to the thermodynamic limit.  Different 
statistical ensembles have been discussed which agree in the 
thermodynamic limit, but not for small Bose condensates.
Condensates with negative scattering length are only stable due to the 
finite size and the zero-point energy associated with it.

\item \textit{Phase of the condensate.}  Its relation to spontaneous symmetry breaking and the quantum 
measurement process could be treated within a microscopic picture.  Bose condensates might become a model system for dissipation, coherence and quantum 
fluctuations in small systems.

\item \textit{Tunable condensates.}  Feshbach resonances can be used to modify interactions between 
atoms and ``design'' quantum matter with novel properties.

\item \textit{Spinor and multi-component condensates.}  These are a new playground with 
prospects to study hydrodynamics of interpenetrating superfluids, complex phase diagrams and 
rich dynamics.

\item \textit{Nanokelvin temperatures and the atom laser.}  New techniques to reach record-low 
temperatures and to generate coherent atomic beams are of general applicability far beyond the 
study of BEC.
\end{itemize}

These aspects give rise to optimism that BEC in dilute atomic gases will 
continue to provide rich new 
physics.  It may follow the tradition of the quantum fluids $^{4}$He and 
$^{3}$He and 
establish a new interdisciplinary field between atomic and condensed-matter 
physics,
allowing many-body 
physics to be studied with the methods and precision of atomic physics!  Furthermore, 
control and manipulation of atoms is central to atomic physics, and coherent atomic beams are 
likely to play a major role in atom optics, atom interferometry, precision measurements and 
beyond.  The last few years have been full of surprises.
The whole summer school and this thick 
book are based on developments of just a few years and demonstrate that there is much more 
excitement to come!

\appendix

\section{Image Processing\label{sec:imageprocessing}}

The imaging techniques discussed in sect.~\ref{probesection} 
are used in an imperfect environment in 
which background light, scattered probe light, and imperfect optics 
obscure the desired signal.  The goal is to determine the integrated 
column density $\tilde{n}(x,y)$ across the 
image.
In this appendix, we discuss the image processing methods by which 
this is done.
These methods differ for absorption and 
phase-contrast imaging, as we discuss below.

\subsection{Absorption image processing}
 
\begin{figure}[htbf]
\epsfxsize=100mm
\centerline{\epsfbox[0 0 570 89]{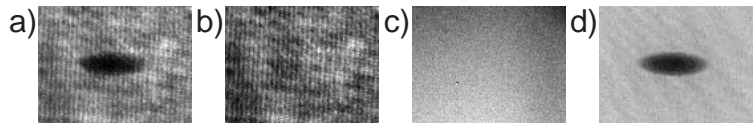}}
\caption[Obtaining a reliable transmission image]{Obtaining a reliable 
transmission image.  Three images are 
collected: a) an image of the probe light after passing through the 
atoms, b) a bright-field image, and c) a dark-field image.
These are processed according to eq.\ (\ref{absequation}) to give a 
transmission image (d).}
\label{absorption}
\end{figure}

An absorption image measures a photon fluence field $F_{I}(x,y) $
which has three components given by
\begin{equation}
     F_{I}(x,y) = F_{I0} \left[ P(x,y) e^{-\tilde{D}(x,y)} +  S(x,y) 
     \right] + N(x,y)
\end{equation}
The first component describes light which passes through the cloud, 
is collected by the imaging system, and imaged onto the camera.  The 
 fluence of the probe beam is given by $F_{I0}$, and $P(x,y)$ 
describes the normalized beam profile in the object plane.  This beam profile is 
typically far from Gaussian due to flawed optical elements and 
multiple reflections in the beam path before the object plane.   The 
quantity $\tilde{D}(x,y)$ is the optical density of the cloud, from which 
the column density $\tilde{n}(x,y)$ can be derived.
The second component, $F_{I0} S(x,y)$ describes probe 
light scattered after the object plane.  The last component, 
$N(x,y)$, describes background light from all sources other than the 
probe beam.
Useful data cannot typically be extracted from the absorption image 
alone (see for example fig.\ \ref{absorption}a).  Thus, we take two additional 
images for each absorption image: a bright-field image $F_{B}(x,y)$ where 
the probe beam is imaged with no absorbing atoms, and a dark-field 
image $F_{D}(x,y)$ with no atoms and no probe light.  The intensity 
measured in these images are 
given by
\begin{eqnarray}
    F_{B}(x,y) & = & F_{B0} \left[ P(x,y) + S(x,y) \right] + N(x,y) \\
    F_{D}(x,y) & = & N(x,y)
\end{eqnarray}

We obtain a transmission image $\tilde{T}(x,y) = t^{2}$ by the normalization
\begin{eqnarray}
\tilde{T}(x,y) &=& \frac{F_{I}(x,y) - F_{D}(x,y)}
   {F_{B}(x,y) - F_{D}(x,y)}\\
&=& \frac{F_{I0}}{F_{B0}} 
  \frac{P(x,y) e^{-\tilde{D}(x,y)} + S(x,y)}{P(x,y) + S(x,y)} 
\label{absequation}
\end{eqnarray}

We can compensate for shot-to-shot probe intensity fluctuations
($F_{I0} \neq F_{B0}$) by examining a portion of each image in which 
there are no atoms.  With this, and assuming that the $ S(x,y) $ 
``noise'' term is small, we get

\begin{equation}
\tilde{T} \simeq  e^{-\tilde{D}(x,y)} + \left[ 1 - 
e^{-\tilde{D}(x,y)} \right] \frac{S(x,y)}{P(x,y)}
\end{equation}

For small $S(x,y)$ and small optical densities,  this 
normalization directly measures the absorption of the probe light by 
the atoms.  An example of this normalization is given in fig.\ 
\ref{absorption}.  
Notice that even with substantial inhomogeneities in the probe beam
a high-quality absorption image is obtained.
This normalization scheme does have its limitations.
For example, $\tilde{D}(x,y) =  - \ln[ \tilde{T}(x,y) ]$, and thus
the error in the optical density is
given by $d \tilde{D} = - d \tilde{T} 
/ \tilde{T}$.  Therefore, measurements of high optical densities (larger than 
about two) are highly sensitive to technical noise.

A second limitation is the difficulty 
of canceling out high-spatial frequency noise in the presence of 
vibrations of the imaging system.  Consider a component in $  P(x,y)$ of 
wavevector ${\bf k}$ (i.e.\ 
$P(x,y) = P_{0}[1 + \epsilon \cos({\bf k} 
\cdot {\bf r})]$ where ${\bf r}$ is the vector $(x,y)$) and  
suppose the imaging system is displaced by a small distance ${\bf d}$ between the absorption 
and bright field image.  Then, neglecting $N(x,y)$ and $S(x,y)$, after 
normalization we obtain
\begin{eqnarray}
\tilde{T}(x,y) & =&  \frac{ \left[ 1 + \epsilon \cos({\bf k} \cdot 
{\bf r}) \right] e^{-\tilde{D}(x,y)} }
 { 1 + \epsilon \cos \left( {\bf k} \cdot ({\bf r}-{\bf d}) \right)  } \\
& \simeq & e^{-\tilde{D}(x,y)} \left[ 1 + \epsilon [ {\bf k} \cdot 
{\bf d} ] \sin({\bf 
k}\cdot {\bf r}) \right]   
\end{eqnarray}
This noise can be minimized by reducing the amplitude $\epsilon$ of the 
noise, decreasing the spatial frequency of the noise, and minimizing the 
vibrations of the imaging system.  It is often possible to ensure that the 
predominant technical noise is at spatial frequencies that are 
orthogonal to the typical displacements or in directions 
that are not relevant to interpreting the images.

\subsection{Phase-contrast image processing\label{sec:pcprocessing}}

The spatial filtering in phase-contrast imaging adds one complication:
high spatial frequency components of the probe beam can miss the 
phase-shifting spot in the Fourier plane (fig.~\ref{fig:pcsetup}).
Therefore, different normalization methods are required for absorption 
and phase contrast images.  In phase contrast imaging 
there are two sources of 
imaging error: (1) inhomogeneities in the probe light, and (2) 
probe light which is scattered onto the camera in a random pattern.  
To simplify the discussion, let us consider the presence of only one 
of these error sources at a time.

For the first case, we consider a beam incident on the atoms with an 
inhomogeneity described by $ \epsilon(x,y) $: 
$E_1(x,y) = E_{0} [ 1 + \epsilon(x,y) ]$.  
After 
passing through the atomic sample and acquiring a phase shift 
$\phi(x,y)$, the electric field becomes $E_2(x,y) = E_{0} [ 1 + (e^{-i 
\phi} - 1) + \epsilon(x,y) e^{-i \phi} ]$.  The uniform part of this light 
comes to a tight focus at the Fourier plane of the imaging system and is 
phase-shifted, resulting in a field at the camera 
$E_3(x,y) = E_{0} [-i + (e^{-i \phi} - 1) + \epsilon e^{-i \phi}  ]$. 
The light fluence collected in the image
is then
\begin{equation}
     F_{I}(x,y)  = {F_{I0}} \left[ 3 - 2 \sqrt{2} \cos(\phi + 
     \frac{\pi}{4})  + 
|\epsilon|^2 + 2 \mbox{Re}\left[ \epsilon e^{-i \phi} (e^{i \phi} 
-1 +i) \right]  \right]+ 
N(x,y)
\end{equation}
To try to 
eliminate the errors in this signal, we collect two more images as 
before, a bright field image $F_{B}(x,y)$ and a dark-field image $F_{D}(x,y)$ 
given by
\begin{eqnarray}
F_{B}(x,y) & = & F_{B0} \left[ 1 + |\epsilon|^2 + 2 \mbox{Re} [i \epsilon 
] \right] + 
N(x,y)   \\ 
F_{D}(x,y) & = & N(x,y)
\end{eqnarray}

We have used two methods to recover the phase-contrast signal.  The 
first method is the same we use for analyzing absorption images.  
This gives a phase-contrast signal
\begin{equation}
PC^{(1)}(x,y) = 1 + 2 \left[ \frac{1 - \sqrt{2} \cos(\phi + \frac{\pi}{4})  +  
\mbox{Re}[ \epsilon (1 - e^{-i \phi})(1 - i)]}
{1 + |\epsilon|^2 - 2 \, \mbox{Im}[\epsilon]} \right]
\end{equation}
In this method the noise $\epsilon$ 
in the probe beam is mixed into the phase-contrast signal.

The second normalization method is to subtract out the dark-field 
images, and separately normalize $F_{I}(x,y)$ and $F_{B}(x,y)$ by
dividing each by the 
average fluence in a portion of the image where the atom cloud is not 
visible (usually a far corner of the image).  We then subtract the 
normalized $F_{B}(x,y)$ from $F_{I}(x,y)$ obtaining
\begin{equation}
PC^{(2)}(x,y) = 2 \left[  1 - \sqrt{2} \cos(\phi + \frac{\pi}{4}) +
 \mbox{Re} [\epsilon (1 - e^{-i \phi})(1 - i) ] \right]    
\end{equation}
Now the mixing of the signal with the noise in the denominator has been eliminated.

Let us now consider the 
noise $\epsilon(x,y)$ to be a speckle pattern which originates from 
scattering off surfaces after the object plane.
Thus, the noise $\epsilon$ is not affected by the atoms, i.e.\ the 
electric field at the camera is 
$E_{3}(x,y) = E_{0} [-i + \epsilon + (e^{-i \phi} - 
1)]$.  The data image collected at the camera is described by
\begin{equation}
F_{I}(x,y) = F_{I0} \left[ 3 - 2 \sqrt{2} \cos(\phi + \frac{\pi}{4}) + 
|\epsilon|^2 + 2 \, \mbox{Re} \left[ \epsilon (e^{i \phi} - 1 + 
i) \right] \right] + N(x,y)
\end{equation}
In this case, the two normalization approaches yield
\begin{eqnarray}
& PC^{(1)}(x,y)  = 1 + 2 \left[ \frac{    1 - \sqrt{2} \cos(\phi + \frac{\pi}{4}) +  
\mbox{Re}[\epsilon (e^{i \phi} - 1)] }{1 + 
\left| \epsilon \right|^2 - 2 \, \mbox{Im}[\epsilon]}  \right] \\
& PC^{(2)}(x,y)  = 2 [ 1 - \sqrt{2} \cos(\phi + \frac{\pi}{4}) + \mbox{Re}[ 
\epsilon 
(e^{i \phi} - 1) ]
\end{eqnarray}
In this case, if the phase of $\epsilon$ varies spatially, the second 
method of normalization is clearly superior since the errors are 
random.

\begin{figure}[htbf]
\epsfxsize=70mm
\centerline{\epsfbox[0 0 792 417]{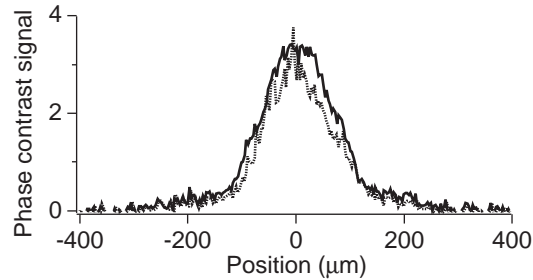}}
\caption[Comparison of two normalization methods for phase-contrast images]{
     Comparison of two normalization methods for phase-contrast images.
Dotted line 
     shows result of normalization $PC^{(1)}$, and the solid line shows 
     $PC^{(2)}$, which gives a more reliable signal.}
\label{pccompare}
\end{figure}

In the end, we turned to our data to answer the question of which 
normalization procedure to use.  Figure \ref{pccompare} compares data from 
images 
of a partly condensed cloud which was normalized using both 
normalization procedures.  The second normalization method gives a 
better fit to the expected parabolic density profile of a condensate, 
and thus is preferable.

\stars Work on BEC at MIT has been a tremendous team effort, and we are 
grateful to the past and present collaborators who have shared both the excitement and the hard 
work: M.R. Andrews, A.P. Chikkatur, K.B. Davis, J. Gore, S. Inouye, M.A. Joffe,
M. K{\"o}hl, C. 
Kuklewicz, A. Martin, M.-O. Mewes, H.-J. Miesner, R. Onofrio, D.E. Pritchard, C. Raman, J. 
Stenger, C.G. Townsend, N.J. van Druten, and J. Vogels.  Special thanks to Dan Kleppner and Tom 
Greytak for inspiration and constant encouragement.  We also acknowledge the fruitful 
interactions with our colleagues who have contributed to this rich and 
exciting field, and last but not least, we are grateful to the organizers and participants of the 
Varenna summer school for creating collective excitement and a lot of (bosonic) stimulation.

We want to thank the Office of Naval Research, the National Science Foundation and the Joint 
Services Electronics Program (ARO), NASA and the David and Lucile Packard Foundation for their 
encouragement and financial support of this work.

\bibliographystyle{v3}

\bibliography{wkrefs}

\begin{thebibliography}{100}

\bibitem{ande95}
 \BY{M.H. Anderson \etal }  \IN{Science}{269}{1995}{198}.

\bibitem{davi95bec}
 \BY{K.B. Davis \etal }  \IN{Phys. Rev. Lett.}{75}{1995}{3969}.

\bibitem{brad97bec}
 \BY{C.C. Bradley, C.A. Sackett, \atque R.G. Hulet}  \IN{Phys. Rev.
  Lett.}{78}{1997}{985}.

\bibitem{frie98}
 \BY{D.G. Fried \etal }  \IN{Phys. Rev. Lett.}{81}{1998}{3811}.

\bibitem{gsuhomepage2}
BEC home page of the Georgia Southern University,
  http://amo.phy.gasou.edu/bec.html .

\bibitem{huan64}
 \BY{K. Huang}  in {\em Studies in Statistical Mechanics}, edited by  \NAME{J.
  de~Boer \atque G.E. Uhlenbeck} (North-Holland, Amsterdam) 1964, Vol.~II, p.\
  3.

\bibitem{grif95}
 \BY{A. Griffin, D.W. Snoke, \atque S. Stringari} {\em Bose-Einstein
  Condensation} (Cambridge University Press, Cambridge) 1995.

\bibitem{kett96phsc}
 \BY{W. Ketterle \etal }  \IN{Phys. Scr.}{T66}{1996}{31}.

\bibitem{vand96}
 \BY{N.J. van Druten \etal }  \IN{Czech. J. Phys.}{46}{1996}{3077}.

\bibitem{town96tops}
 \BY{C.G. Townsend \etal }  in {\em Ultracold Atoms and
  Bose-Einstein-Condensation, 1996}, {\em OSA Trends in Optics and Photonics
  Series, Vol. 7}, edited by  \NAME{K. Burnett} (Optical Society of America,
  Washington D.C.) 1996, p.\ 2.

\bibitem{town97icap}
 \BY{C.G. Townsend \etal }  in {\em Atomic Physics 15}, {\em Fifteenth
  International Conference on Atomic Physics, Amsterdam, August 1996} (World
  Scientific, Singapore) 1997, p.\ 192.

\bibitem{andr98}
 \BY{M.R. Andrews \etal }  \IN{J. Low Temp. Phys.}{110}{1998}{153}.

\bibitem{sten98odt}
 \BY{J. Stenger \etal }  \IN{J. Low Temp. Phys.}{113}{1998}{167}.

\bibitem{mies98rev}
 \BY{H.-J. Miesner \atque W. Ketterle}  \IN{Solid State Comm.}{107}{1998}{629}.

\bibitem{kett96evap}
 \BY{W. Ketterle \atque N.J. van Druten}  in {\em Advances in Atomic,
  Molecular, and Optical Physics}, edited by  \NAME{B. Bederson \atque H.
  Walther} (Academic Press, San Diego) 1996, Vol.~37, p.\ 181.

\bibitem{kett96leos}
 \BY{W. Ketterle \atque M.-O. Mewes} LEOS Newsletter, IEEE, August 1996, pp.
  18-21 (1996).

\bibitem{town97phwo}
 \BY{C.G. Townsend, W. Ketterle, \atque S. Stringari}  \IN{Physics
  World}{March}{1997}{29}.

\bibitem{durf98}
 \BY{D.S. Durfee \atque W. Ketterle}  \IN{Optics Express}{2}{1998}{299}.

\bibitem{kett99atla}
 \BY{W. Ketterle}  in {\em McGraw-Hill 1999 Yearbook of Science and
  Technology}, {\em companion volume to Encyclopedia of Science and Technology}
  (McGraw-Hill, New York) 1999.

\bibitem{huan87}
 \BY{K. Huang} {\em Statistical Mechanics} (Wiley, New York) 1987.

\bibitem{stam98rev}
 \BY{D.M. Stamper-Kurn \etal }  \IN{Phys. Rev. Lett.}{81}{1998}{2194}.

\bibitem{ehre31}
 \BY{P. Ehrenfest \atque J.R. Oppenheimer}  \IN{Phys. Rev.}{37}{1931}{333}.

\bibitem{free80}
 \BY{J.H. Freed}  \IN{Journal of Chemical Physics}{72}{1980}{1414}.

\bibitem{nozi95}
 \BY{P. Nozi{\`e}res}  in {\em Bose-Einstein Condensation}, edited by  \NAME{A.
  Griffin, D.W. Snoke, \atque S. Stringari} (Cambridge University Press,
  Cambridge) 1995, p.\ 15.

\bibitem{myat97}
 \BY{C.J. Myatt \etal }  \IN{Phys. Rev. Lett.}{78}{1997}{586}.

\bibitem{zeld72}
 \BY{Y.B. Zel'dovich \atque R.A. Sunyaev}  \IN{Sov. Phys. JETP}{81}{1972}{153}.

\bibitem{penr56}
 \BY{O. Penrose \atque L. Onsager}  \IN{Phys. Rev.}{104}{1956}{576}.

\bibitem{andr97prop}
 \BY{M.R. Andrews \etal }  \IN{Phys. Rev. Lett.}{79}{1997}{553}.

\bibitem{mies98form}
 \BY{H.-J. Miesner \etal }  \IN{Science}{279}{1998}{1005}.

\bibitem{stam98coll}
 \BY{D.M. Stamper-Kurn \etal }  \IN{Phys. Rev. Lett.}{81}{1998}{500}.

\bibitem{pais82}
 \BY{A. Pais} {\em Subtle is the Lord, The Science and the Life of Albert
  Einstein} (Clarendon Press, Oxford) 1982.

\bibitem{gavr95}
 \BY{K. Gavroglu} {\em Fritz London: A Scientific Biography} (Cambridge
  University Press, Cambridge) 1995.

\bibitem{eins25qua2}
 \BY{A. Einstein}  \IN{Sitzungsber. Preuss. Akad. Wiss.}{Bericht 1}{1925}{3}.

\bibitem{croo83}
 \BY{B.C. Crooker \etal }  \IN{Phys. Rev. Lett.}{51}{1983}{666}.

\bibitem{repp84}
 \BY{J.D. Reppy}  \IN{Physica B}{126}{1984}{335}.

\bibitem{raso84}
 \BY{M. Rasolt, M.H. Stephen, M.E. Fisher, \atque P.B. Weichman}  \IN{Phys.
  Rev. Lett.}{53}{1984}{798}.

\bibitem{wolf95}
 \BY{J.P. Wolfe, J.L. Lin, \atque D.W. Snoke}  in {\em Bose-Einstein
  Condensation}, edited by  \NAME{A. Griffin, D.W. Snoke, \atque S. Stringari}
  (Cambridge University Press, Cambridge) 1995, p.\ 281.

\bibitem{fort95}
 \BY{E. Fortin, E. Benson, \atque A. Mysyrowicz}  in {\em Bose-Einstein
  Condensation}, edited by  \NAME{A. Griffin, D.W. Snoke, \atque S. Stringari}
  (Cambridge University Press, Cambridge) 1995, p.\ 519.

\bibitem{shen97}
 \BY{M.Y. Shen, T. Yokouchi, S. Koyama, \atque T. Goto}  \IN{Phys. Rev.
  B}{56}{1997}{13066}.

\bibitem{lin93}
 \BY{J.L. Lin \atque J.P. Wolfe}  \IN{Phys. Rev. Lett.}{71}{1993}{1222}.

\bibitem{hech59}
 \BY{C.E. Hecht}  \IN{Physica}{25}{1959}{1159}.

\bibitem{stwa76}
 \BY{W.C. Stwalley \atque L.H. Nosanow}  \IN{Phys. Rev. Lett.}{36}{1976}{910}.

\bibitem{silv80}
 \BY{I.F. Silvera \atque J.T.M. Walraven}  \IN{Phys. Rev.
  Lett.}{44}{1980}{164}.

\bibitem{clin80}
 \BY{R.W. Cline, D.A. Smith, T.J. Greytak, \atque D. Kleppner}  \IN{Phys. Rev.
  Lett.}{45}{1980}{2117}.

\bibitem{grey95bec}
 \BY{T.J. Greytak}  in {\em Bose-Einstein Condensation}, edited by  \NAME{A.
  Griffin, D.W. Snoke, \atque S. Stringari} (Cambridge University Press,
  Cambridge) 1995, p.\ 131.

\bibitem{grey84}
 \BY{T.J. Greytak \atque D. Kleppner}  in {\em New Trends in Atomic Physics},
  {\em Les Houches Summer School1982}, edited by  \NAME{G. Grynberg \atque R.
  Stora} (North-Holland, Amsterdam) 1984, p.\ 1125.

\bibitem{silv86}
 \BY{I.F. Silvera \atque J.T.M. Walraven}  in {\em Progress in Low Temperature
  Physics}, edited by  \NAME{D.F. Brewer} (Elsevier, Amsterdam) 1986, Vol.~X,
  p.\ 139.

\bibitem{walr96}
 \BY{J.T.M. Walraven}  in {\em Quantum Dynamics of Simple Systems}, edited by
  \NAME{G.L. Oppo, S.M. Barnett, E. Riis, \atque M. Wilkinson} (Institute of
  Physics Publ., London) 1996, p.\ 315.

\bibitem{gold81}
 \BY{V.V. Goldman, I.F. Silvera, \atque A.J. Leggett}  \IN{Phys. Rev.
  B}{24}{1981}{2870}.

\bibitem{huse82}
 \BY{D.A. Huse \atque E. Siggia}  \IN{J. Low Temp. Phys.}{46}{1982}{137}.

\bibitem{oliv89}
 \BY{J. Oliva}  \IN{Phys. Rev. B}{39}{1989}{4197}.

\bibitem{stoo88}
 \BY{H.T.C. Stoof, J.M.V.A. Koelman, \atque B.J. Verhaar}  \IN{Phys. Rev.
  B}{38}{1988}{4688}.

\bibitem{hess86}
 \BY{H.F. Hess}  \IN{Phys. Rev. B}{34}{1986}{3476}.

\bibitem{masu88}
 \BY{N. Masuhara \etal }  \IN{Phys. Rev. Lett.}{61}{1988}{935}.

\bibitem{silv95bec}
 \BY{I.F. Silvera}  in {\em Bose-Einstein Condensation}, edited by  \NAME{A.
  Griffin, D.W. Snoke, \atque S. Stringari} (Cambridge University Press,
  Cambridge) 1995, p.\ 160.

\bibitem{mosk98}
 \BY{A.P. Mosk, M.W. Reynolds, T.W. Hijmans, \atque J.T.M. Walraven}  \IN{J.
  Low Temp. Phys.}{113}{1998}{217}.

\bibitem{mats95}
 \BY{A. Matsubara \etal }  in {\em Bose-Einstein Condensation}, edited by
  \NAME{A. Griffin, D.W. Snoke, \atque S. Stringari} (Cambridge University
  Press, Cambridge) 1995, p.\ 478.

\bibitem{safo98prl}
 \BY{A.I. Safonov \etal }  \IN{Phys. Rev. Lett.}{81}{1998}{4545}.

\bibitem{arim92}
 \BY{E. Arimondo, W.D. Phillips, \atque F. Strumia} {\em Laser Manipulation of
  Atoms and Ions} (North-Holland, Amsterdam) 1992.

\bibitem{metc94}
 \BY{H. Metcalf \atque P. van~der Straten}  \IN{Phys. Rep.}{244}{1994}{203}.

\bibitem{adam97}
 \BY{C.S. Adams \atque E. Riis}  \IN{Progress in Quantum
  Electronics}{21}{1997}{1}.

\bibitem{chu98nob}
 \BY{S. Chu}  \IN{Rev. Mod. Phys.}{70}{1998}{685}.

\bibitem{cohe98nob}
 \BY{C.N. Cohen-Tannoudji}  \IN{Rev. Mod. Phys.}{70}{1998}{707}.

\bibitem{phil98nob}
 \BY{W.D. Phillips}  \IN{Rev. Mod. Phys.}{70}{1998}{721}.

\bibitem{leto80}
 \BY{V.S. Letokhov \atque V.G. Minogin}  \IN{Optics Comm.}{35}{1980}{199}.

\bibitem{chu85}
 \BY{S. Chu \etal }  \IN{Phys. Rev. Lett.}{55}{1985}{48}.

\bibitem{prit86trap}
 \BY{D.E. Pritchard}  in {\em Electronic and atomic collisions : invited papers
  of the XIV International Conference on the Physics of Electronic and Atomic
  Collisions, Palo Alto, California, 24-30 July, 1985}, edited by  \NAME{D.C.
  Lorents, W.E. Meyerhof, \atque J.R. Peterson} (Elsevier, New York) 1986, p.\
  593.

\bibitem{walk94}
 \BY{T. Walker \atque P. Feng}  in {\em Advances in Atomic, Molecular, and
  Optical Physics}, edited by  \NAME{B. Bederson \atque H. Walther} (Academic
  Press, San Diego) 1994, Vol.~34, p.\ 125.

\bibitem{wein95}
 \BY{J. Weiner}  in {\em Advances in Atomic, Molecular, and Optical Physics},
  edited by  \NAME{B. Bederson \atque H. Walther} (Academic Press, San Diego)
  1995, Vol.~35, p.\ 45.

\bibitem{walk90}
 \BY{T. Walker, D. Sesko, \atque C. Wieman}  \IN{Phys. Rev.
  Lett.}{64}{1990}{408}.

\bibitem{adam95}
 \BY{C.S. Adams \etal }  \IN{Phys. Rev. Lett.}{74}{1995}{3577}.

\bibitem{boir98}
 \BY{D. Boiron \etal }  \IN{Phys. Rev. A}{57}{1998}{R4106}.

\bibitem{depu98}
 \BY{M.T. DePue \etal } preprint. (1998).

\bibitem{chu86}
 \BY{S. Chu, J.E. Bjorkholm, A. Ashkin, \atque A. Cable}  \IN{Phys. Rev.
  Lett.}{57}{1986}{314}.

\bibitem{drew94}
 \BY{M. Drewsen \etal }  \IN{Appl. Phys. B}{59}{1994}{283}.

\bibitem{town95}
 \BY{C.G. Townsend \etal }  \IN{Phys. Rev. A}{52}{1995}{1423}.

\bibitem{town96spot}
 \BY{C.G. Townsend \etal }  \IN{Phys. Rev. A}{53}{1996}{1702}.

\bibitem{vigu86}
 \BY{J. Vigu{\'e}}  \IN{Phys. Rev. A}{34}{1986}{4476}.

\bibitem{corn91}
 \BY{E.A. Cornell, C. Monroe, \atque C.E. Wieman}  \IN{Phys. Rev.
  Lett.}{67}{1991}{2439}.

\bibitem{kett92stat}
 \BY{W. Ketterle \atque D.E. Pritchard}  \IN{Appl. Phys. B}{54}{1992}{403}.

\bibitem{prit89}
 \BY{D.E. Pritchard, K. Helmerson, \atque A.G. Martin}  in {\em Atomic Physics
  11}, edited by  \NAME{S. Haroche, J.C. Gay, \atque G. Grynberg} (World
  Scientific, Singapore) 1989, p.\ 179.

\bibitem{migd85}
 \BY{A.L. Migdall \etal }  \IN{Phys. Rev. Lett.}{54}{1985}{2596}.

\bibitem{bagn87stop}
 \BY{V.S. Bagnato \etal }  \IN{Phys. Rev. Lett.}{58}{1987}{2194}.

\bibitem{helm92cool}
 \BY{K. Helmerson, A. Martin, \atque D.E. Pritchard}  \IN{J. Opt. Soc. Am.
  B}{9}{1992}{1988}.

\bibitem{monr90}
 \BY{C. Monroe, W. Swann, H. Robinson, \atque C. Wieman}  \IN{Phys. Rev.
  Lett.}{65}{1990}{1571}.

\bibitem{monr92}
 \BY{C. Monroe, E. Cornell, \atque C. Wieman}  in {\em Laser Manipulation of
  Atoms and Ions}, {\em Proceedings of the International School of Physics
  ``Enrico Fermi'', Course CXVIII}, edited by  \NAME{E. Arimondo, W.D.
  Phillips, \atque F. Strumia} (North-Holland, Amsterdam) 1992, p.\ 361.

\bibitem{kett93spot}
 \BY{W. Ketterle \etal }  \IN{Phys. Rev. Lett.}{70}{1993}{2253}.

\bibitem{davi94icap}
 \BY{K.B. Davis, M.O. Mewes, M.A. Joffe, \atque W. Ketterle}  in {\em
  Fourteenth International Conference on Atomic Physics, Boulder, Colorado,
  1994, Book of Abstracts, 1-M3} (University of Colorado, Boulder, Colorado)
  1994.

\bibitem{petr94icap}
 \BY{W. Petrich, M.H. Anderson, J.R. Ensher, \atque E.A. Cornell}  in {\em
  Fourteenth International Conference on Atomic Physics, Boulder, Colorado,
  1994, Book of Abstracts, 1M-7} (University of Colorado, Boulder, Colorado)
  1994.

\bibitem{brad95bec}
 \BY{C.C. Bradley, C.A. Sackett, J.J. Tollet, \atque R.G. Hulet}  \IN{Phys.
  Rev. Lett.}{75}{1995}{1687}.

\bibitem{brad97erra}
 \BY{C.C. Bradley, C.A. Sackett, J.J. Tollet, \atque R.G. Hulet}  \IN{Phys.
  Rev. Lett.}{79}{1997}{1170}.

\bibitem{petr94cmot}
 \BY{W. Petrich, M.H. Anderson, J.R. Ensher, \atque E.A. Cornell}  \IN{J. Opt.
  Soc. Am. B}{11}{1994}{1332}.

\bibitem{phil85}
 \BY{W.D. Phillips, J.V. Prodan, \atque H.J. Metcalf}  \IN{J. Opt. Soc. Am.
  B}{2}{1985}{1751}.

\bibitem{phil92}
 \BY{W.D. Phillips}  in {\em Laser Manipulation of Atoms and Ions}, {\em
  Proceedings of the International School of Physics ``Enrico Fermi'', Course
  CXVIII}, edited by  \NAME{E. Arimondo, W.D. Phillips, \atque F. Strumia}
  (North-Holland, Amsterdam) 1992, p.\ 289.

\bibitem{prod82}
 \BY{J.V. Prodan, W.D. Phillips, \atque H. Metcalf}  \IN{Phys. Rev.
  Lett.}{49}{1982}{1149}.

\bibitem{joff93}
 \BY{M.A. Joffe, W. Ketterle, A. Martin, \atque D.E. Pritchard}  \IN{J. Opt.
  Soc. Am. B}{10}{1993}{2257}.

\bibitem{chu92}
 \BY{S. Chu}  in {\em Laser Manipulation of Atoms and Ions}, {\em Proceedings
  of the International School of Physics ``Enrico Fermi'', Course CXVIII},
  edited by  \NAME{E. Arimondo, W.D. Phillips, \atque F. Strumia}
  (North-Holland, Amsterdam) 1992, p.\ 239.

\bibitem{raab87}
 \BY{E.L. Raab \etal }  \IN{Phys. Rev. Lett.}{59}{1987}{2631}.

\bibitem{prit92}
 \BY{D.E. Pritchard \atque W. Ketterle}  in {\em Laser Manipulation of Atoms
  and Ions}, {\em Proceedings of the International School of Physics ``Enrico
  Fermi'', Course CXVIII}, edited by  \NAME{E. Arimondo, W.D. Phillips, \atque
  F. Strumia} (North-Holland, Amsterdam) 1992, p.\ 473.

\bibitem{ande94}
 \BY{M.H. Anderson, W. Petrich, J.R. Ensher, \atque E.A. Cornell}  \IN{Phys.
  Rev. A}{50}{1994}{R3597}.

\bibitem{han98}
 \BY{D.J. Han, R.H. Wynar, P. Courteille, \atque D.J. Heinzen}  \IN{Phys. Rev.
  A}{57}{1998}{R4114}.

\bibitem{hau98}
 \BY{L.V. Hau \etal }  \IN{Phys. Rev. A}{58}{1998}{R54}.

\bibitem{ande98pra}
 \BY{B.P. Anderson \atque M. Kasevich} {\it Phys.\ Rev.\ A}, in press (1998).

\bibitem{lett88}
 \BY{P.D. Lett \etal }  \IN{Phys. Rev. Lett.}{61}{1988}{169}.

\bibitem{lett89}
 \BY{P.D. Lett \etal }  \IN{J. Opt. Soc. Am. B}{6}{1989}{2084}.

\bibitem{weis89}
 \BY{D.S. Weiss \etal }  \IN{J. Opt. Soc. Am. B}{6}{1989}{2072}.

\bibitem{coop94}
 \BY{C.J. Cooper \etal }  \IN{Europhys. Lett.}{28}{1994}{397}.

\bibitem{cabl90}
 \BY{A. Cable, M. Prentiss, \atque N.P. Bigelow}  \IN{Optics
  Lett.}{15}{1990}{507}.

\bibitem{stea95}
 \BY{A. Steane, P. Szriftgiser, P. Desbiolles, \atque J. Dalibard}  \IN{Phys.
  Rev. Lett.}{74}{1995}{4972}.

\bibitem{myat96}
 \BY{C.J. Myatt \etal }  \IN{Optics Lett.}{21}{1996}{290}.

\bibitem{gibb95}
 \BY{K. Gibble, S. Chang, \atque R. Legere}  \IN{Phys. Rev.
  Lett.}{75}{1995}{2666}.

\bibitem{lu96lvis}
 \BY{Z.T. Lu \etal }  \IN{Phys. Rev. Lett.}{77}{1996}{3331}.

\bibitem{swan96}
 \BY{T.B. Swanson \etal }  \IN{J. Opt. Soc. Am. B}{13}{1996}{1833}.

\bibitem{diec98}
 \BY{K. Dieckmann, R.J.C. Spreeuw, M. Weidem{\"u}ller, \atque J.T.M. Walraven}
  \IN{Phys. Rev. A}{58}{1998}{3891}.

\bibitem{kase92}
 \BY{M. Kasevich \atque S. Chu}  \IN{Phys. Rev. Lett.}{69}{1992}{1741}.

\bibitem{reic95}
 \BY{J. Reichel \etal }  \IN{Phys. Rev. Lett.}{75}{1995}{4575}.

\bibitem{aspe88}
 \BY{A. Aspect \etal }  \IN{Phys. Rev. Lett.}{61}{1988}{826}.

\bibitem{lawa95}
 \BY{J. Lawall \etal }  \IN{Phys. Rev. Lett.}{75}{1995}{4194}.

\bibitem{lee96}
 \BY{H.J. Lee, C.S. Adams, M. Kasevich, \atque S. Chu}  \IN{Phys. Rev.
  Lett.}{76}{1996}{2658}.

\bibitem{lee98}
 \BY{H.J. Lee \atque S. Chu}  \IN{Phys. Rev. A}{57}{1998}{2905}.

\bibitem{hess87}
 \BY{H. Hess \etal }  \IN{Phys. Rev. Lett.}{59}{1987}{672}.

\bibitem{vanr88}
 \BY{R. van Roijen, J.J. Berkhout, S. Jaakkola, \atque J.T.M. Walraven}
  \IN{Phys. Rev. Lett.}{61}{1988}{931}.

\bibitem{wein98}
 \BY{J.D. Weinstein \etal }  \IN{Nature}{395}{1998}{148}.

\bibitem{kim97}
 \BY{J. Kim \etal }  \IN{Phys. Rev. Lett.}{78}{1997}{3665}.

\bibitem{spre94}
 \BY{R.J.C. Spreeuw \etal }  \IN{Phys. Rev. Lett.}{72}{1994}{3162}.

\bibitem{ovch97}
 \BY{Y.B. Ovchinnikov, I. Manek, \atque R. Grimm}  \IN{Phys. Rev.
  Lett.}{79}{1997}{2225}.

\bibitem{ovch98}
 \BY{Y.B. Ovchinnikov \etal }  \IN{Europhys. Lett.}{43}{1998}{510}.

\bibitem{take95cs}
 \BY{T. Takekoshi \atque R.J. Knize}  \IN{Optics Lett.}{21}{1995}{77}.

\bibitem{frie98co2}
 \BY{S. Friebel \etal }  \IN{Phys. Rev. A}{57}{1998}{R20}.

\bibitem{sava97}
 \BY{T.A. Savard, K.M. O'Hara, \atque J.E. Thomas}  \IN{Phys. Rev.
  A}{56}{1997}{R1095}.

\bibitem{agos89}
 \BY{C.C. Agosta, I.F. Silvera, H.T.C. Stoof, \atque B.J. Verhaar}  \IN{Phys.
  Rev. Lett.}{62}{1989}{2361}.

\bibitem{love85}
 \BY{R.V.E. Lovelace, C. Mahanian, T.J. Tommila, \atque D.M. Lee}
  \IN{Nature}{318}{1985}{30}.

\bibitem{riis93}
 \BY{E. Riis \atque S.M. Barnett}  \IN{Europhys. Lett.}{21}{1993}{533}.

\bibitem{shim92}
 \BY{F. Shimizu \atque M. Morinaga}  \IN{Jpn. J. Appl. Phys.}{31}{1992}{L1721}.

\bibitem{newb96}
 \BY{N.R. Newbury \atque C.E. Wieman}  \IN{Am. J. Phys.}{64}{1996}{18}.

\bibitem{berg87}
 \BY{T. Bergeman, G. Erez, \atque H. Metcalf}  \IN{Phys. Rev.
  A}{35}{1987}{1535}.

\bibitem{kugl85}
 \BY{K.-J. K{\"u}gler, K. Moritz, W. Paul, \atque U. Trinks}  \IN{Nuclear
  Instruments and Methods in Physics Research}{228}{1985}{240}.

\bibitem{wing84}
 \BY{W.H. Wing}  \IN{Prog. Quant. Electr.}{8}{1984}{181}.

\bibitem{majo32}
 \BY{E. Majorana}  \IN{Nuovo Cimento}{9}{1932}{43}.

\bibitem{schw37}
 \BY{J. Schwinger}  \IN{Phys. Rev.}{51}{1937}{648}.

\bibitem{berg89}
 \BY{T.H. Bergeman \etal }  \IN{J. Opt. Soc. Am. B}{6}{1989}{2249}.

\bibitem{davi95evap}
 \BY{K.B. Davis \etal }  \IN{Phys. Rev. Lett.}{74}{1995}{5202}.

\bibitem{petr95}
 \BY{W. Petrich, M.H. Anderson, J.R. Ensher, \atque E.A. Cornell}  \IN{Phys.
  Rev. Lett.}{74}{1995}{3352}.

\bibitem{shap96}
 \BY{V.E. Shapiro}  \IN{Phys. Rev. A}{54}{1996}{R1018}.

\bibitem{prit83}
 \BY{D.E. Pritchard}  \IN{Phys. Rev. Lett.}{51}{1983}{1336}.

\bibitem{gott62}
 \BY{Y.V. Gott, M.S. Ioffe, \atque V.G. Tel'kovskii}  \IN{Nuclear Fusion
  Supplement}{3}{1962}{1045}.

\bibitem{erns98}
 \BY{U. Ernst \etal }  \IN{Europhys. Lett.}{41}{1998}{1}.

\bibitem{mewe96bec}
 \BY{M.-O. Mewes \etal }  \IN{Phys. Rev. Lett.}{77}{1996}{416}.

\bibitem{essl98}
 \BY{T. Esslinger, I. Bloch, \atque T.W. H{\"{a}}nsch}  \IN{Phys. Rev.
  A}{58}{1998}{R2664}.

\bibitem{sodi98}
 \BY{J. S{\"{o}}ding \etal }  \IN{Phys. Rev. Lett.}{80}{1998}{1869}.

\bibitem{toll95}
 \BY{J.J. Tollet, C.C. Bradley, C.A. Sackett, \atque R.G. Hulet}  \IN{Phys.
  Rev. A}{51}{1995}{R22}.

\bibitem{desr97}
 \BY{B. Desruelle \etal }  \IN{European Physical Journal D}{1}{1998}{255}.

\bibitem{mewe97thes}
 \BY{M.-O. Mewes} Ph.D. thesis, Massachusetts Institute of Technology, 1997.

\bibitem{pink97}
 \BY{P.W.H. Pinkse \etal }  \IN{Phys. Rev. Lett.}{78}{1997}{990}.

\bibitem{suku97}
 \BY{C.V. Sukumar \atque D.M. Brink}  \IN{Phys. Rev. A}{56}{1997}{2451}.

\bibitem{doyl91stic}
 \BY{J.M. Doyle \etal }  \IN{Phys. Rev. Lett.}{67}{1991}{603}.

\bibitem{hijm89}
 \BY{T.W. Hijmans, O.J. Luiten, I.D. Setija, \atque J.T.M. Walraven}  \IN{J.
  Opt. Soc. Am. B}{6}{1989}{2235}.

\bibitem{kett93evap}
 \BY{W. Ketterle \etal } OSA Annual Meeting, Toronto, Canada, October 3-8
  (1993).

\bibitem{wu96sim}
 \BY{H. Wu \atque C.J. Foot}  \IN{J. Phys. B}{29}{1996}{L321}.

\bibitem{holl96traj}
 \BY{M. Holland, J. Williams, K. Coakley, \atque J. Cooper}  \IN{Quantum and
  Semiclassical Optics}{8}{1996}{571}.

\bibitem{wu97}
 \BY{H. Wu, E. Arimondo, \atque C.J. Foot}  \IN{Phys. Rev. A}{56}{1997}{560}.

\bibitem{sack97evap}
 \BY{C.A. Sackett, C.C. Bradley, \atque R.G. Hulet}  \IN{Phys. Rev.
  A}{55}{1997}{3797}.

\bibitem{arim98}
 \BY{E. Arimondo, E. Cerboneschi, \atque H. Wu} this volume (1998).

\bibitem{pink98}
 \BY{P.W.H. Pinkse \etal }  \IN{Phys. Rev. A}{57}{1998}{4747}.

\bibitem{phil98var}
 \BY{W.D. Phillips} this volume (1998).

\bibitem{myat97thes}
 \BY{C.J. Myatt} Ph.D. thesis, University of Colorado, 1997.

\bibitem{corn98var}
 \BY{E. Cornell, J.R. Ensher, \atque C.E. Wieman} this volume (1998).

\bibitem{andr96}
 \BY{M.R. Andrews \etal }  \IN{Science}{273}{1996}{84}.

\bibitem{andr97int}
 \BY{M.R. Andrews \etal }  \IN{Science}{275}{1997}{637}.

\bibitem{stam98odt}
 \BY{D.M. Stamper-Kurn \etal }  \IN{Phys. Rev. Lett.}{80}{1998}{2072}.

\bibitem{mewe97}
 \BY{M.-O. Mewes \etal }  \IN{Phys. Rev. Lett.}{78}{1997}{582}.

\bibitem{hall98dyn}
 \BY{D.S. Hall \etal }  \IN{Phys. Rev. Lett.}{81}{1998}{4531}.

\bibitem{matt98}
 \BY{M.R. Matthews \etal }  \IN{Phys. Rev. Lett.}{81}{1998}{243}.

\bibitem{hall98phas}
 \BY{D.S. Hall, M.R. Matthews, C.E. Wieman, \atque E.A. Cornell}  \IN{Phys.
  Rev. Lett.}{81}{1998}{1543}.

\bibitem{kozu98}
 \BY{M. Kozuma \etal } preprint (1998).

\bibitem{jin96coll}
 \BY{D.S. Jin \etal }  \IN{Phys. Rev. Lett.}{77}{1996}{420}.

\bibitem{mewe96coll}
 \BY{M.-O. Mewes \etal }  \IN{Phys. Rev. Lett.}{77}{1996}{988}.

\bibitem{jin97}
 \BY{D.S. Jin \etal }  \IN{Phys. Rev. Lett.}{78}{1997}{764}.

\bibitem{verh95}
 \BY{B.J. Verhaar}  in {\em Atomic Physics}, edited by  \NAME{D.J. Wineland,
  C.E. Wieman, \atque S.J. Smith} (AIP, New York) 1995, Vol.~14, p.\ 351.

\bibitem{guer98}
 \BY{D. Gu{\'e}ry-Odelin, J. S{\"{o}}ding, P. Desbiolles, \atque J. Dalibard}
  \IN{Europhys. Lett.}{44}{1998}{25}.

\bibitem{arnd97}
 \BY{M. Arndt \etal }  \IN{Phys. Rev. Lett.}{79}{1997}{625}.

\bibitem{corn98priv}
 \BY{E. Cornell} private communication (1998).

\bibitem{sack97bec}
 \BY{C.A. Sackett, C.C. Bradley, M. Welling, \atque R.G. Hulet}  \IN{Bra. J.
  Phys.}{27}{1997}{154}.

\bibitem{poli97}
 \BY{H.D. Politzer}  \IN{Phys. Rev. A}{55}{1997}{1140}.

\bibitem{cohe92}
 \BY{C. Cohen-Tannoudji, J. Dupont-Roc, \atque G. Grynberg} {\em Atom-Photon
  Interactions} (Wiley, New York) 1992.

\bibitem{mori95}
 \BY{O. Morice, Y. Castin, \atque J. Dalibard}  \IN{Phys. Rev.
  A}{51}{1995}{3896}.

\bibitem{alle75}
 \BY{L. Allen \atque J.H. Eberly} {\em Optical Resonance and Two-Level Atoms}
  (Dover Publications, New York) 1975.

\bibitem{andr98thes}
 \BY{M.R. Andrews} Ph.D. thesis, Massachusetts Institute of Technology, 1998.

\bibitem{luit93lym}
 \BY{O.J. Luiten \etal }  \IN{Phys. Rev. Lett.}{70}{1993}{544}.

\bibitem{bagn87ext}
 \BY{V. Bagnato, D.E. Pritchard, \atque D. Kleppner}  \IN{Phys. Rev.
  A}{35}{1987}{4354}.

\bibitem{baym96}
 \BY{G. Baym \atque C.J. Pethick}  \IN{Phys. Rev. Lett.}{76}{1996}{6}.

\bibitem{edwa95}
 \BY{M. Edwards \atque K. Burnett}  \IN{Phys. Rev. A}{51}{1995}{1382}.

\bibitem{dalf98}
 \BY{F. Dalfovo, S. Giorgini, L.P. Pitaevskii, \atque S. Stringari} {\it Rev.\
  Mod.\ Phys.\ }, in press (1998).

\bibitem{cast96}
 \BY{Y. Castin \atque R. Dum}  \IN{Phys. Rev. Lett.}{77}{1996}{5315}.

\bibitem{kaga96evol}
 \BY{Y. Kagan, E.L. Surkov, \atque G.V. Shlyapnikov}  \IN{Phys. Rev.
  A}{54}{1996}{R1753}.

\bibitem{dalf97}
 \BY{F. Dalfovo, C. Minniti, S. Stringari, \atque L. Pitaevskii}  \IN{Physics
  Letters A}{227}{1997}{259}.

\bibitem{holl97int}
 \BY{M. Holland, D.S. Jin, M.L. Chiofalo, \atque J. Cooper}  \IN{Phys. Rev.
  Lett.}{78}{1997}{3801}.

\bibitem{gior97}
 \BY{S. Giorgini, L.P. Pitaevskii, \atque S. Stringari}  \IN{Phys. Rev.
  Lett.}{78}{1997}{3987}.

\bibitem{gior97jltp}
 \BY{S. Giorgini, L.P. Pitaevskii, \atque S. Strigari}  \IN{J. Low Temp.
  Phys.}{109}{1997}{309}.

\bibitem{dodd98two}
 \BY{R.J. Dodd, K. Burnett, M. Edwards, \atque C.W. Clark}  \IN{Acta Physica
  Polonica}{93}{1998}{45}.

\bibitem{nara98semi}
 \BY{M. Naraschewski \atque D.M. Stamper-Kurn}  \IN{Phys. Rev.
  A}{58}{1998}{2423}.

\bibitem{dodd97coh}
 \BY{R.J. Dodd, C.W. Clark, M. Edwards, \atque K. Burnett}  \IN{Optics
  Express}{1}{1997}{282}.

\bibitem{wu98exp1}
 \BY{H. Wu \atque E. Arimondo}  \IN{Phys. Rev. A}{58}{1998}{3822}.

\bibitem{wu98exp2}
 \BY{H. Wu \atque E. Arimondo}  \IN{Europhys. Lett.}{43}{1998}{141}.

\bibitem{ande98priv}
 \BY{B. Anderson \atque M. Kasevich} private communication (1998).

\bibitem{bijl96}
 \BY{M. Bijlsma \atque H.T.C. Stoof}  \IN{Phys. Rev. A}{54}{1996}{5085}.

\bibitem{grut98}
 \BY{P. Gr{\"{u}}ter, D. Ceperley, \atque F. Lalo{\"{e}}}  \IN{Phys. Rev.
  Lett.}{79}{1998}{3549}.

\bibitem{gior96}
 \BY{S. Giorgini, L.P. Pitaevskii, \atque S. Stringari}  \IN{Phys. Rev.
  A}{54}{1996}{R4633}.

\bibitem{gros95lamb}
 \BY{S. Grossman \atque M. Holthaus}  \IN{Zeitschrift f{\"u}r
  Naturforschung}{50a}{1995}{921}.

\bibitem{kett96fin}
 \BY{W. Ketterle \atque N.J. van Druten}  \IN{Phys. Rev. A}{54}{1996}{656}.

\bibitem{ensh96}
 \BY{J.R. Ensher \etal }  \IN{Phys. Rev. Lett.}{77}{1996}{4984}.

\bibitem{hutc97}
 \BY{D.A.W. Hutchinson, E. Zaremba, \atque A. Griffin}  \IN{Phys. Rev.
  Lett.}{78}{1997}{1842}.

\bibitem{dodd97}
 \BY{R.J. Dodd, M. Edwards, C.W. Clark, \atque K. Burnett}  \IN{Phys. Rev.
  A}{57}{1998}{R32}.

\bibitem{ming97}
 \BY{A. Minguzzi, S. Conti, \atque M.P. Tosi}  \IN{Journal of Physics:
  Condensed Matter}{9}{1997}{L33}.

\bibitem{mcal95}
 \BY{W.I. McALexander \etal }  \IN{Phys. Rev. A}{51}{1995}{R871}.

\bibitem{nozi90}
 \BY{P. Nozi{\`e}res \atque D. Pines} {\em The Theory of Quantum Liquids}
  (Addison-Wesley, Redwood City, CA) 1990.

\bibitem{grif93}
 \BY{A. Griffin} {\em Excitations in a Bose-condensed liquid} (Cambridge
  University Press, Cambridge) 1993.

\bibitem{bogo47}
 \BY{N.N. Bogoliubov}  \IN{J. Phys. (USSR)}{11}{1947}{23}.

\bibitem{lee59}
 \BY{T.D. Lee \atque C.N. Yang}  \IN{Phys. Rev.}{113}{1959}{1406}.

\bibitem{stri98cigar}
 \BY{S. Stringari}  \IN{Phys. Rev. A}{58}{1998}{2365}.

\bibitem{edwa96coll}
 \BY{M. Edwards \etal }  \IN{Phys. Rev. Lett.}{77}{1996}{1671}.

\bibitem{stri96coll}
 \BY{S. Stringari}  \IN{Phys. Rev. Lett.}{77}{1996}{2360}.

\bibitem{andr98erra}
 \BY{M.R. Andrews \etal }  \IN{Phys. Rev. Lett.}{80}{1998}{2967}.

\bibitem{zare97}
 \BY{E. Zaremba, A. Griffin, \atque T. Nikuni}  \IN{Phys. Rev.
  A}{57}{1998}{4695}.

\bibitem{kavo98}
 \BY{G.M. Kavoulakis \atque C.J. Pethick}  \IN{Phys. Rev. A}{58}{1998}{1563}.

\bibitem{morg97}
 \BY{S.A. Morgan, R.J. Ballagh, \atque K. Burnett}  \IN{Phys. Rev.
  A}{55}{1997}{4338}.

\bibitem{rein97}
 \BY{W.P. Reinhardt \atque C.W. Clark}  \IN{J. Phys. B}{30}{1997}{L785}.

\bibitem{jack98soli}
 \BY{A.D. Jackson, G.M. Kavoulakis, \atque C.J. Pethick}  \IN{Phys. Rev.
  A}{58}{1998}{2417}.

\bibitem{gay85}
 \BY{C. Gay \atque A. Griffin}  \IN{J. Low Temp. Phys.}{58}{1985}{479}.

\bibitem{popo87}
 \BY{V.N. Popov} {\em Functional Integrals and Collective Modes} (Cambridge
  University Press, New York) 1987.

\bibitem{grif96cons}
 \BY{A. Griffin}  \IN{Phys. Rev. B}{53}{1996}{9341}.

\bibitem{hutc98}
 \BY{D.A.W. Hutchinson, R.J. Dodd, \atque K. Burnett}  \IN{Phys. Rev.
  Lett.}{81}{1998}{2198}.

\bibitem{olsh98coll}
 \BY{M. Olshanii} preprint, cond-mat/9807412 (1998).

\bibitem{bijl98coll}
 \BY{M.J. Bijlsma \atque H.T.C. Stoof} preprint, cond-mat/9807051. (1998).

\bibitem{liu97}
 \BY{W.V. Liu}  \IN{Phys. Rev. Lett.}{79}{1997}{4056}.

\bibitem{pita97}
 \BY{L.P. Pitaevskii \atque S. Stringari}  \IN{Physical Letters
  A}{235}{1997}{398}.

\bibitem{gior98}
 \BY{S. Giorgini}  \IN{Phys. Rev. A}{57}{1998}{2949}.

\bibitem{fedi98}
 \BY{P.O. Fedichev, G.V. Shlyapnikov, \atque J.T.M. Walraven}  \IN{Phys. Rev.
  Lett.}{80}{1998}{2269}.

\bibitem{pita97phen}
 \BY{L.P. Pitaevskii}  \IN{Physics Letters A}{229}{1997}{406}.

\bibitem{kukl97}
 \BY{A.B. Kuklov \etal }  \IN{Phys. Rev. A}{55}{1997}{R3307}.

\bibitem{grah97}
 \BY{R. Graham, D.F. Walls, M.J. Collett, \atque E.M. Wright}  \IN{Phys. Rev.
  A}{57}{1998}{484}.

\bibitem{sina98}
 \BY{A. Sinatra \etal } preprint, cond-mat/9809061 (1998).

\bibitem{grif97}
 \BY{A. Griffin \atque E. Zaremba}  \IN{Phys. Rev. A}{56}{1997}{4839}.

\bibitem{shen98}
 \BY{V.B. Shenoy \atque T.-L. Ho}  \IN{Phys. Rev. Lett.}{80}{1998}{3985}.

\bibitem{grif96}
 \BY{A. Griffin, W.-C. Wu, \atque S. Stringari}  \IN{Phys. Rev.
  Lett.}{78}{1996}{1836}.

\bibitem{kavo98damp}
 \BY{G.M. Kavoulakis, C.J. Pethick, \atque H. Smith}  \IN{Phys. Rev.
  A}{57}{1998}{2938}.

\bibitem{kavo98rela}
 \BY{G.M. Kavoulakis, C.J. Pethick, \atque H. Smith}  \IN{Phys. Rev.
  Lett.}{81}{1998}{4036}.

\bibitem{ties96}
 \BY{E. Tiesinga \etal }  \IN{J. Res. Natl. Inst. Stand.
  Technol.}{101}{1996}{505}.

\bibitem{pita98}
 \BY{L.P. Pitaevskii \atque S.Stringari}  \IN{Phys. Rev.
  Lett.}{81}{1998}{4541}.

\bibitem{flie98}
 \BY{M. Fliesser \atque R. Graham} preprint, cond-mat/9806115 (1998).

\bibitem{holl96exp}
 \BY{M. Holland \atque J. Cooper}  \IN{Phys. Rev. A}{53}{1996}{R1954}.

\bibitem{nara96}
 \BY{M. Naraschewski \etal }  \IN{Phys. Rev. A}{54}{1996}{2185}.

\bibitem{rokh97}
 \BY{D.S. Rokhsar}  \IN{Phys. Rev. Lett.}{79}{1997}{2164}.

\bibitem{muel98}
 \BY{E.J. Mueller, P.M. Goldbart, \atque Y. Lyanda-Geller}  \IN{Phys. Rev.
  A}{57}{1998}{R1505}.

\bibitem{java98pers}
 \BY{J. Javaneinen, S.M. Paik, \atque S.M. Yoo}  \IN{Phys. Rev.
  A}{58}{1998}{580}.

\bibitem{hule98}
 \BY{R.G. Hulet} Invited talk at ICAPXVI, Windsor, Canada, August 3-7 (1998).

\bibitem{burt97}
 \BY{E.A. Burt \etal }  \IN{Phys. Rev. Lett.}{79}{1997}{337}.

\bibitem{sodi98deca}
 \BY{J. S{\"{o}}ding \etal } preprint, cond-mat/9811339 (1998).

\bibitem{burk97}
 \BY{J.P.J. Burke, J.L. Bohn, B.D. Esry, \atque C.H. Greene}  \IN{Phys. Rev.
  A}{55}{1997}{R2511}.

\bibitem{kokk97coll}
 \BY{S.J.J.M.F. Kokkelmans, H.M.J.M. Boesten, \atque B.J. Verhaar}  \IN{Phys.
  Rev. A}{55}{1997}{R1589}.

\bibitem{juli97stab}
 \BY{P.S. Julienne, F.H. Mies, E. Tiesinga, \atque C.J. Williams}  \IN{Phys.
  Rev. Lett.}{78}{1997}{1880}.

\bibitem{sten98stro}
 \BY{J. Stenger \etal } preprint (1998).

\bibitem{wise96}
 \BY{H. Wiseman, A. Martins, \atque D. Walls}  \IN{Quantum and Semiclassical
  Optics}{8}{1996}{737}.

\bibitem{holl96alas}
 \BY{M. Holland \etal }  \IN{Phys. Rev. A}{54}{1996}{R1757}.

\bibitem{wise97}
 \BY{H.M. Wiseman}  \IN{Phys. Rev. A}{56}{1997}{2068}.

\bibitem{wise95atla}
 \BY{H.M. Wiseman \atque M.J. Collett}  \IN{Physics Letters A}{202}{1995}{246}.

\bibitem{spre96}
 \BY{R.J.C. Spreeuw, T. Pfau, U. Janicke, \atque M. Wilkens}  \IN{Europhys.
  Lett.}{32}{1996}{469}.

\bibitem{olsh96}
 \BY{M. Olshanii, Y. Castin, \atque J. Dalibard}  in {\em Laser Spectroscopy
  XII}, edited by  \NAME{M. Inguscio, M. Allegrini, \atque A. Sasso} (World
  Scientific, Singapore) 1996, p.\ 7.

\bibitem{klep97}
 \BY{D. Kleppner} Physics Today, Aug. 1997, p. 11; Jan. 1998, p. 90 .

\bibitem{grif95qm}
 \BY{D.J. Griffiths} {\em Introduction to Quantum Mechanics} (Prentice Hall,
  Englewood Cliffs) 1995.

\bibitem{toda91}
 \BY{M. Toda, R. Kubo, N. Saito, \atque N. Hashitsume} {\em Statistical Physics
  I} (Springer-Verlag, New York) 1991.

\bibitem{levi77}
 \BY{E. Levich \atque V. Yakhot}  \IN{Phys. Rev. B}{15}{1977}{243}.

\bibitem{stoo91}
 \BY{H.T.C. Stoof}  \IN{Phys. Rev. Lett.}{66}{1991}{3148}.

\bibitem{tikh90}
 \BY{S.G. Tikhodeev}  \IN{Sov. Phys. JETP}{70}{1990}{380}.

\bibitem{stoo95}
 \BY{H.T.C. Stoof}  in {\em Bose-Einstein Condensation}, edited by  \NAME{A.
  Griffin, D.W. Snoke, \atque S. Stringari} (Cambridge University Press,
  Cambridge) 1995, p.\ 226.

\bibitem{snok89}
 \BY{D.W. Snoke \atque J.P. Wolfe}  \IN{Phys. Rev. B}{39}{1989}{4030}.

\bibitem{ecke84}
 \BY{U. Eckern}  \IN{J. Low Temp. Phys.}{54}{1984}{333}.

\bibitem{svis91}
 \BY{B.V. Svistunov}  \IN{J. Moscow Phys. Soc.}{1}{1991}{373}.

\bibitem{semi95}
 \BY{D.V. Semikoz \atque I.I. Tkachev}  \IN{Phys. Rev. Lett.}{74}{1995}{3093}.

\bibitem{stoo97}
 \BY{H.T.C. Stoof}  \IN{Phys. Rev. Lett.}{78}{1997}{768}.

\bibitem{kaga92}
 \BY{Y. Kagan, B.V. Svistunov, \atque G.V. Shlyapnikov}  \IN{Sov. Phys.
  JETP}{75}{1992}{387}.

\bibitem{kaga94}
 \BY{Y. Kagan \atque B.V. Svistunov}  \IN{Sov. Phys. JETP}{78}{1994}{187}.

\bibitem{kaga97}
 \BY{Y. Kagan \atque B.V. Svistunov}  \IN{Phys. Rev. Lett.}{79}{1997}{3331}.

\bibitem{kaga95}
 \BY{Y. Kagan}  in {\em Bose-Einstein Condensation}, edited by  \NAME{A.
  Griffin, D.W. Snoke, \atque S. Stringari} (Cambridge University Press,
  Cambridge) 1995, p.\ 202.

\bibitem{gard97}
 \BY{C.W. Gardiner, P. Zoller, R.J. Ballagh, \atque M.J. Davis}  \IN{Phys. Rev.
  Lett.}{79}{1997}{1793}.

\bibitem{gard98}
 \BY{C.W. Gardiner \etal }  \IN{Phys. Rev. Lett.}{31}{1998}{5266}.

\bibitem{jose62}
 \BY{B.D. Josephson}  \IN{Phys. Lett.}{1}{1962}{251}.

\bibitem{feyn64}
 \BY{R.P. Feynman, R.B. Leighton, \atque M. Sands} {\em The Feynman Lectures on
  Physics} (Addison-Wesley, Reading,MA) 1964.

\bibitem{pere97}
 \BY{S.V. Pereversev \etal }  \IN{Nature}{388}{1997}{449}.

\bibitem{java96phas}
 \BY{J. Javanainen \atque S.M. Yoo}  \IN{Phys. Rev. Lett.}{76}{1996}{161}.

\bibitem{cira96}
 \BY{J.I. Cirac, C.W. Gardiner, M. Naraschewski, \atque P. Zoller}  \IN{Phys.
  Rev. A}{54}{1996}{R3714}.

\bibitem{cast97}
 \BY{Y. Castin \atque J. Dalibard}  \IN{Phys. Rev. A}{55}{1997}{4330}.

\bibitem{wong96int}
 \BY{T. Wong, M.J. Collett, \atque D.F. Walls}  \IN{Phys. Rev.
  A}{54}{1996}{R3718}.

\bibitem{ande86}
 \BY{P.W. Anderson}  in {\em The Lesson of Quantum Theory}, edited by
  \NAME{J.d. Boer, E. Dal, \atque O. Ulfbeck} (Elsevier, Amsterdam) 1986, p.\
  23.

\bibitem{pfle67}
 \BY{R.L. Pfleegor \atque L. Mandel}  \IN{Phys. Rev.}{159}{1967}{1084}.

\bibitem{rohr97}
 \BY{A. R{\"o}hrl, M. Naraschewski, A. Schenzle, \atque H. Wallis}  \IN{Phys.
  Rev. Lett.}{78}{1997}{4143}.

\bibitem{wall98}
 \BY{H. Wallis \atque H. Steck}  \IN{Europhys. Lett.}{41}{1998}{477}.

\bibitem{host96}
 \BY{W. Hoston \atque L. You}  \IN{Phys. Rev. A}{53}{1996}{4254}.

\bibitem{wall97phas}
 \BY{H. Wallis, A. R{\"o}hrl, M. Naraschewski, \atque A. Schenzle}  \IN{Phys.
  Rev. A}{55}{1997}{2109}.

\bibitem{java97spli}
 \BY{J. Javanainen \atque M. Wilkens}  \IN{Phys. Rev. Lett.}{78}{1997}{4675}.

\bibitem{helm98}
 \BY{K. Helmerson} private communication (1998).

\bibitem{ande98atla}
 \BY{B.P. Anderson \atque M.A. Kasevich}  \IN{Science}{282}{1998}{1686}.

\bibitem{lewe96phas}
 \BY{M. Lewenstein \atque L. You}  \IN{Phys. Rev. Lett.}{77}{1996}{3489}.

\bibitem{wrig96col1}
 \BY{E.M. Wright, D.F. Walls, \atque J.C. Garrison}  \IN{Phys. Rev.
  Lett.}{77}{1996}{2158}.

\bibitem{molm98}
 \BY{K. M{\o}lmer}  \IN{Phys. Rev. A}{58}{1998}{566}.

\bibitem{loud83}
 \BY{R. Loudon} {\em The Quantum Theory of Light} (Clarendon, Oxford) 1983.

\bibitem{kuo91}
 \BY{S.J. Kuo, D.T. Smithey, \atque M.G. Raymer}  \IN{Phys. Rev.
  A}{43}{1991}{4083}.

\bibitem{yasu96}
 \BY{M. Yasuda \atque F. Shimizu}  \IN{Phys. Rev. Lett.}{77}{1996}{3090}.

\bibitem{kett97}
 \BY{W. Ketterle \atque H.-J. Miesner}  \IN{Phys. Rev. A}{56}{1997}{3291}.

\bibitem{kaga85}
 \BY{Y. Kagan, B.V. Svistunov, \atque G.V. Shlyapnikov}  \IN{JETP
  Lett.}{42}{1985}{209}.

\bibitem{jack98flow}
 \BY{B. Jackson, J.F. McCann, \atque C.S. Adams} preprint, cond-mat/9804038
  (1998).

\bibitem{zhan98grav}
 \BY{W. Zhang \atque D.F. Walls}  \IN{Phys. Rev. A}{57}{1998}{1248}.

\bibitem{bloc98}
 \BY{I. Bloch, T.W. H{\"{a}}nsch, \atque T. Esslinger} preprint,
  cond-mat/9812258. (1998).

\bibitem{hagl98}
 \BY{E. Hagley \etal } preprint (1998).

\bibitem{mill93}
 \BY{J.D. Miller, R.A. Cline, \atque D.J. Heinzen}  \IN{Phys. Rev.
  A}{47}{1993}{R4567}.

\bibitem{kuga97}
 \BY{T. Kuga \etal }  \IN{Phys. Rev. Lett.}{78}{1997}{4713}.

\bibitem{kuhn96}
 \BY{A. Kuhn, H. Perrin, W. H{\"a}nsel, \atque C. Salomon}  in {\em Ultracold
  Atoms and Bose-Einstein-Condensation, 1996}, {\em OSA Trends in Optics and
  Photonics Series, Vol. 7}, edited by  \NAME{K. Burnett} (Optical Society of
  America, Washington D.C.) 1996, p.\ 58.

\bibitem{houb97}
 \BY{M. Houbiers, H.T.C. Stoof, \atque E.A. Cornell}  \IN{Phys. Rev.
  A}{56}{1997}{2041}.

\bibitem{gard98quas}
 \BY{C.W. Gardiner \etal } preprint, cond-mat/980101 (1998).

\bibitem{mari98}
 \BY{M. Marinescu \atque L. You}  \IN{Phys. Rev. Lett.}{81}{1998}{4596}.

\bibitem{fesh62}
 \BY{H. Feshbach}  \IN{Annals of Physics}{19}{1962}{287}.

\bibitem{ties92bec}
 \BY{E. Tiesinga, A.J. Moerdijk, B.J. Verhaar, \atque H.T.C. Stoof}  \IN{Phys.
  Rev. A}{46}{1992}{R1167}.

\bibitem{ties93}
 \BY{E. Tiesinga, B.J. Verhaar, \atque H.T.C. Stoof}  \IN{Phys. Rev.
  A}{47}{1993}{4114}.

\bibitem{brya77}
 \BY{H.C. Bryant \etal }  \IN{Phys. Rev. Lett.}{38}{1977}{228}.

\bibitem{moer95res}
 \BY{A.J. Moerdijk, B.J. Verhaar, \atque A. Axelsson}  \IN{Phys. Rev.
  A}{51}{1995}{4852}.

\bibitem{voge97}
 \BY{J.M. Vogels \etal }  \IN{Phys. Rev. A}{56}{1997}{R1067}.

\bibitem{boes96}
 \BY{H.M.J.M. Boesten, J.M. Vogels, J.G.C. Tempelaars, \atque B.J. Verhaar}
  \IN{Phys. Rev. A}{54}{1996}{R3726}.

\bibitem{fedi96}
 \BY{P.O. Fedichev, Y. Kagan, G.V. Shlyapnikov, \atque J.T.M. Walraven}
  \IN{Phys. Rev. Lett.}{77}{1996}{2913}.

\bibitem{bohn97}
 \BY{J.L. Bohn \atque P.S. Julienne}  \IN{Phys. Rev. A}{56}{1997}{1486}.

\bibitem{moer96dres}
 \BY{A.J. Moerdijk, B.J. Verhaar, \atque T.M. Nagtegaal}  \IN{Phys. Rev.
  A}{53}{1996}{4343}.

\bibitem{inou98}
 \BY{S. Inouye \etal }  \IN{Nature}{392}{1998}{151}.

\bibitem{cour98fesh}
 \BY{P. Courteille \etal }  \IN{Phys. Rev. Lett.}{81}{1998}{69}.

\bibitem{abee98}
 \BY{F.A. van Abeelen, D.J. Heinzen, \atque B.J. Verhaar}  \IN{Phys. Rev.
  A}{57}{1998}{R4102}.

\bibitem{robe98}
 \BY{J.L. Roberts \etal }  \IN{Phys. Rev. Lett.}{81}{1998}{5109}.

\bibitem{abee97priv}
 \BY{F.A. van Abeelen \atque B.J. Verhaar} private communication (1997).

\bibitem{ueda98}
 \BY{M. Ueda \atque A.J. Legget}  \IN{Phys. Rev. Lett.}{80}{1998}{1576}.

\bibitem{kaga98scat}
 \BY{Y. Kagan, A.E. Muryshev, \atque G.V. Shlyapnikov}  \IN{Phys. Rev.
  Lett.}{81}{1998}{933}.

\bibitem{sack98}
 \BY{C.A. Sackett, H.T.C. Stoof, \atque R.G. Hulet}  \IN{Phys. Rev.
  Lett.}{80}{1998}{2031}.

\bibitem{tomm98}
 \BY{P. Tommasini, E. Timmermans, M. Hussein, \atque A. Kerman} preprint,
  cond-mat/9804015 (1998).

\bibitem{ho98}
 \BY{T.-L. Ho}  \IN{Phys. Rev. Lett.}{81}{1998}{742}.

\bibitem{ohmi98}
 \BY{T. Ohmi \atque K. Machida}  \IN{Journal of the Physical Society of
  Japan}{67}{1998}{1822}.

\bibitem{law98}
 \BY{C.K. Law, H. Pu, \atque N.P. Bigelow} preprint, cond-mat/9807258 (1998).

\bibitem{land77qm}
 \BY{L.D. Landau \atque E.M. Lifshitz} {\em Quantum Mechanics: Non-Relativistic
  Theory} (Pergamon Press, New York) 1977.

\bibitem{petr69}
 \BY{M.I. Petrashen \atque E.D. Trifonov} {\em Applications of Group Theory in
  Quantum Mechanics} (MIT Press, Cambridge) 1969.

\bibitem{tink64}
 \BY{M. Tinkham} {\em Group Theory and Quantum Mechanics} (McGraw-Hill, New
  York) 1964.

\bibitem{corn98jltp}
 \BY{E.A. Cornell, D.S. Hall, M.R. Matthews, \atque C.E. Wieman}  \IN{J. Low
  Temp. Phys.}{113}{1998}{151}.

\bibitem{sten98spin}
 \BY{J. Stenger \etal }  \IN{Nature}{396}{1998}{345}.

\bibitem{burk98}
 \BY{J.P. J.~Burke, C.H. Greene, \atque J.L. Bohn}  \IN{Phys. Rev.
  Lett.}{81}{1998}{3355}.

\bibitem{mies98meta}
 \BY{H.-J. Miesner \etal } preprint, cond-mat/9811161 (1998).

\bibitem{gutt53}
 \BY{L. Guttman \atque J.R. Arnold}  \IN{Phys. Rev.}{92}{1953}{547}.

\bibitem{khal57}
 \BY{I.M. Khalatnikov}  \IN{Soviet Physics JEPT}{5}{1957}{542}.

\bibitem{cols78}
 \BY{W.B. Colson \atque A.L. Fetter}  \IN{J. Low Temp. Phys.}{33}{1978}{231}.

\bibitem{sigg80}
 \BY{E.D. Siggia \atque E.A. Ruckenstein}  \IN{Phys. Rev.
  Lett.}{44}{1980}{1423}.

\bibitem{ho96bin}
 \BY{T.-L. Ho \atque V.B. Shenoy}  \IN{Phys. Rev. Lett.}{77}{1996}{3276}.

\bibitem{esry97hf}
 \BY{B.D. Esry, C.H. Greene, J.P.J. Burke, \atque J.L. Bohn}  \IN{Phys. Rev.
  Lett.}{78}{1997}{3594}.

\bibitem{gold97}
 \BY{E.V. Goldstein \atque P. Meystre}  \IN{Phys. Rev. A}{55}{1997}{2935}.

\bibitem{law97}
 \BY{C.K. Law, H. Pu, N.P. Bigelow, \atque J.H. Eberly}  \IN{Phys. Rev.
  Lett.}{79}{1997}{3105}.

\bibitem{ao98}
 \BY{P. Ao \atque S.T. Chui}  \IN{Phys. Rev. A}{58}{1998}{4836}.

\bibitem{pu98two}
 \BY{H. Pu \atque N.P. Bigelow}  \IN{Phys. Rev. Lett.}{80}{1998}{1130}.

\end{thebibliography}
\end{document}